\documentclass[a4paper,fleqn,usenatbib]{mnras}

\usepackage{newtxtext,newtxmath}

\usepackage[T1]{fontenc}
\usepackage{ae,aecompl}

\usepackage{graphicx}	
\usepackage{amsmath}	
\usepackage{amssymb}	

\usepackage{verbatim}
\usepackage[utf8]{inputenc}
\usepackage{hyperref}
\usepackage[dvipsnames]{xcolor}
\usepackage{units}
\usepackage{textcomp}
\usepackage{colortbl}

\usepackage{pgfplots}
\usepackage{filecontents}

\allowdisplaybreaks

\title[Dust survival rates for the Cas A reverse shock]{Dust survival rates in clumps passing through the Cas~A reverse shock I: results for a range of clump densities}

\author[F. Kirchschlager et al.]{Florian Kirchschlager$^{1}$\thanks{E-mail: f.kirchschlager@ucl.ac.uk}\href{https://orcid.org/0000-0002-3036-0184}{\includegraphics[trim=0cm -35cm 0cm 0cm, clip=true,width=0.27cm]{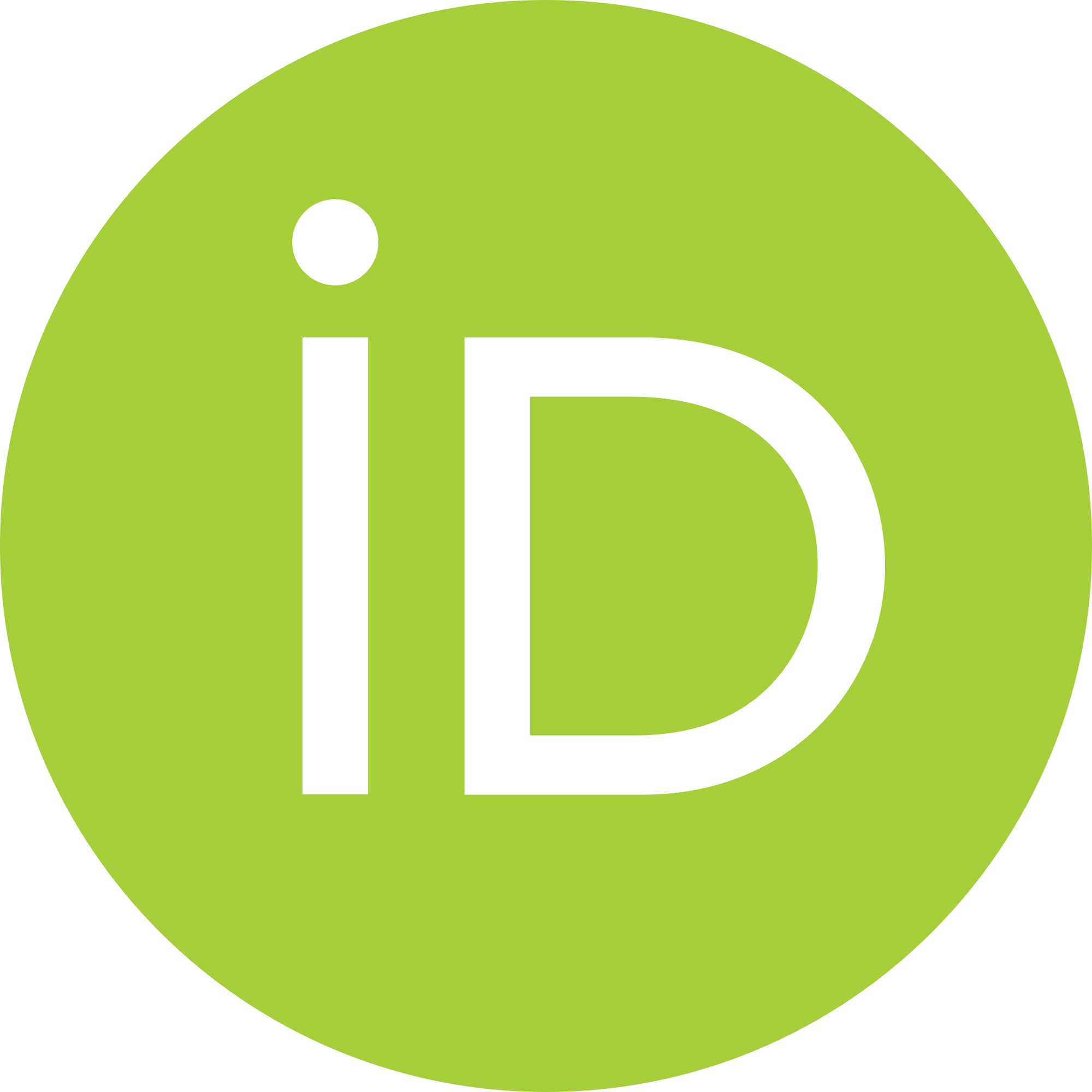}}, Franziska~D.~Schmidt$^{1}$, M.~J.~Barlow$^{1}$, Erica~L.~Fogerty$^{2}$,\newauthor Antonia Bevan$^{1}$ and  Felix~D.~Priestley$^{1,3}$\\
 $^{1}$Department of Physics and Astronomy, University College London, Gower Street, London WC1E 6BT, United Kingdom\\
 $^{2}$Center for Theoretical Astrophysics, Los Alamos National Lab, Los Alamos, NM 87545, United States\\
 $^{3}$School of Physics and Astronomy, Cardiff University, Queen's Buildings, The Parade, Cardiff, CF24 3AA, United Kingdom
}

\date{Accepted 2019 August 28. Received 2019 August 28; in original form 2019 May 13}

\pubyear{2019}

\begin{document}
\label{firstpage}
\pagerange{\pageref{firstpage}--\pageref{lastpage}}
\maketitle

 \begin{abstract}
The reverse shock  in the ejecta of core-collapse supernovae is potentially able to destroy newly formed dust material. In order to determine dust survival rates, we have performed a set of hydrodynamic simulations using the grid-based code  \textsc{\mbox{AstroBEAR}} in order to model a shock wave interacting with clumpy supernova ejecta. Dust motions and destruction rates were computed using our newly developed external, post-processing code \textsc{\mbox{Paperboats}}, which includes gas drag, grain charging, sputtering and grain-grain collisions. We have determined  dust destruction rates for the oxygen-rich supernova remnant Cassiopeia~A as a function of initial grain sizes and clump gas density. We found that up to $\unit[30]{\%}$ of the carbon dust mass is able to survive the passage of the reverse shock if the initial grain size distribution is narrow with radii around $\sim\unit[10 - 50]{nm}$ for high gas densities, or with radii around $\sim \unit[0.5-1.5]{\text{\textmu} m}$ for low and medium gas densities. Silicate grains with initial radii around $\unit[10-30]{nm}$ show survival rates of up to $\unit[40]{\%}$ for medium and high density contrasts, while silicate material with micron sized distributions is mostly destroyed. For both materials, the surviving dust mass is rearranged into a new size distribution that can be approximated by two components: a power-law distribution of small grains and a log-normal distribution of grains having the same size range as the initial distribution. Our results show that grain-grain collisions and sputtering are synergistic and that grain-grain collisions can play a crucial role in determining the surviving dust budget in supernova remnants.
 \end{abstract}

 


\section{Introduction}
\label{101}
Dust is omnipresent in the Universe and plays a key role across the  astrophysical spectrum: from galaxy evolution to star and planet formation. Yet, the origin of dust, as well as its initial physical properties remains a matter of debate. Generally, there are believed to be two main stellar production sites of cosmic dust. First, dust has been shown to form in the ejecta of supernova explosions (\citealt{Barlow2010,Gall2011,Matsuura2011,Gomez2012,Wesson2015,Bevan2017,DeLooze2017}). Second, dust is produced in the winds and outer shells of evolved stars such as asymptotic giant branch stars (AGBs; \citealt{Woitke2006, Zhukovska2008, Matsuura2009, Olofsson2010, Schneider2014, DellAgli2015, Maercker2018}). 
 
Significant quantities of dust have been observed in galaxies and quasars in the early universe (\citealt{Pettini1994, Bertoldi2003, Watson2015}).  ALMA observations have recently revealed a dusty galaxy at redshift $8.38$, emitting only $\unit[\sim200]{Myr}$ after the onset of cosmic reionisation (\citealt{Laporte2017}). 
Given the short evolution timescale of massive stars, core collapse supernovae \mbox{(CCSNe)} are assumed to be significant producers of dust in the early Universe.

It is well established that dust grains can form in the ejecta of CCSNe (e.g.~\citealt{Lucy1989, Wooden1993, Meikle1993, Bouchet1993}). Classical nucleation theory (\citealt{Kozasa1989,Schneider2004}) and the chemical kinetic approach under non-equilibrium conditions followed by subsequent coalescence and coagulation of clusters (\citealt{Cherchneff2008, Sarangi2013}) are the most common theories to form dust grains in the ejecta. However, the radii of the newly formed grains are not well determined. For a progenitor mass of $\unit[15-25]{M_\odot}$, carbon grains are predicted to have radii of $\unit[\sim1-150]{nm}$, forsterite grains to have a size of $\unit[\sim1-30]{nm}$, and MgSiO${}_3$, Mg${}_2$SiO${}_4$, and SiO${}_2$ grains $\unit[\sim0.5-100]{nm}$ (\citealt{Todini2001, Nozawa2003, Bianchi2007, Bocchio2014, Marassi2015, Sarangi2015,Biscaro2016}).  On the other hand, the dust grain radii derived from modelling infra-red continuum observations as well as by modelling the red-blue asymmetries of SN optical line profiles are of the order of $\unit[0.1]{\text{\textmu} m}$ up to a few micrometres (\citealt{Stritzinger2012,Gall2014,Owen2015,Fox2015,Wesson2015,Bevan2016,Bevan2017, Priestley2019b}). Therefore, grain size ranges of up to three to four orders of magnitude should be considered when studying dust in the ejecta of supernova remnants (SNRs). It is commonly assumed that the dust grains formed in over-dense gas clumps in the ejecta instead of a uniform distribution of dust residing in a homogeneous ejecta medium (\citealt{Lagage1996, Rho2008, Lee2015}).  
  
While supernovae (SNe) can be significant producers of dust, a large fraction of the dust can potentially be destroyed by the reverse shock. Moreover, the forward shock can trigger the destruction of interstellar dust grains. The net dust survival rate is crucial for determining whether or not SNe significantly contribute to the dust budget in the interstellar medium (ISM). This is in particular important for galaxies in the early Universe where large amounts of dust have been observed and where CCSNe are assumed to be significant dust producers. In this paper, we focus on the dust survival rate in the reverse shock. For supernova triggered shock waves in the ISM we refer to the studies of \cite{Nozawa2006}, \cite{Bocchio2014} and \cite{Slavin2015}.

Several previous studies have investigated the dust survival rate in the reverse shock for a wide range of conditions. \cite{Nozawa2007} found that depending on the energy of the explosion, between 0 and $\unit[80]{\%}$ of the initial dust mass can survive. The survival rate derived by \cite{Bianchi2007} was between 2 and $\unit[20]{\%}$, depending on the density of the surrounding ISM. \cite{Nath2008} found a survival rate between 80 and $\unit[99]{\%}$, \cite{Silvia2010} between 0 and $\unit[99]{\%}$, depending on the shock velocities and on the grain species, \cite{Biscaro2016} between 6 and $\unit[11]{\%}$, \cite{Bocchio2014} $\unit[1-8]{\%}$, and  \cite{Micelotta2016}  $\unit[10-13]{\%}$ and $\unit[13-17]{\%}$ for silicate and carbon dust, respectively. The different survival rates show a wide diversity and emphasize the strong dependence on initial dust properties such as grain size and the dust material. Furthermore, the survival rate depends on properties of the ejecta such as the shock velocity and the gas densities in the clumps (\citealt{Biscaro2016}).

In this paper, we focus on the effect of different ejecta clump gas densities and on the requirements for the initial dust  properties to enable the survival of a significant fraction of the dust mass. We have developed the code \textsc{\mbox{Paperboats}} to study the processing of dust grains in a SNR. Unlike many other studies, both sputtering \textit{and} grain-grain collisions are considered as destruction processes, providing a more complete picture of the dust evolution in SNRs. We perform hydrodynamical simulations followed by dust post-processing to calculate the dust destruction in Cassiopeia~A (Cas~A), a dusty SNR that has been studied extensively (e.g. \citealt{Dwek1987a, Lagage1996, Gotthelf2001, Fesen2006,  Rho2008, Barlow2010, Arendt2014, Micelotta2016, DeLooze2017}) and which provides a unique laboratory to investigate the destruction of dust by a reverse shock. 

Cas~A has a highly clumped structure (e.g.~\citealt{Milisavljevic2013}), with most of its dust mass located in central regions that have yet to encounter the reverse shock (\citealt{DeLooze2017}). The survival prospects  of this dust when it encounters the reverse shock is of significant interest -- to assess these prospects we have chosen to model the dust destruction processes for a single represenative clump, with typical clump parameters drawn from the work of \cite{Docenko2010}, \cite{Fesen2011} and \cite{Priestley2019a}.

This paper is the first of a series aiming to understand the influence of various properties of the ejecta on the dust destruction rate and to quantify dust masses and grain sizes that are able to survive this ejecta phase. The present paper is organised as follows: In Section~\ref{seccasa} the Cas~A SNR and its properties are introduced. In Section~\ref{sec_hydro} we describe the hydrodynamical simulations that have been performed to simulate the reverse shock impacting an over-dense clump of gas and dust in the ejecta. Section~\ref{secpaper} describes the dust physics required to achieve our scientific goals and used in our post-processing code: the dust advection by collisional and plasma drag are outlined, the comprehensive models for grain-grain collisions and for sputtering are presented, as well as the grain charge calculation is described. We then conduct simulations for different gas densities in the clumps and different initial dust properties and present the results along with computed dust survival rates in Section~\ref{sec_results}. After a detailed comparison of our results with that of previous studies in Section~\ref{sec_compar}, we conclude with a summary of our findings in Section~\ref{sec_conc}.

\section{The supernova remnant Cas A }
\label{seccasa}
Cas~A is a Galactic remnant of a SN explosion of a massive progenitor star $\sim340-350$ years ago, at a distance of $\unit[3.4]{kpc}$ and with a radius of $\unit[1.7]{pc}$ (\citealt{Reed1995, Thorstensen2001, Fesen2006}). 
Based on spectra of optical light echoes, it was classified as a hydrogen-poor type IIb core-collapse SN (\citealt{Krause2008}) with an explosion energy of $\unit[(1-4)\times10^{51}]{erg}$ (\citealt{Willingale2003, Laming2003}). The main-sequence mass of the progenitor is estimated to be $\unit[15-25]{M_\odot}$ and the mass at explosion to be $\unit[4-6]{M_\odot}$ (\citealt{Young2006}). The stellar wind of the progenitor formed circumstellar (CS) material (\citealt{Hwang2009}) into which the SN explosion has driven a forward shock, sweeping up the CS material ($\unit[8-10]{M_\odot}$; (\citealt{Borkowski1996, Chevalier2003, Favata1997, Vink1996,Willingale2003}) and generating a reverse shock (\citealt{McKee1974,Truelove1999}).  

The SN ejecta has been estimated to have a total mass of $\unit[2-4]{M_\odot}$ (\citealt{Borkowski1996, Chevalier2003, Favata1997}), mostly composed of oxygen (\citealt{Chevalier1979, Willingale2002}). Observations reveal a complex structure in the ejecta (e.g. \citealt{Ennis2006, Smith2009, Milisavljevic2013}) with material covering a wide range of densities and temperatures. Dense gas clumps and knots are observed which are associated with the location of freshly produced dust material (\citealt{Lagage1996, Arendt1999, Hines2004, Rho2008, Rho2009, Rho2012}). The total dust mass in the ejecta has been derived by different observations and strategies to be $\unit[\sim1]{M_\odot}$ (\citealt{Dunne2009}), $\unit[\sim0.06]{M_\odot}$ (\citealt{Sibthorpe2010}), and (post-\textit{Herschel}) $\unit[\sim0.075]{M_\odot}$ (\citealt{Barlow2010}), $\unit[0.3-0.6]{M_\odot}$ (\citealt{DeLooze2017}), $\unit[\sim1.1]{M_\odot}$ (\citealt{Bevan2017}), and $\unit[\sim0.6]{M_\odot}$ (\citealt{Priestley2019a}), while a theoretical study of dust formation and evolution in Cas~A predicted masses of the order of $\unit[0.08]{M_\odot}$ (\citealt{Nozawa2010}). In order to be released by the SN and to contribute to the dust budget of the ISM, the dust material has to survive the passage of the reverse shock. To simulate a shock wave impacting on an ejecta clump composed of gas and dust, several physical parameters are required that are given in the next two sections.
 
\subsection{Reverse shock and ejecta properties}
\label{sec_casa_prop}
X-ray observations by \cite{Gotthelf2001} have resolved the radius of the reverse shock to be $R_\text{rev} = \unit[1.57 \pm 0.17]{pc}$ ($\unit[95]{''}\pm\unit[10]{''}$). The relative velocity between the reverse shock and the ejecta is constrained to $\unit[1000-2000]{km/s}$ (\citealt{Laming2003,Morse2004,Docenko2010}) while \cite{Micelotta2016} derived $\unit[1586]{km/s}$. The over-dense clumps in the ejecta with pre-shock gas density $\unit[20-1000]{cm^{-3}}$ (\citealt{Sutherland1995,Docenko2010, Silvia2010, Silvia2012, Biscaro2014, Biscaro2016, Micelotta2016}) are embedded in an ambient (inter-clump) medium with a pre-shock gas density of $\unit[0.1-10]{cm^{-3}}$ (\citealt{Borkowski1990,Morse2004, Nozawa2003, Micelotta2016}). The observed clump radii are in the range $\unit[(0.5-2.5)\times10^{16}]{cm}$ (\citealt{Fesen2011}), significantly larger than the knots located outside the ejecta, at or ahead of the forward shock front (\citealt{Fesen2006,Hammell2008}). The ambient medium and the clump gas abundances are dominated by oxygen (\citealt{Chevalier1979, Willingale2002, Docenko2010}). The gas-to-dust mass ratio in the clumps is $5-10$ as derived from modelling of the dust continuum emission (\citealt{Priestley2019a}).

\subsection{The electron density in Cas~A}
\label{sec_Electrondensity}
\begin{figure}
 \centering
\includegraphics[trim=2.6cm 2.5cm 2.15cm 2.5cm, clip=true,page=1,width=1.0\linewidth]{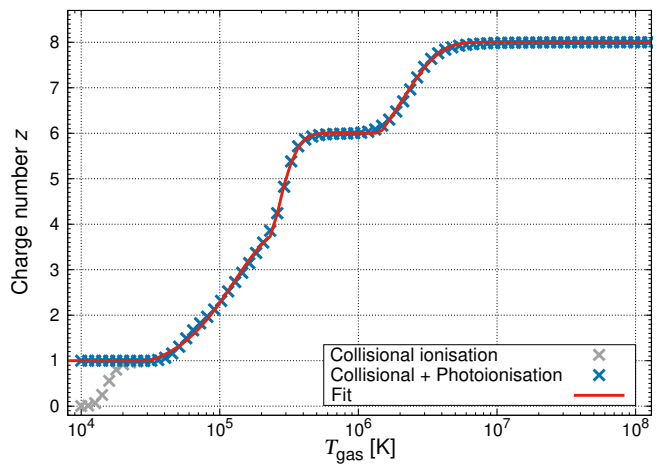}
\caption{Charge number $z$ as a function of gas temperature $T_\text{gas}$ for pure oxygen. The grey points show $z$ calculated with CHIANTI (only collisional ionisation), while the blue points consider also photoionisation, setting a lower limit of $z=1$. The red solid line is a fit to the collisional ionisation and photoionisation data (equations~\ref{f_1}-\ref{f_all}).}
\label{Chianti} 
\end{figure}
The electron density $n_\text{e}$ in a gas depends on the average charge number $z$ of the gas particles. For a collisionally ionised, pure oxygen gas, the charge number is calculated as a function of gas temperature using CHIANTI\footnote{\href{http://www.chiantidatabase.org/}{http://www.chiantidatabase.org/}}, a database of assessed atomic parameters and transition rates needed for the calculation of the line and continuum emission of optically thin, collisionally-dominated plasma (\citealt{DelZanna2015}; Fig.~\ref{Chianti}). The gas is mostly neutral ($z=0$) for temperatures below  $\unit[\sim10^4]{K}$ and fully ionised for $T_\text{gas}\gtrsim\unit[6\times10^6]{K}$, with O$^{8+}$ as the dominant gas species at these temperatures. Photoionisation by shock-emitted radiation can become important for temperatures around (and below) $\unit[2\times10^4]{K}$, however, CHIANTI considers only collisional ionisation. Therefore, a lower limit of $z=1$ is adopted to take this into account. Our charge numbers are similar to the values obtained by \cite{Sutherland1995}\footnote{If a misprint for the ion spectroscopic symbols is considered in Fig.~3 of \cite{Sutherland1995}, as suggested by \cite{Docenko2010}.}, \cite{Boehringer1998} and \cite{Docenko2010}. The dominant oxygen ion in Cas~A is predicted to be O$^{+}$ (\citealt{Priestley2019a}).
\begin{table}
\centering
\caption{Parameters for the analytic function to calculate the charge number $z$ of the gas particles (equations~\ref{f_1}-\ref{f_all}) considering collisional ionisation and photoionisation.}
\begin{tabular}{ l l l}
\hline\hline
$\phantom{c\,c}f_1$&$\phantom{c\,c}f_2$&$\phantom{c\,c}f_3$\\\hline
$c_1=3.345$&$c_4= 7.213$&$c_7=3.245$\\
$c_2= 2.91\text{E-29}$&$c_5=1.6\text{E-30}$&$c_8=1.084\text{E-21}$\\
$c_3=24.531$&$c_6= 27.353$&$c_9=19.567$\\\hline
\end{tabular}
\label{Para_list_ion}
\end{table}

The charge number is calculated for each cell and at each time-step of the simulation as a function of temperature. To reduce calculation times, we fit three exponential functions to the CHIANTI data set using a least squares approximation and obtain an analytical expression for the charge number as a function of the gas temperature $T_\text{gas}$:
\begin{align}
 f_1(T_\text{gas}) =& \max{\left(0, 2 - c_{1} \exp{\left[-c_2 (\ln(T_\text{gas}/\text{K}))^{c_3}\right]}\right)}\label{f_1},\\
 f_2(T_\text{gas}) =& \max{\left(0, 2 - c_{4} \exp{\left[-c_{5} (\ln(T_\text{gas}/\text{K}))^{c_{6}}\right]}\right)},\\
 f_3(T_\text{gas}) =& \max{\left(1, 4 - c_{7} \exp{\left[-c_8 (\ln(T_\text{gas}/\text{K}))^{c_{9}}\right]}\right)},\\[0.3cm] 
 z(T_\text{gas}) =&\, f_1(T_\text{gas}) + f_2(T_\text{gas}) + f_3(T_\text{gas})   \label{f_all}.
\end{align}
The nine fitting parameters $c_{i},\,i\in\mathbb{N}_{\le 9}$, are listed in Table~\ref{Para_list_ion}. Finally, the electron density is calculated for each cell and at each time-step as $n_\text{e}=z(T_\text{gas})\,n_\text{gas}$, where $n_\text{gas}$ the number density of the gas (oxygen ions).

\section{Hydrodynamical setup}
\label{sec_hydro}

In this section, we  describe the set of initial conditions used to simulate the dynamical evolution of a SNR reverse shock impacting a clump of ejecta material. For this purpose, the hydrodynamic code \textsc{\mbox{AstroBEAR}}\footnote{\href{https://www.pas.rochester.edu/astrobear/}{https://www.pas.rochester.edu/astrobear/}} (\citealt{Cunningham2009, Carroll-Nellenback2013}) was employed, a highly parallelised, multidimensional adaptive mesh refinement code designed for astrophysical contexts. It solves the conservative equations of hydrodynamics and magnetohydrodynamics on a Cartesian grid and includes a wide-range of multiphysics solvers. \textsc{\mbox{AstroBEAR}} is well tested (see for example \citealt{Poludnenko2002, Cunningham2009, Kaminski2014, Fogerty2016,Fogerty2017}), is under active development, and is maintained by the University of Rochester's computational astrophysics group. 

The \textsc{\mbox{AstroBEAR}}  simulations model only the gas phase of the ejecta environment. For the current analysis of dust destruction by the reverse shock, dust advection and processing have been handled externally, utilizing the density, velocity and temperature fields given by the hydrodynamical simulations (see Section~\ref{secpaper}). 

\subsection{Model setup}
\label{sec_mod_setup}
In order to investigate the temporal and spatial ejecta evolution when the reverse shock passes through the SNR, two different approaches exist: The first one examines the entire three-dimensional remnant in which the shock impacts the ejecta material, including over-dense gas and dust clumps, and the second investigates a section of the remnant in which one or several clumps are impacted by the reverse shock. While the first approach is able to explore the global evolution of the remnant, the second has the advantage of being able to investigate the destruction of the clumps at higher resolution.  As we are interested in the evolution of the dust, which might be highly affected by the local gas density distribution, we pursue here the simulation of a section of the remnant. This kind of problem is called a cloud-crushing scenario (\citealt{Woodward1976}) and was already applied by \cite{Silvia2010,Silvia2012} to investigate dust survival in SNRs.

\begin{figure}
 \centering
\includegraphics[trim=0.3cm 6.3cm 0.3cm 6.3cm, clip=true,page=1,width=1.0\linewidth]{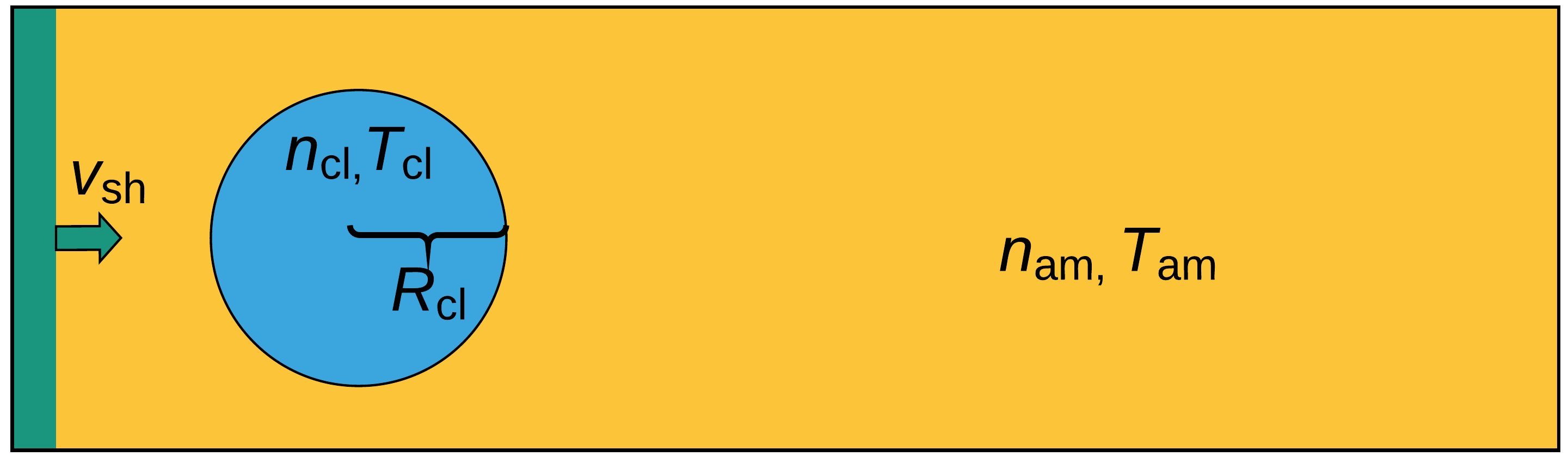}
\caption{Sketch of the cloud-crushing model and its relevant initial parameters: reverse shock (left, green), impacting on an over-dense clump of gas (blue) embedded in a low-density gaseous medium (yellow).}
\label{sketch_clump} 
\end{figure}

In our particular problem a planar shock is driven into an over-dense clump of gas which is embedded in a low-density gaseous medium (Fig.~\ref{sketch_clump}). At the beginning of the simulation, the ambient medium has a number density $n_\text{am}=\unit[1]{cm^{-3}}$ of gas particles (oxygen) and a temperature $T_\text{am}=\unit[10^4]{K}$. The embedded clump has a spherical shape with radius $R_\text{cl} = \unit[10^{16}]{cm}\approx\unit[668.5]{au}$, a uniform gas number density of $n_\text{cl}=\chi n_\text{am}$, and a temperature of $T_\text{cl}=\unit[10^2]{K}$. We vary the initial density contrast $\chi=n_\text{cl}/n_\text{am}$, adopting $\chi=100,200,300,400,600$ and 1000. For $\chi=100$, clump and ambient medium are in pressure equilibrium. The shock velocity in the ambient medium is adopted to be $v_\text{sh} = \unit[1600]{km/s}$ following the analytical result of \cite{Micelotta2016}. The shock velocity in the ambient medium is fixed for each simulation, independent of the density contrast\footnote{The value of $\unit[1600]{km/s}$ corresponds to the shock velocity in the ambient medium, while the velocity is decelerated in the over-dense clump to $\sim\!\chi^{-0.5} \unit[1600]{km/s}$.}. The mean molecular weight of the pre-shock gas is set to \mbox{$\mu=16.0$,} corresponding to a pure oxygen gas, and the adiabatic exponent is $\gamma_\text{hydro}=5/3$. 

The presented parameters are consistent with a clump and the reverse shock in Cas~A as outlined in Section~\ref{seccasa}. For a density contrast of $\chi=100$ ($\chi=1000$), the dust mass in a single clump amounts to $\unit[\sim5.3\times10^{-7}]{M_\odot}$ ($\sim\unit[5.3\times10^{-6}]{M_\odot}$). In order to obtain a total dust mass of $\unit[\sim0.6]{M_\odot}$ as derived from modelling of thermal infra-red emission of Cas~A (\citealt{DeLooze2017, Priestley2019b}), at least $\sim10^6$ ($\sim10^5$) of these clumps located in the ejecta are needed. The impact of the reverse shock on a single clump as simulated in our study is assumed to happen for all the ejecta clumps so that our results can be applied and projected to them.

Amongst crucial parameters for the simulation of the cloud-crushing scenario are the size of the computational domain and the simulation time. At the beginning of the simulation ($t=0$), the clump midpoint is placed at a distance of $2\,R_\text{cl}$ in front of the shock front to ensure that material swept up by the bow shock (after the first contact of the shock with the clump) and temporarily transported in the direction contrary to the shock propagation can stay in the domain. The simulation is executed for a time $3\,\tau_\text{cc}$ after the first contact of the shock with the clump, where
\begin{align}
\tau_\text{cc}=\chi^{0.5}R_\text{cl}/v_\text{sh} \label{cloudcrushingtime}                                                                                                                                                                                                                                                                                                                                                                                                                                                                                                                                                                                                                                                                                                                                                                                                                                                                                                                                                                                                                                                                                                                                \end{align}
is the cloud-crushing time as defined by \cite{Klein1994} which gives the characteristic time for the clump to be crushed by the shock. $3\,\tau_\text{cc}$ is a commonly used value to investigate post-shock structures. In total, the simulation time amounts to $t_\text{sim}=R_\text{cl}/v_\text{sh}+ 3\tau_\text{cc}=(3\chi^{0.5}+1)R_\text{cl}/v_\text{sh}$. The simulation time for $\chi=100$ is then $\sim \unit[61.5]{yr}$ which is roughly $\unit[20]{\%}$ of the total age of Cas A. 

The Rankine-Hugoniot jump conditions constrain the post-shock gas velocity to be $3/4\,v_\text{sh}$. Dust grains in the clump can move, at most, with the gas velocity. In order to ensure that the dust does not flow out of the domain at the back end during the simulation time $t_\text{sim}$, the length of the domain has to be at most \mbox{$(3/4\,v_\text{sh}) (t_\text{sim} -3R_\text{cl} / v_\text{sh})+3\,R_\text{cl}=3/4 R_\text{cl} (3\chi^{0.5}+2)$.}  Test simulations showed that using \mbox{$l_\text{box}= 3/4 R_\text{cl} (3\chi^{0.5}-2)$} as the length of the domain, as well as \mbox{$w_\text{box}= l_\text{box}/3$} as the domain width (perpendicular to the shock propagation),  are sufficient to keep the dust in the domain. Typical values are $l_\text{box}=21\,R_\text{cl} = \unit[0.068]{pc}$
and $w_\text{box}= 7\,R_\text{cl} = \unit[0.023]{pc}$ for $\chi=100$.  
 
In principle, the hydrodynamical simulations as well as the dust post-processing can be conducted in 1D, 2D, or 3D. 
However, in this paper we consider only 2D simulations\footnote{The clump has a circular shape in 2D.} due to the large computational effort for highly resolved 3D post-processing simulations. The computational domain consists of $420\times140$ cells such that there are 20 cells per clump radius. This yields a physical resolution  of $\Delta_\text{cell}=\unit[5\times10^{14}]{cm}$ ($\sim\unit[33]{au}$) per cell (for $\chi=100$).  Outflow boundary conditions are used on all sides of the domain, with the exception of the lower x-boundary, which used an inflow boundary for injecting a continuous post-shock wind into the domain. Since the shock width (parallel to the shock direction) is much larger than the clump radius, the shock is generated by the constant inflow of material. The \mbox{Harten-Lax-van~Leer} method  (HLL; \citealt{Harten1983}) is used by \textsc{AstroBEAR} to solve the hydrodynamic equations. Note that magnetic fields are ignored in this work and will be examined in a future work. 

%
 

\subsection{Gas cooling}
For most of our hydrodynamical simulations, radiative cooling is considered. The cooling function $\Lambda$ is equal to the total emitted power divided by the product of the ion and electron number densities and is calculated using CHIANTI (\citealt{DelZanna2015}) for a gas of pure oxygen in ionisation equilibrium in the temperature range $T_\text{gas}=\unit[10^4{-}10^9]{K}$. 

\begin{figure}
 \centering
\includegraphics[trim=2.4cm 2.5cm 2.1cm 2.5cm, clip=true,page=1,width=1.0\linewidth]{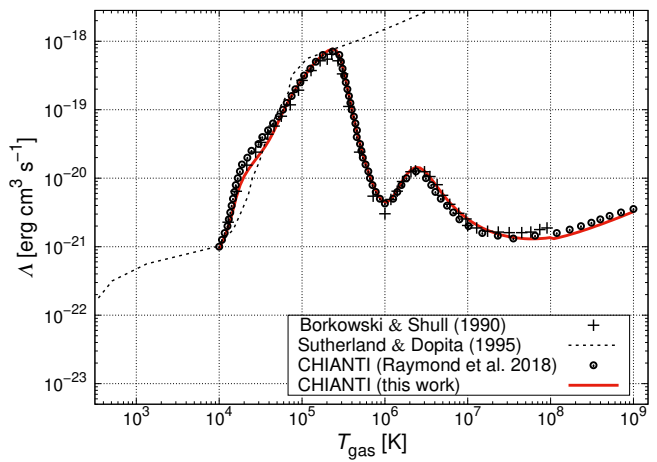}
\caption{Cooling function $\Lambda$ for a gas of pure oxygen as a function of gas temperature $T_\text{gas}$ under the assumption of collisional ionisation equilibrium (red solid line). For comparison we show the oxygen-rich cooling functions of \mbox{\citet{Borkowski1990}} (crosses), \citet{Sutherland1995} (dashed line), and \citet{Raymond2018} (circles).}
\label{Cooling_fct} 
\end{figure}

The calculated cooling function (Fig.~\ref{Cooling_fct}) shows a drop between $\unit[2\times 10^5]{K}$ and $\unit[10^6]{K}$, caused by the dominant O$^{6+}$ and O$^{7+}$ ions which have no easily excitable electrons. This can be also seen in the plateau of the 6th charge number in the same temperature range (Fig.~\ref{Chianti}). The cooling at lower temperatures is dominated by line emission and at higher temperatures by collisional ionisation while the increasing slope at the highest temperatures is given by bremsstrahlung emission plus contributions from radiative recombination (\citealt{Raymond2018}). We find good agreement over the whole temperature range between our cooling function and the oxygen-dominated values computed by \cite{Raymond2018} who also used CHIANTI, as well as with that of \cite{Borkowski1990}. To our knowledge the only other available data for oxygen-rich shocked gas is from \cite{Sutherland1995} who calculated the cooling function in self-consistent shock models for a shock velocity of $\unit[150]{km/s}$ (within the clump). Their values show good agreement with our function for temperatures between $T_\text{gas}=\unit[10^4]{K}$ and the first peak at $\unit[2\times 10^5]{K}$, however, their values and the \mbox{CHIANTI} results diverge widely above $\unit[2\times 10^5]{K}$. Since their calculated cooling function also covers lower temperatures, we adopt it for $T_\text{gas}<\unit[10^4]{K}$ while we use the \mbox{CHIANTI} results for higher temperatures.

We note that gas cooling due to thermal emission of the dust grains (\citealt{Dwek1987, Hirashita2015}), which are embedded in and can be heated up by the gas, is not considered due to the nature of the dust post-processing.

\section{The new external dust-processing code \textsc{\mbox{Paperboats}}}
\label{secpaper}
To investigate dust advection and dust destruction, as well as potential dust growth in a gas, we have developed the parallelised 3D external dust-processing code \textsc{\mbox{Paperboats}}. \textsc{\mbox{Paperboats}} utilises the time- and spatially-resolved density, velocity and temperature output of the grid-based hydrodynamical code \textsc{AstroBEAR} to calculate the spatial distribution of the dust particles\footnote{In this study, the hydrodynamical simulations are in 2D, however, it is important to consider grain-grain collisions in 3D as this will affect the grain cross sections and collision probabilities. The 2D hydro simulations are extended here to 3D assuming a single cell in the z-direction as well as no gas velocity in the z-direction.}. It makes use of an approach we have called ``dusty-grid approach'' and where the dust location is discretised to spatial cells in the domain. The dust mass (partially) moves to (an)other cell(s) in a discretised time-step according to the gas conditions (density, velocity and temperature). Furthermore, the dust in each cell is apportioned in different grain size bins for each dust material species. The dust grains can move both spatially as well as between the grain size bins as a result of dust destruction or growth during a time-step.  Due to the nature of the post-processing, the dust medium can not alter the state of the surrounding gas medium and no feedback is considered. However, in Section~\ref{binning} we will introduce a ``dusty gas'' (gas particles from the grains) that is composed of the solid dust material which was destroyed by sputtering, vaporisation or grain shattering.

In this section, the code \textsc{\mbox{Paperboats}} is introduced, with the implementation of the dusty-grid approach and the comprehensive dust physics described in detail. Section~\ref{grains} covers the initial grain size distribution and location of the dust grains. The grid and size bins of the dust grains are presented in Section~\ref{binning} and the dust acceleration by gas and plasma drag is described in Section~\ref{advec}. Finally, the processes of grain charging, sputtering and  grain-grain collisions are outlined in Sections~\ref{sec_char}--\ref{sec_sput}.

\subsection{Initial dust grain size distribution and gas-to-dust mass ratio}
\label{grains}
We assume that the dust is located in over-dense clumps in the ejecta. In our model, we initially assume a homogeneous dust distribution within the clump, while the ambient medium is dust-free. The gas-to-dust mass ratio in the clump is set to $\Delta_\text{gd}=10$ which was obtained from SED modelling of the infra-red continuum emission of Cas A (\citealt{Priestley2019a}).

\begin{figure}
 \centering
\includegraphics[trim=1.85cm 1.85cm 1.65cm 0.8cm, clip=true,page=1,width=1.0\linewidth]{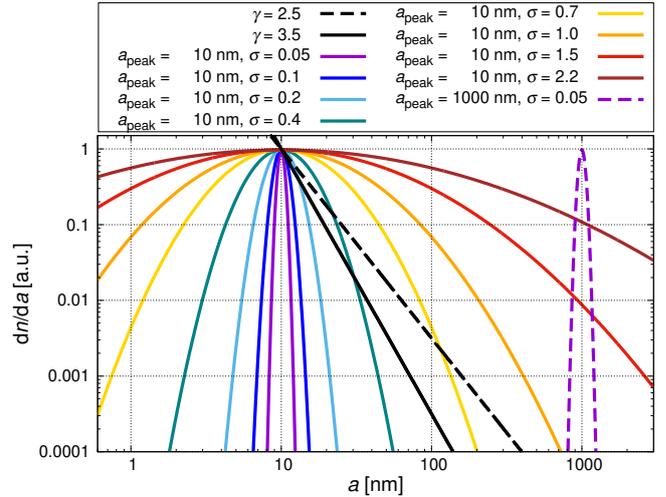}
\caption{Different initial dust distributions (log-normal and power-law), normalized to 1 at $a=\unit[10]{nm}$. The parameters $a_\text{peak}$ and $\sigma$ describe the peak radius and the width, respectively, of the log-normal distribution, and $\gamma$ the slope of the power-law distribution.}
\label{init_dust_fig} 
\end{figure}

The dust grains are assumed to be compact, homogeneous and spherical with radius $a$, material density $\rho_\text{bulk}$ and mass $m = \frac{4\,\pi}{3}a^3\,\rho_\text{bulk}$. The number of dust grains with radii between $a$ and $a+\text{da}$ is denoted as $\tilde{n}(a)\,\text{d}a$ and is defined between a minimum and maximum dust grain size $a_\text{min}$ and $a_\text{max}$, respectively.  We investigate the following size distributions:
\begin{align}
&\text{Power-law distribution:\hspace*{0.2cm}}\nonumber\\
&\hspace*{0.9cm}\tilde{n}(a)\,\text{d}a  \,\propto\, a^{-\gamma}\,\text{d}a,\label{120519}\\
&\text{Log-normal distribution:} \nonumber\\
&\hspace*{0.9cm}\tilde{n}(a)\,\text{d}a  \,\propto\,  \frac{1}{a}\exp{\left(-\frac{\left[\ln{\left(a/a_\text{peak}\right)}-\sigma^2\right]^2}{2\,\sigma^2}\right)}          \,\text{d}a,
\end{align}
where $\gamma$ is the grain size exponent that is usually between 2 and 4 (e.g.\,\citealt{Dohnanyi1969,Jones1996}). For the log-normal distribution, $a_\text{peak}$ is the grain radius at the maximum of the distribution and $\sigma$ is the parameter that defines the width of the distribution, so that for increasing $\sigma$, the width is increasing\footnote{It should be noted, that our choice of parameters describing the log-normal distribution, $a_\text{peak}$ and $\sigma$, is slightly different from \cite{Bocchio2012,Bocchio2014}, where the parameters are $a_0$ and $\sigma$ with a different definition. Moreover, we note that $\frac{\text{d}n}{\text{d}a} = \frac{1}{a}\,\frac{\text{d}n}{\text{d}\ln{(a)}}$.} (see Fig.~\ref{init_dust_fig}).

\textsc{\mbox{Paperboats}} enables one to model silicate and carbon grains individually or simultaneously, with different proportions, size distributions and minimum and maximum grain radii for each material. The material parameters required for the dust post-processing are given in Table~\ref{mat_para}.
\begin{table*}
\caption{Properties of the carbon (C) and silicate (MgSiO$_3$) dust components: bulk density $\rho_\text{bulk}$, vaporisation and fragmentation threshold velocities $v_\text{vapo}$ and $v_\text{frag}$, speed of sound $c_0$ in the grain, the dimensionless fragmentation constant $s$, the critical pressures for vaporisation $P_{\text{v}}$ and fragmentation $P_{\text{l}}$, the surface energy per unit area  $\gamma_\text{A}$, Young's modulus $E_\text{Y}$, Poisson's ratio $\nu_\text{Poi}$, the surface binding energy $U_0$, the average atomic number and mass of the grain atoms, $\left\langle Z_\text{atom}\right\rangle$  and $\left\langle M_\text{atom}\right\rangle$, and the dimensionless sputtering constant $k_\text{sput}$. The density of $\unit[2.2]{g\,\text{cm}^{-3}}$ for carbonaceous grains corresponds to graphite (see e.g. \citealt{Bocchio2014}). In this study, we will use the terms carbon and graphite synonymously.}
\begin{tabular}{l c c c c c c c c}
\hline\hline\\[-0.25cm]
         &$\rho_\text{bulk}\,[\text{g}\,\text{cm}^{-3}]$& $v_\text{vapo}\,[\unit[]{km\,s^{-1}}]$&$v_\text{frag}\,[\unit[]{km\,s^{-1}}]$&$c_0\,[\unit[]{km\,s^{-1}}]$  &$s$          & $P_{\text{v}}\,[\unit[]{kg\,m^{-1}s^{-2}}]$&$P_{\text{l}}\,[\unit[]{kg\,m^{-1}s^{-2}}]$\\[0.05cm]\hline
 carbon	 & $2.2\,(a)$                     		& $23\,(b)$				& $1.2\,(c)$			       &$1.8\,(b)$		      &$ 1.9\phantom{0}\,(b)$&$580\times10^9{}\,(b)$		     &$\phantom{0}4\times10^9{}\,(c)$			     \\
 silicate& $3.3\,(b)$                     		& $19\,(b)$ 				& $2.7\,(c)$			       &$5.0\,(b)$		      &$1.23\,(b)$&$540\times10^9{}\,(b)$                    &$30\times10^9{}\,(c)$			     \\[0.2cm] 
         &$\gamma_\text{A}\,[\unit[]{kg\,s^{-2}}]$&$E_\text{Y}\,[\unit[]{kg\,m^{-1}s^{-2}}]$&$\nu_\text{Poi}$&  $U_0\,[\unit[]{eV}]$&$\left\langle Z_\text{atom}\right\rangle$&$\left\langle M_\text{atom}\right\rangle\,[m_\text{amu}]$&$k_\text{sput}$\\[0.05cm]\hline
 carbon	 &$0.075\,(a)$		             &$10\times10^9{}\,(a)$		       &$0.32\,(a)$    &  $4.0\,(b)$		& $\phantom{0}6\,(b)$				   &$12\,(b)$					    &$0.65\,(b)$\\
 silicate&$0.025\,(a)$		             &$54\times10^9{}\,(a)$		       &$0.17\,(a)$    &  $5.7\,(b)$		& $10\,(d)$					  &$20\,(d)$					    &$ 0.1\phantom{0}\,(b)$\\\hline
\end{tabular}\\[0.15cm]
Note: $(a)$ \cite{Chokshi1993} (silicate: quartz), $(b)$ \cite{Tielens1994,Jones1994},\\  $(c)$ \cite{Jones1996},  $(d)$ \cite{Nozawa2006} 
\label{mat_para}
\end{table*}

Dust destruction processes such as shattering or grain growth by gas accretion can produce dust grains which are smaller than the minimum $a_\text{min}$ or larger than the maximum dust grain size $a_\text{max}$ of the initial distribution. For this reason, absolute values for the minimum and maximum grain radius, $a_\text{min,abs}$ and $a_\text{max,abs}$, are defined. The question of the size of the smallest possible dust grain is philosophical as there is a smooth transition between solid grains and molecules/atoms. We set $a_\text{min,abs}=\unit[0.6]{nm}$. Carbon (silicate) grains of this size contain 100 atoms (78 averaged\footnote{Silicate dust is composed of several elements (Si, Mg and O) and the term ``averaged atom'' denotes that the mass-weighted mean of these elements regarding their abundance is taken.} atoms) which depicts an appropriate minimum size similar to that of fullerenes (e.g. C$_{60}$, buckminsterfullerene). For comparison, \cite{Silvia2010} used $\unit[0.5]{nm}$ as the minimum dust grain radius. The maximum grain radius $a_\text{max,abs}\ge a_\text{max}$ is adjusted for each simulation to ensure simultaneously a high bin size resolution, a limited number of grain sizes (computational effort) and the opportunity to investigate dust growth effects.

\subsection{The dust grain size bins}
\label{binning}
For the numerical calculations, $N_\text{grain}$ discrete log-spaced bins are considered for the grain radius, where bin \mbox{$i\in\mathbb{N}_{\le N_\text{grain}}$} contains dust grains with radius 
\begin{align}
a_i = a_\text{min,abs}\,\Delta_a^{i-1},\label{eq_177}
\end{align}
and $\Delta_a$ specifies the width of the bins and is given as
\begin{align}
\Delta_a = \begin{cases}
           (a_\text{max,abs}/a_\text{min,abs})^{1/(N_\text{grain}-1)}\hspace{0.3cm}\text{if}\hspace{0.1cm}N_\text{grain}\in \mathbb{N}_{ \ge 2},\\
                                                                        1\hspace{4.51cm}\text{if}\hspace{0.1cm}N_\text{grain}=1.
          \end{cases}
\end{align}
A grain in the $i$-th bin has mass $m_i=4/3\pi a_i^3\rho_\text{bulk}$ and will be referred to as `grain $i$'. Increasing the number of bins leads to a size distribution that is defined with more and more precision, but the computing time roughly increases as the square of the number of bins if grain-grain collisions are evaluated. In this paper, $N_\text{grain}$ is set to 40 which has proven to be sufficient in previous studies (cf. \citealt{Hirashita2009} (40~bins),  \citealt{Bocchio2014} (9, 15 and 25~bins)). Furthermore, two additional bins are defined: Due to e.g. fragmentation, dust grains with radii below $a_1$ can be produced, or a dust particle can be completely destroyed by, e.g., vaporisation. To take into account these small grains or completely destroyed, obliterated dust masses, an additional bin $i=0$ is defined for each cell, the so-called ``collector bin", which represents ``dusty gas". The material of the dusty gas is completely atomic and composed of the removed dust material, i.e. C atoms in the case of graphite dust and mass averaged atoms of Mg, Si and O in the case of silicate. The dusty gas is not processed further by sputtering or grain-grain collisions, but advected. Here, it is assumed that the dusty gas has the same velocity as the regular gas derived by the hydrodynamical simulation. At the beginning of the simulation, the number densities of the collector bin are set to 0. It should be noted that no feedback of the dusty gas on the regular gas medium is considered. The dusty gas contributes to the sputtering and can also be (re-)accreted by dust grains of the same dust composition, thus leading to grain growth (Section~\ref{sec_sput}). The charge number of the dusty gas is set equal to the charge number of the regular gas, although this is only a rough approximation, but a more accurate calculation of the charge number of a carbon gas and especially of a mixture of Si, Mg and O is beyond the scope of this paper. Furthermore, as the dusty gas density is low compared to the density of the regular gas, the dusty gas is not considered as a relevant component of the collisional or plasma drag for the dust advection (Section~\ref{advec}).

On the other end of the grain size distribution, dust growth by gas accretion and sticking of dust particles in a grain-grain collision can produce grains with radii larger than the total maximum grain radius $a_\text{max, abs}$. To consider these large grains, the quantity $M_\text{large}$ is defined which represents the total dust mass of all grains in the domain with radii larger than $a_\text{max, abs}$. At the beginning of the simulation, $M_\text{large}$ is 0. The dust mass  $M_\text{large}$ is neither advected, sputtered, nor colliding with other dust grains, and hence $M_\text{large}$ is increasing with time. However, in all of our conducted simulations, $M_\text{large}$ is much smaller than $\unit[1]{\%}$ of the initial dust mass. The dust mass in all size bins integrated over all cells, the mass of the dusty gas and the mass $M_\text{large}$ of the large dust grains enable mass conservation during advection and dust-processing (see Section~\ref{sec_12234}).

The boundary between the size bins $i$ and $i+1$ ($i\in\mathbb{N}_{\le N_\text{grain}-1}$) is defined as $a_i \left(\frac{1+\Delta_a^3}{2} \right)^{1/3}$, which represents the mass-related mean of $a_i$  and $a_{i+1}$. The boundary between size bin $i=0$ and $i=1$ is $a_1 \left(\frac{1+\Delta_a^{-3}}{2} \right)^{1/3}$ and the boundary between size bin $i=N_\text{grain}$ and the grains representing the dust mass $M_\text{large}$ is $a_{N_\text{grain}} \left(\frac{1+\Delta_a^3}{2} \right)^{1/3}$.

Based on the gas number density, gas-to-dust mass ratio and the dust grain size distribution at the beginning of the simulation, the number density of dust particles $n_i,\, i\in\mathbb{N}_{\le N_\text{grain}}$, is calculated for each cell, and subsequently (considering changes in $n_i$ due to dust advection and destruction/growth) also for later time-points in each cell.

\subsection{Dust advection}
\label{advec}
The dust velocity $\mathbf{v_\text{dust}}(t+\Delta t)$ at time $t+\Delta t$ is determined by its velocity $\mathbf{v_\text{dust}}(t)$ at time $t$ and the acceleration experienced during the time interval $\Delta t$.  Here, $\Delta t$ is the time-step given by the output of the hydrodynamical simulations and we assume that the conditions of the surrounding gas are constant during $\Delta t$. The acceleration depends on the current dust velocity, and for the sake of higher velocity accuracy, the time interval $\Delta t$ is divided into ten equally-sized intervals in which the acceleration is calculated. The dust velocity $\mathbf{v_\text{dust}}(t+\Delta t)$  at time $t+\Delta t$ is then given by 
\begin{align}
\mathbf{v_\text{dust}}(t+\Delta t) = \mathbf{v_\text{dust}}(t) + \sum_{i=1}^{10} \frac{\mathbf{F}_\text{drag}\left(t'\right)}{m} \,\frac{\Delta t}{10},
\end{align}
where the drag force $\mathbf{F}_\text{drag}\left(t'\right)$ at time $t'=t+\Delta t(i-1)/10$ is a function of $\mathbf{v_\text{gas}}\left(t'\right) - \mathbf{v_\text{dust}}\left(t'\right)$.
  
The drag is caused by the relative velocity between the dust and surrounding gas, $\mathbf{v}_\text{rel} = |\mathbf{v_\text{gas}} - \mathbf{v_\text{dust}}|$, and decreases with decreasing $v_\text{rel} = |\mathbf{v}_\text{rel}|$. In general, there are two different types of gas drag: The classical drag is evoked by collisions of the dust with gas particles, and the plasma drag by the Coulomb interchange between the charged grains and ionised gas. In the following we omit the vector notation of the forces, but it should be kept in mind that the acceleration of the grains is in the direction of  $\mathbf{v_\text{rel}}$. Following  \cite{Baines1965} and \cite{Draine1979}, the net drag caused by collisional drag and by plasma drag is given as (in cgs units)
\begin{align}
F_\text{drag} &= F_\text{col} + F_\text{pla}\\
 &= 2 \sqrt{\pi} k_\text{B} T_\text{gas} a^{ 2} \sum_j n_{\text{gas},j} \left(\mathcal{F}_\text{col,j}+ \mathcal{F}_\text{pla,j}\right), \label{drag_force}
\end{align}
with the ``Collisional term"
\begin{align}
 &\mathcal{F}_\text{col,j} = \left(S_j + \frac{1}{2S_j}\right) \exp{\left[ -S_j^2\right]} +\sqrt{\pi}\left(S_j^2 + 1 - \frac{1}{4\,S_j^2}\right)\,\text{erf}\left[S_j\right]\\
  &\text{and the ``Plasma term"}\nonumber\\
  &\mathcal{F}_\text{pla,j} = z_j^2 \phi^2 \ln{\!\left[\frac{\Lambda_\text{Cou}}{Z_j}\right]}\left(\sqrt{\pi} \frac{\text{erf}\left[S_j\right]}{S_j^2}-\frac{2\exp{\left[-S_j^2\right]}}{S_j}\right).
\end{align}
The drag force in equation~(\ref{drag_force}) is summed over all plasma species $j$ within the gas (atoms, molecules, ions and electrons), each with number density $n_{\text{gas},j}$, particle mass $m_{\text{gas},j}$, particle charge number $z_j$ and velocity parameter $S_j~=~\sqrt{\frac{m_{\text{gas},j}}{2\,k_\text{B}\,T_\text{gas}}} v_\text{rel}$. For our model of Cas~A, oxygen ions and electrons are considered ($j\in\mathbb{N}_{\le2}$). $Z_\text{grain}$ is the charge number of the grain (see Section~\ref{sec_char}),  the grain potential parameter is $\phi = \frac{Z_\text{grain} \, e^2}{a\,k_\text{B}\,T_\text{gas}}$, the Coulomb ``cutoff factor'' is $\Lambda_\text{Cou} = \frac{3}{2\,a\,e\,\phi}\left(\frac{k_\text{B}\,T_\text{gas}}{\pi n_\text{e}}\right)^{0.5}$ (e.g. \citealt{Dwek1992})\footnote{$\ln{\!(\Lambda_\text{Cou})}$ is also called the Coulomb logarithm (e.g. \citealt{McKee1987}).}, $k_\text{B}$ is the Boltzmann-constant, $n_\text{e}$ is the electron density, $e$ is the elementary
charge, and $\text{erf}(S_j)$ is the error function. We assume that all species in the plasma have the same temperature, $T_\text{gas}$.
 
Plasma drag has a negligible effect on the dynamics of small grains for high gas temperatures and high relative velocities, while it exceeds collisional drag for large grains at low gas temperatures and small relative velocities (see Fig.~2 of \citealt{Bocchio2016}). In this paper, we will ignore magnetic fields and the potential (betatron) acceleration by the Lorentz force on charged grains, which we will examine in a future work. 

\subsection{Grain charging}
\label{sec_char}
\begin{figure}
 \centering
\includegraphics[trim=1.77cm 3.55cm 1.65cm 2.3cm, clip=true,page=1,width=1.0\linewidth]{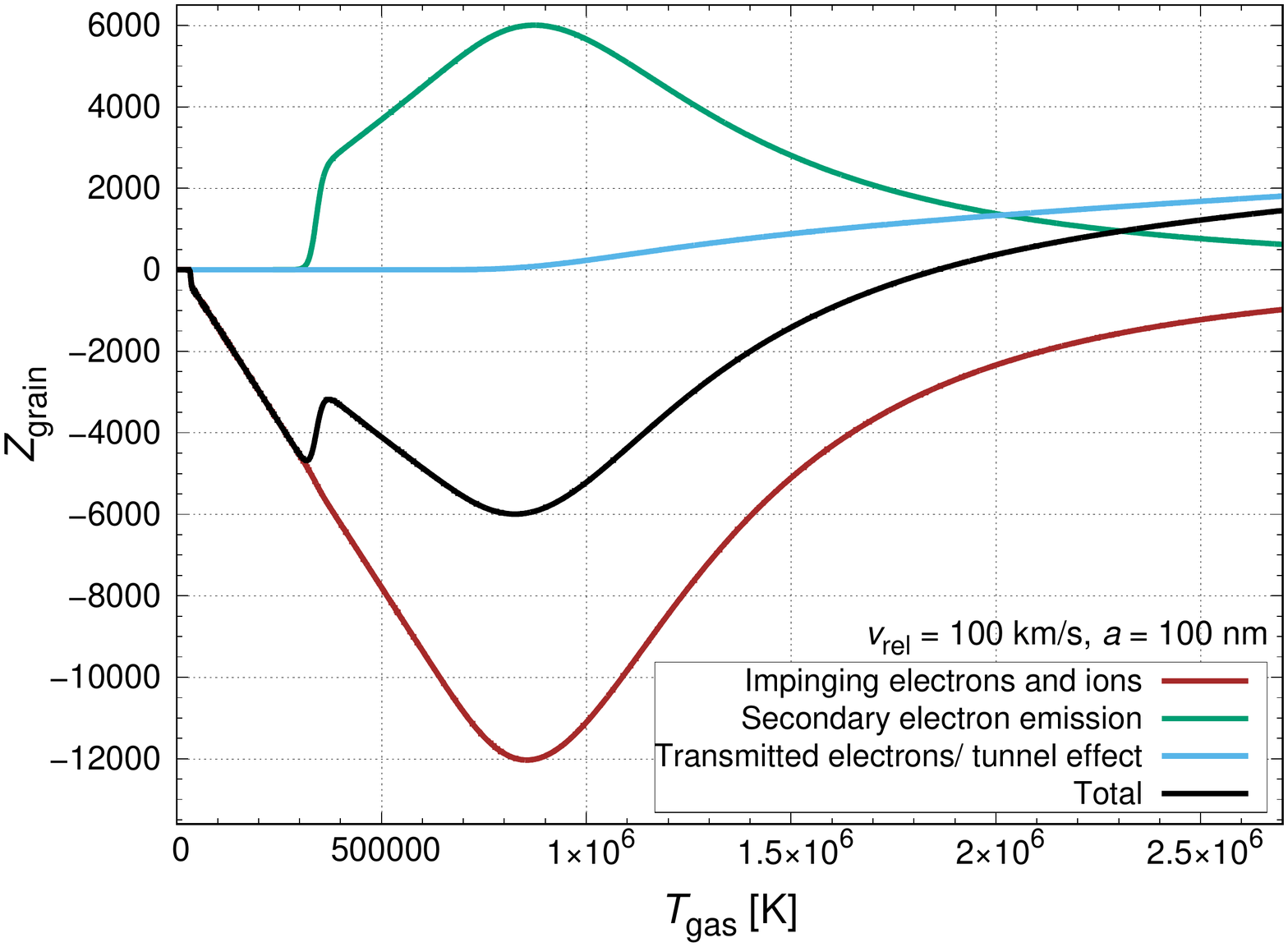}\\[-0.49cm]
\includegraphics[trim=1.61cm 1.5cm 1.65cm 0.99cm, clip=true,page=1,width=1.0\linewidth]{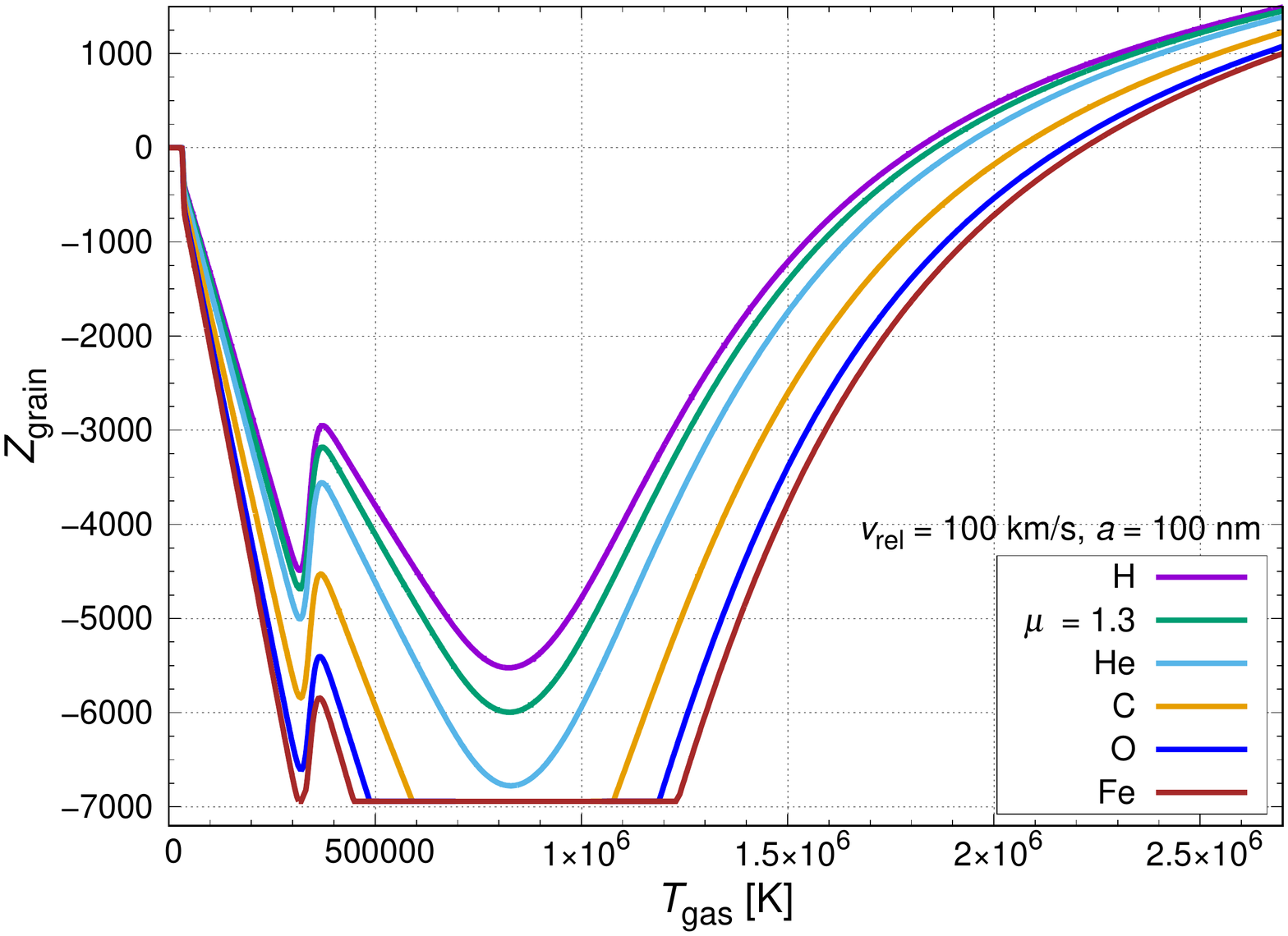}
\caption{Grain charge number $Z_\text{grain} = Q_\text{grain}/e$ as a function of gas temperature $T_\text{gas}$ for a grain radius $a=\unit[100]{nm}$ and a relative velocity $v_\text{rel}=\unit[100]{km/s}$  between gas and dust grain. \textit{Top}: The effect of the secondary electron emission and the transmitted electrons on the total charge (field emission is not shown) for a gas with mean molecular weight $\text{\textmu}=1.3$ (solar abundance). \textit{Bottom}: $Z_\text{grain}$ for different gas species (including oxygen), considering field emission as a lower charge limit.}
\label{charge_fig} 
\end{figure}

Dust grains within the SNR are electrically charged by the impacts of plasma particles (ions and electrons). Several processes can influence the total charge of the grain such as the kind of the impinging plasma particles, associated secondary electrons, transmitted plasma particles, and field emission (e.g. \citealt{Kimura1998}). Here, we ignore photoelectron emission. Numerical calculations of the remaining charging processes based on detailed modelling of the atomic physics are very computationally-intensive. In order to simplify the calculations, we apply the analytical description of the charging processes derived by \cite{Fry2018}\footnote{This approach was introduced by \cite{Shull1978} and \cite{McKee1987}. Multi-valued potentials at a given temperature are ignored, as is the cooling and heating rate of the dust grains.}, where the grain potential $\Phi_\text{total}$ is numerically solved for the steady-state value and then fitted as a function of gas temperature $T_\text{gas}$, grain size $a$ and relative velocity $v_\text{rel}$. The applied fitting function for the grain potential $\Phi_\text{total}$ is given in Appendix \ref{app_grain_pot}. The dust grain charge is then (in cgs-units\footnote{In SI-units, equation~(\ref{final_charge}) would transform to\\
 $Q_\text{grain} = 4\pi\epsilon_0\frac{a\,\Phi_\text{total}\,k_\text{B}\,T_\text{gas}}{e}$.})
 \begin{align}
  Q_\text{grain} = \frac{a\,\Phi_\text{total}\,k_\text{B}\,T_\text{gas}}{e}.\label{final_charge}
 \end{align}
 Following equation~(\ref{final_charge}), $Q_\text{grain}$ is calculated for each dust grain species, cell and time-step in the domain. Apart from the escape length $\lambda_\text{esc}$, the treatment of the grain charge is independent of the dust material but depends on the gas temperature $T_\text{gas}$, grain radius $a$ and relative velocity $v_\text{rel}$ as well as on the gas species. On the other hand, the grain charge has an impact on the dust advection (Section~\ref{advec}), grain-grain collisions (Section~\ref{sec_ggcoll}), sputtering (Section~\ref{sec_sput}), and gas accretion (Section~\ref{sec_gas_acc}). The grain charge number ($Z_\text{grain} = Q_\text{grain}/e$) is shown in Fig.~\ref{charge_fig} for several gas types, including pure oxygen, as well as the impact of different effects (e.g. secondary electron emission) on the total dust grain charge. For the modelling of a clump in Cas A, we evaluated the grain charge in a pure oxygen gas.

\subsection{Grain-grain collisions}
\label{sec_ggcoll}
Collisions between dust grains of different sizes can occur in a SNR due to the relative velocities between them which are caused by the size-dependent gas drag (Section~\ref{advec}). The timescale and the probability for grain-grain collisions as well as the collisional outcome are discussed in this Section.

\subsubsection{Collisional timescale}
\label{coltime}
Grain-grain collisions are neglected in many studies that investigate dust destruction in SNRs. To show the importance of this process, we determine here the grain-grain collisional timescale $\tau_\text{col}$. 

For the sake of simplicity, we assume a population of dust grains with a single grain size $a$, a mean number density $\overline{n}$, and a mean relative velocity $\overline{v}$ between the grains. The mean free path of a particle is then $\lambda_\text{path}=1/(\pi a^2\,\overline{n})$ and the collisional timescale  (e.g.~\citealt{Bocchio2016})  $\tau_\text{col}=\lambda_\text{path}/\,\overline{v} = 1/(\pi a^2\,\overline{n}\,\overline{v})$. The  mean number density of dust grains and gas particles, $\overline{n}$ and $\overline{n}_\text{gas}$, respectively, are related by the gas-to-dust mass ratio $\Delta_\text{gd}$ by $\overline{n}=\overline{n}_\text{gas} \frac{\mu\,m_\text{amu}}{\Delta_\text{gd} 4/3\pi\rho_\text{bulk}a^3}$, with $m_\text{amu}$ as the atomic mass unit. It follows that the collisional timescale is
\begin{align}
 \tau_\text{col}&= \frac{4\,\Delta_\text{gd}\rho_\text{bulk}}{3\,\mu\,m_\text{amu}}\frac{a}{\overline{n}_\text{gas} \,\overline{v}},\\
 &\approx 35000\frac{\left(a/\text{nm}\right)}{\left(\overline{n}_\text{gas}/\text{cm}^{-3}\right) \left(\overline{v}/\text{(km/s)}\right)}\unit[]{yr}.\label{taucol}
\end{align}
Considering a typical gas density of $\overline{n}_\text{gas}=\unit[100]{cm^{-3}}$ ($\unit[1000]{cm^{-3}}$), grains with radius $a=\unit[10]{nm}$ and a mean velocity $\overline{v}=\unit[100]{km/s}$, the timescale between grain-grain collisions is $\tau_\text{col}\approx\unit[35]{yr}$ ($\unit[3.5]{yr}$) which is roughly half ($2\,\%$) of the simulation time ($t_\text{sim}\approx\unit[61.5]{yr}$ and $\unit[190.2]{yr}$, resp.). The shock wave increases the dust and the gas number densities, making collisions even likelier. On the other hand, if we consider grains with radius $a=\unit[1000]{nm}$, the timescales derived with eq.~(\ref{taucol}) are a factor of 100 larger, making grain-grain collisions significantly less likely. We note, that $\overline{v}$ was fixed but that the largest relative velocities will occur between small and large grains. In summary, we expect that grain-grain collisions are important for at least some size populations and have the potential to influence the dust survival rates in SNRs such as Cas~A.
%

\subsubsection{Collision probability}
\label{sec_col_prob}
\begin{figure}
 \centering
 \fbox{\includegraphics[trim=1.5cm 1.26cm 0cm 4.4cm, clip=true,page=1,height=0.35\linewidth]{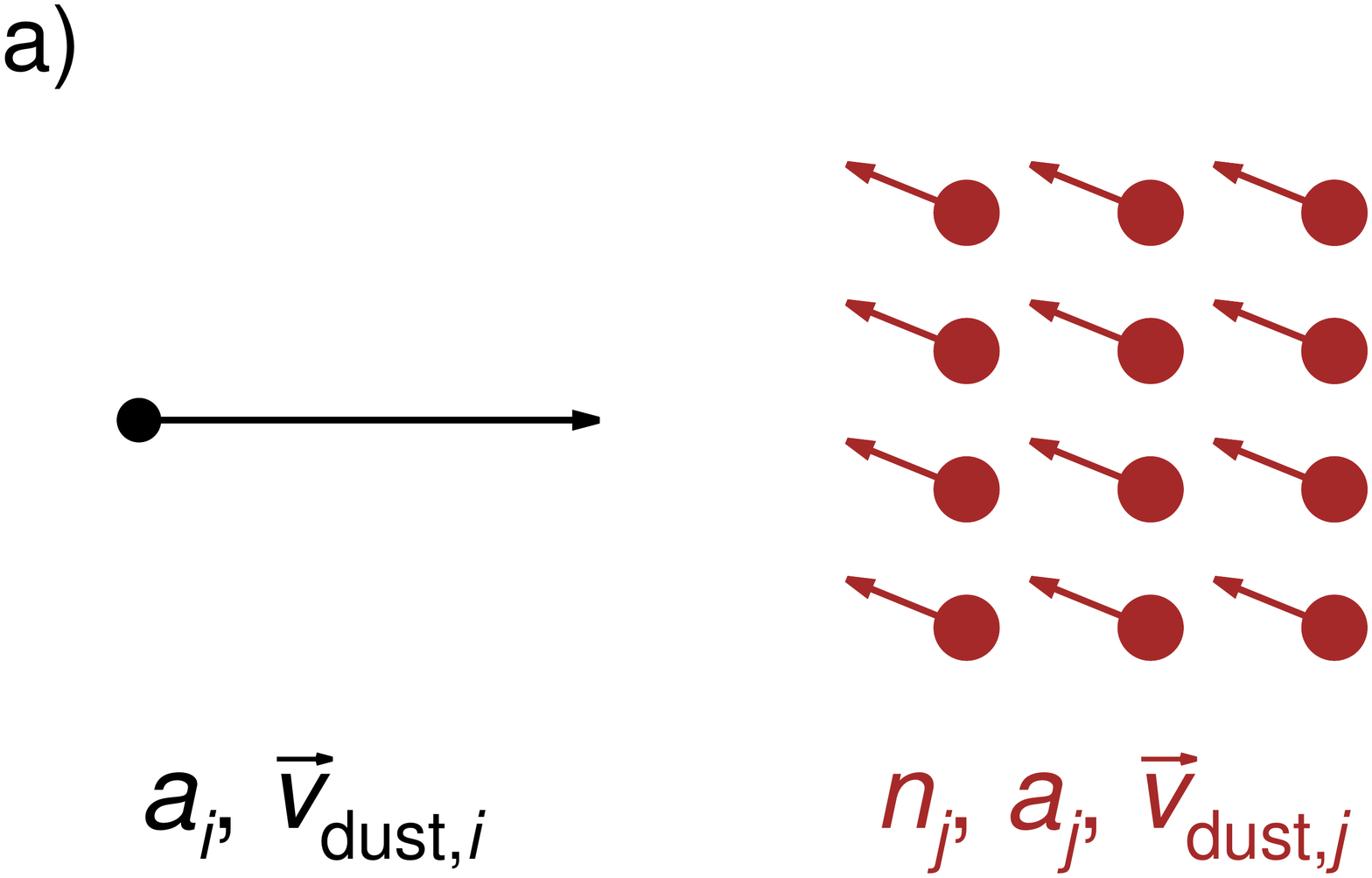}}\hspace*{-0.01cm}
 \fbox{\includegraphics[trim=8.0cm 4.1cm 6.5cm 3.8cm, clip=true,page=1,height=0.35\linewidth]{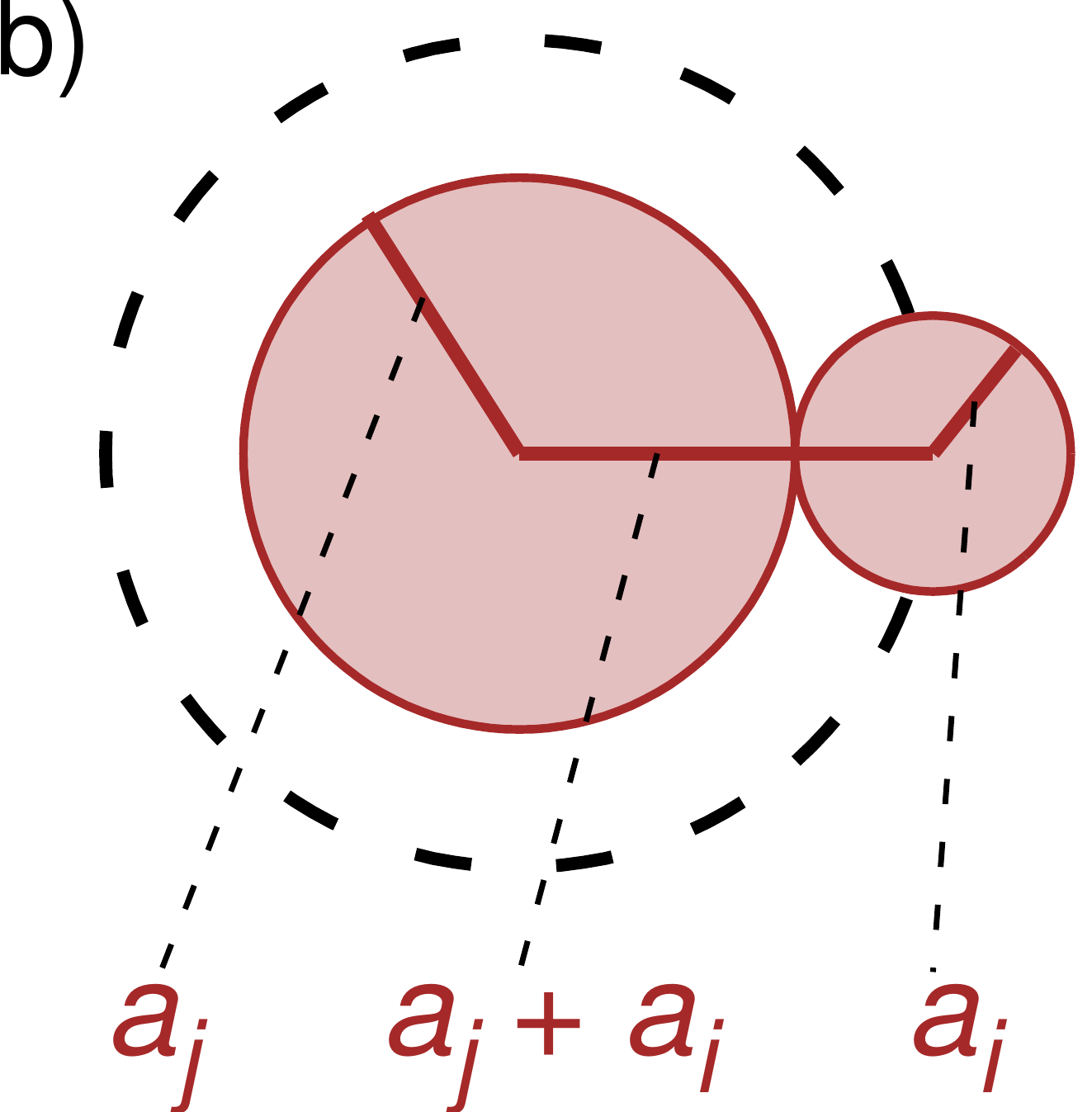}}
\caption{\textit{Left:} Collision of a dust grain $i$ (grain radius $a_i$) with any grain $j$ (grain radius $a_j$) for a homogeneous, parallel directed distribution with number density $n_j$. \textit{Right:} Collision cross section $\sigma_\text{col}$ of particles with radii $a_i$ and  $a_j$, respectively, resulting in $\sigma_\text{col} = \pi\left(a_i+a_j\right)^2$ for neutral charged grains.}
\label{fig_sigma_coll} 
\end{figure}

We consider a dust grain $i$ with radius $a_i$ and a dust velocity $\mathbf{v}_{\text{dust},i}$ that is constant during the time interval $\Delta t$. Furthermore, we assume a homogeneous distribution of dust grains $j$ with radius $a_j$ and number density $n_j$ that all have the velocity  $\mathbf{v}_{\text{dust},j}$ in the same direction (Fig.~\ref{fig_sigma_coll}, a). We want to calculate the probability $P_{ij}$ that a single dust grain $i$ collides with any other dust grain $j$.
 \begin{figure}
 \centering
 \fbox{\includegraphics[trim=2.6cm 2.16cm 1.75cm 1.78cm, clip=true,page=1,height=0.35\linewidth, page =1]{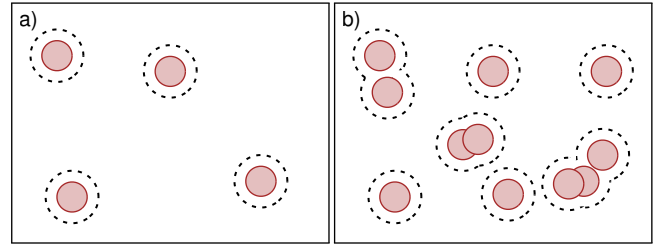}}\hspace*{-0.01cm}
 \fbox{\includegraphics[trim=2.6cm 2.16cm 1.75cm 1.78cm,  clip=true,page=1,height=0.35\linewidth, page =2]{Pics/fraction_gg.pdf}}
\caption{Dust grains in the projected area. The dashed circles are the same as in Fig.~\ref{fig_sigma_coll} b) and represent the cross section $\sigma_\text{col}$. \textit{Left:} For $\tau\ll1$, the fraction of the area covered by the dust grains is simply the number of dust grains per unit area multiplied by $\sigma_\text{col}$. \textit{Right:} For $\tau\nll1$, self-shielding of the dust grains occurs and the fraction  of the area covered by the dust grains is given by $1-\exp{\left[-\tau\right]}$ (see equation~\ref{lab_P2}).}
\label{lab_ggc} 
\end{figure}

For uncharged grains, the collision velocity of a potential projectile and target is equal to their relative velocity, $v_\text{col} = |\mathbf{v}_{\text{dust},i} - \mathbf{v}_{\text{dust},j}|$, and the geometrical cross section for a single collision is $\sigma_\text{col}=\pi\left(a_i+a_j \right)^2$ (Fig.~\ref{fig_sigma_coll},~b). We ignore the Brownian motion of the dust grains which is negligible compared to the high dust velocities in a shock-impacted ejecta clump. Considering the electric charges $Q_i$ and $Q_j$ of the grains, respectively, the grains can be attracted or repulsed and the actual collision velocity and the cross section are changed. Setting $\alpha_q= \frac{2\,Q_i\,Q_j\,(m_i+m_j)}{(a_i+a_j)m_im_j|\mathbf{v}_{\text{dust},i} - \mathbf{v}_{\text{dust},j}|^2}$, we obtain for the collision velocity (see Appendix \ref{gg_ec} for derivation)
\begin{align}
v_\text{col}=\left(1-\alpha_q\right)^{0.5}|\mathbf{v}_{\text{dust},i} - \mathbf{v}_{\text{dust},j}|                                                                                                                                                                                                                                                                                                                                                                                                                                                                                                                                                                                                                                                                                                                                                                                                                                                                                                                                                                                                                                                                                                                                                                                                                                                                       \end{align}                                                                                                                                                                                                                                                                                                                                                                                                                                                                                                                                                                                                                                                      and for the collision cross section 
\begin{align}
\sigma_\text{col}=\left(1-\alpha_q\right)\pi\left(a_i+a_j \right)^2. \label{siggg}
\end{align}

In the rest frame of the dust grains $j$ with size $a_j$, the dust grain $i$ travels along a length $\Delta l=v_\text{col} \Delta t$. We define $A~=~N/(n_j\,\Delta l)$ as the area projected along the propagation direction of $i$ in the rest frame of $j$ which contains $N$ dust grains of radii $a_j$. The probability $\tilde{P}$ for a collision of $i$ with one of the $N$ grains with radius $a_j$ is then the ratio of $\sigma_\text{col}$ to $A$, $\tilde{P}=\sigma_\text{col}/A$. The dimensionless quantity 
\begin{align}
\tau=n_j\,\Delta l\,\sigma_\text{col}\,\left(=N\, \sigma_\text{col}/A\right)
\end{align}
gives the number of cross sections $\sigma_\text{col}$ per unit area $A$. For $\tau=0$, grain $i$ ``sees'' no dust grain $j$ in $A$ and hence no collision can occur, while for $\tau=1$ the projected area $A$ is completely covered by grains $j$ and the probability for a collision is $\unit[100]{\%}$. 
We differentiate between two cases:\\
(i) $\tau \ll 1$ for low number densities of grains $j$. The probability $P_{ij}$ for a collision of $i$ with any of the $N$ grains $j$ is then the sum of all probabilities $\tilde{P}$ (Fig.~\ref{lab_ggc},~a):
\begin{align}
 P_{ij} = N\, \tilde{P}=N\, \sigma_\text{col}/A =  n_j\,\Delta l\,\sigma_\text{col} = \tau.\label{lab_P}
\end{align}
(ii) The continuous increase of $\tau$ in equation~(\ref{lab_P}) would inevitably result in a probability $P_{ij}$ larger than $\unit[100]{\%}$, and it becomes already inaccurate for $\tau \nll 1$ (large number density, large collision velocity or large grains). The reason is the enhanced occurrence of self-shielding dust grains (Fig.~\ref{lab_ggc}, b). This problem can be solved when the individual probabilities $\tilde{P}$ for a collision of $i$ with one of the $N$ grains with radius $a_j$ are not just summed up, but when instead the counter-probabilities $(1-\tilde{P})$ are multiplied to get $(1-P_{ij})$. For $N$ dust grains $j$ in the area $A$ it follows:
\begin{align}
 1- P_{ij} &= \left(1- \tilde{P}\right)^N = \left(1-\frac{\sigma_\text{col}}{A}\right)^N\nonumber\\
   & = 1 - N \left(\frac{\sigma_\text{col}}{A}\right) + {N \choose 2} \left(\frac{\sigma_\text{col}}{A}\right)^2 - {N \choose 3} \left(\frac{\sigma_\text{col}}{A}\right)^3 + ...\nonumber\\
   & \approx 1 - \left(n_j\,\Delta l\right)\sigma_\text{col} + \left(n_j\,\Delta l\right)^2\frac{\sigma_\text{col}^2}{2}-\left(n_j\,\Delta l\right)^3\frac{\sigma_\text{col}^3}{6} + ...\nonumber\\
   &  = \sum^N_{k=0} \left(-1\right)^k \frac{1}{k!} \left(n_j\,\Delta l\,\sigma_\text{col} \right)^k =\exp{\left[-\tau\right]}\nonumber.\\
   &\hspace{-0.7cm}\text{Finally, we get}\nonumber\\
   &P_{ij}= 1-\exp{\left[-\tau\right]}.  \label{lab_P2}
\end{align}
We want to highlight the similarity of equation~(\ref{lab_P2}) to a completely different astrophysical problem, the intensity of the radiation of an optically thick dust accumulation (e.g. in the ISM or in a protoplanetary disk). Here, $\tau$ describes the optical depth and the intensity is proportional to $\exp{(-\tau)}$ (cf. Beer-Lambert law). In that scenario, Fig.~\ref{lab_ggc} can be interpreted as an optically thin (a) or optically thick (b) system and the ``collisions'' occur between photons and dust grains instead of collisions between grains.

Based on the local dust velocities and dust number density $n_j$, the collision probability $P_{ij}$ for grain $i$ to collide with any grain of size $a_j$ is calculated in each cell and for each time-step for which the dust velocities have been calculated (see Section~\ref{advec}). It should be noted, that $P_{ij}$ in both equations~(\ref{lab_P}) and (\ref{lab_P2}) is independent of the number density $n_i$, and in general $P_{ij}\neq P_{ji}$. Since the bulk density of carbon grains is lower than that of silicate grains, their acceleration and their grain number densities are higher (for a fixed total dust mass) resulting in an enhanced collision probability for carbon grains. 

Finally, the number density of colliding dust grains $i$ with grains $j$ is $n_\text{col}=P_{ij}\,n_i$. Depending on the collision velocity, dust grain sizes and material properties, $n_\text{col}$ dust particles vaporise, fragment, bounce or stick with their collisional counterpart (from high to low energy) and the grain size distribution is redistributed (e.g.~\citealt{Borkowski1995}). The different collisional processes are described in the following subsections. 
 
\subsubsection{Vaporisation}
\label{sec_vapo}
Vaporisation of the $n_\text{col}$ dust grains in bin $i$ is assumed to occur if the collision velocity $v_\text{col}$ between grain $i$ and $j$ is above the vaporisation threshold velocity, 
\begin{align}
 v_\text{col}\ge v_\text{vapo}.\label{vapothre}
\end{align}
$v_\text{vapo}$ is a function of the dust material only and is given in Table~\ref{mat_para} for carbon and silicate materials. Although the threshold velocity for carbon is larger than that of silicates, this does not inevitably mean that the vaporisation of silicate grains is more efficient. The bulk density of amorphous carbon is smaller by a factor of $1.5$ which causes a greater acceleration of these grains, and the vaporisation threshold is reached at an earlier stage. If the vaporisation condition is fulfilled (equation~\ref{vapothre}), $n_\text{col}$ dust grains are removed from bin $i$ and $n_\text{col} (4/3) \pi a_i^3 \rho_\text{bulk}/(\mu\,m_\text{amu})$ particles (atoms/averaged atoms) are placed in the collector bin 0 (dusty gas). Note, that only particles from bin $i$ are removed, bin $j$ is evaluated when $i$ and $j$ are exchanged.

\begin{figure}
 \centering
\includegraphics[trim=2.35cm 2.1cm 2.1cm 2.5cm,  clip=true,page=1,width=1.0\linewidth, page =1]{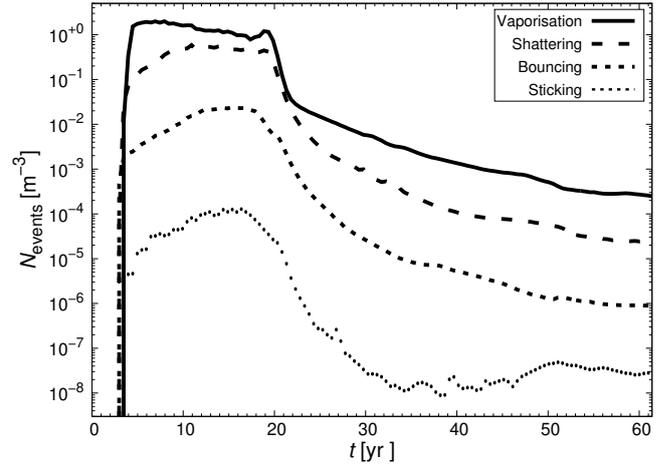}
\caption{Frequency of vaporisation, shattering, bouncing and sticking events as a function of time for an example simulation (shock velocity $v_\text{sh}=\unit[1600]{km/s}$, density contrast $\chi=100$). When the shock impacts the dust-filled clump ($t\unit[\sim3]{yr}$), the number density of particles in collisions rapidly increases, reaching a maximum ($t\unit[\sim5-20]{yr}$) before starting to decrease slowly. Bouncing and sticking events are 2 and 5 orders of magnitude, respectively, rarer than vaporisation and shattering events. The adopted initial grain size distribution of carbon is a power-law with $a_\text{min}=\unit[10]{nm}$, $a_\text{max}=\unit[200]{nm}$, and $\gamma=3.5$. The dust composition is carbon.}
\label{fig_counter} 
\end{figure}

In Fig.~\ref{fig_counter} we show the frequency of colliding particles resulting in vaporisation, fragmentation, bouncing and sticking as a function of simulation time, integrated over the entire simulation domain. Obviously, vaporisation and fragmentation are the dominant processes during the whole simulation, while sticking and bouncing are less frequent.

\subsubsection{Fragmentation}
\label{sec_frag}
A dust grain $i$ is assumed to be shattered by collisions with grains $j$ if the collision velocity $v_\text{col}$ between grain $i$ and $j$ is below the vaporisation threshold velocity and above the fragmentation threshold velocity,
\begin{align}
v_\text{vapo} > v_\text{col}\ge v_\text{frag}.\label{fragthre}
\end{align}
As $v_\text{vapo}$, $v_\text{frag}$ is a function of dust materials only and is given in Table~\ref{mat_para}. Since the fragmentation threshold velocity and bulk density of carbon are smaller than those of silicates, carbonaceous grains tend to faster fragmentation. 

For the description of the fragmentation of grain $i$, we follow \cite{Hirashita2009} whose work is based on \cite{Tielens1994} and \cite{Jones1996}. Although already described in detail by \cite{Hirashita2009}, we give the procedure in Appendix~\ref{sec_frag_app} again for the sake of completeness since some inconsistencies between equations and parameters in their work and that of \cite{Jones1996} appear to be present.

\subsubsection{Grain bouncing}
Collisions between grains $i$ and $j$ result in bouncing if the collision velocity $v_\text{col}$ is below the fragmentation threshold velocity (equation~\ref{vapothre}) and above the coagulation threshold velocity (see equation~\ref{coagthre}),
\begin{align}
v_\text{frag} > v_\text{col}> v_\text{coag}.\label{fragboun}
\end{align}
The size distribution of the dust grains is not affected by bouncing, but bounced grains might take a new speed and in particular a new propagation direction.  We ignore this new velocity direction for two reasons: 
Firstly, each grain's bouncing collisions would result in its own velocity distribution, and the additional computational effort would be immense. Secondly, bouncing is not a frequent event in the simulations (Fig.~\ref{fig_counter}), and a more sophisticated bouncing description is not expected to lead to a very different outcome. Instead, we assume that the post-bounce grains instantaneously have the same velocity and velocity direction as before the bounce, caused by the continuous gas stream. In summary, we assume that the bouncing changes neither the velocities nor the number densities of the grains.

\subsubsection{Grain sticking}
\label{sec_coag}
The two colliding dust grains $i$ and $j$ are assumed to stick together if their collision velocity $v_\text{col}$ is below the coagulation threshold velocity 
\begin{align}
v_\text{coag}\ge v_\text{col},
\end{align}
where $v_\text{coag}$ is given by (\citealt{Chokshi1993, Dominik1997})
\begin{align}
 v_\text{coag} &= 2.14\,F_\text{stick}\sqrt{\frac{a_i^3 + a_j^3}{(a_i+a_j)^3}} \frac{\gamma_\text{A}^{5/6}}{E_\text{PoiY}^{1/3} R_\text{ij}^{5/6}\rho^{1/2}_\text{bulk}}.\label{coagthre}
\end{align}
In order to avoid the complexity of compound species, only sticking between the same dust materials is treated. Based on experimental work by \cite{Blum2000}, $F_\text{stick}$ is set to 10 (see \citealt{Yan2004}). $\gamma_\text{A}$ is the surface energy per unit area, $R_\text{ij}= a_i a_j/(a_i+a_j)$ is the reduced grain radius and $E_\text{PoiY}=0.5 E_\text{Y}/(1-\nu_\text{Poi})^2$ is a dust material quantity that is related to Poisson's ratio $\nu_\text{Poi}$ and Young's modulus $E_\text{Y}$, listed in Table~\ref{mat_para}. Following equation~(\ref{coagthre}), the coagulation threshold velocity of equal-sized grains of carbonaceous (silicate) material is $\unit[99]{m/s}$ ($\unit[21]{m/s}$) for $\unit[10]{nm}$ grains, and $\unit[2.12]{m/s}$ ($\unit[0.45]{m/s}$) for $\unit[1]{\text{\textmu} m}$ grains.

Dust growth by coagulation has been observed in many astrophysical environments, e.g. in dense molecular clouds (e.g.~\citealt{Stepnik2003}) or protoplanetary disks (e.g.~\citealt{Kirchschlager2016}). If the collision velocity between grain $i$ and $j$ is lower than $v_\text{coag}$,  $n_\text{col}/2$ dust grains are removed from both bin $i$ and $j$\footnote{The remaining $n_\text{col} -n_\text{col}/2=n_\text{col}/2$ dust grains of bin $i$ are removed if $i$ and $j$ are exchanged.}. For the sake of simplicity, the newly formed dust aggregate is assumed to have a spherical shape with radius $a_\text{coag}=\left(a_i^3+a_j^3\right)^{1/3}$, and $n_\text{col}/2$ dust grains with size $a_\text{coag}$ are placed in the corresponding bin taking into account mass conservation.

Although the above form for the coagulation threshold velocity is based on both physical and experimental grounds, there could be significant uncertainties (\citealt{Hirashita2009}). However, sticking has a low occurrence in the simulations (Fig.~\ref{fig_counter}), as the gas velocities and hence dust velocities are too high, and as bouncing is also a rare process, the impact of a higher or lower coagulation threshold can be ignored.

\subsection{Sputtering}
\label{sec_sput}
Sputtering is a destruction process whereby grain atoms are ejected due to bombardment by gas particles (atoms, ions or molecules). The rate at which a dust grain is sputtered is influenced by its relative motion through the gas, also known as kinematic, kinetic, inertial or non-thermal sputtering, and by the thermal motions of the gas particles, known as thermal sputtering. 

\subsubsection{Kinematic sputtering}
The rate of decrease of grain radius $a$ due to kinematic (inertial, non-thermal) sputtering can be expressed as (e.g.~\citealt{Dwek1992}),
\begin{align}
 \frac{\text{d}a}{\text{d}t} = \frac{\left\langle M_\text{atom}\right\rangle}{2\,\rho_\text{bulk}}v_\text{rel} \sum_k n_{\text{gas},k} Y_k \left(E\right), \label{isput}
\end{align}
where $\frac{\text{d}a}{\text{d}t} $ is the reduction of grain radius per unit time, $\left\langle M_\text{atom}\right\rangle$ is the average atomic mass of the grain atoms, $v_\text{rel}$ is the grain velocity relative to the ambient gas, $n_{\text{gas},k}$ is the number density of gas species $k$, and $Y_k(E)$ is the sputtering yield, which is the number of ejected grain atoms per incident projectile of species $k$. The sum runs over all gas species, including the dusty gas.\footnote{By considering the dusty gas in the sputtering process we mean that sputtered atoms from the dust grains can subsequently themselves sputter atoms from the grains.} The sputtering yield is a function of the kinetic energy $E = m_{\text{gas},k} v_\text{rel}^2/2$, where 
$m_{\text{gas},k}$ is the particle mass of gas species $k$ (\citealt{Tielens1994, Nozawa2006}). A factor of 2 is included in equation~(\ref{isput}) to correct the yield which is generally measured for normally incident projectiles on a target material (e.g.~\citealt{Micelotta2016}).

\subsubsection{Thermal sputtering}
The thermal sputtering rate is a function of the velocity of the gas particles which is determined by the temperature of the ambient gas (thermal motion). It is defined as (\citealt{Barlow1978,Draine1979})
\begin{align}
 \frac{\text{d}a}{\text{d}t} = \frac{\left\langle M_\text{atom}\right\rangle}{2\,\rho_\text{bulk}}\sum_k n_{\text{gas},k} \left\langle Y_k v \right\rangle, \label{tsput}
\end{align}
where  $\left\langle Y_k v \right\rangle$ is the sputtering yield of gas species $k$ (including the dusty gas) averaged over the Maxwellian distribution $f_\text{M}$, 
\begin{align}
\left\langle Y_k v \right\rangle =\int Y_k(E) \,v\, f_\text{M}(v) \,\text{d}v, \label{Maxwell}
\end{align}
 $v$ is the thermal velocity of a gas particle of species $k$ and 
\begin{align} 
 E = \frac{m_{\text{gas},k}}{2} v^2\label{energy_gas}
\end{align}
the energy of this gas particle.

\subsubsection{Skewed Maxwellian distribution}
Equation~(\ref{isput}) for kinematic sputtering is an approximation which gives good results for $T\lesssim \unit[10^4]{K}$ (\citealt{Bocchio2014}). However, for higher temperatures the relative velocity between a grain and the surrounding gas is not unimodal but is a combination of the thermal motion of the gas particles and the motion of the grain relative to the gas. These two motions can be combined using a
skewed Maxwellian distribution (\citealt{Barlow1978, Shull1978, Bocchio2014}),
\begin{align}
\begin{split}
 f_\text{skM}(v) &= \sqrt{\frac{m_{\text{gas},k}}{2\pi k_\text{B} T_\text{gas}}}\frac{v}{v_\text{rel}}
  \left[ \exp{\left(-\frac{m_{\text{gas},k}}{2\pi k_\text{B} T_\text{gas}} (v-v_\text{rel})^2\right)} \right. \\
 & \hspace*{2.3cm}\left. - \exp{\left(-\frac{m_{\text{gas},k}}{2\pi k_\text{B} T_\text{gas}} (v+v_\text{rel})^2\right)} \right]
\end{split}\label{eq_skm}
\end{align}
where $f_\text{skM}$ is the velocity probability function. Note, that $f_\text{skM}$ converges to $f_\text{M}$ for $v_\text{rel}\longrightarrow 0$.

Replacing $f_\text{M}$ by $f_\text{skM}$ in equation~(\ref{Maxwell}) and inserting this expression into (\ref{tsput}), we obtain the erosion rate of a dust grain that is moving through a gas whose particles are in random thermal motion at a temperature $T_\text{gas}$.

\subsubsection{Sputtering yields and parameters}
For the sputtering yield $Y_k$  of gas species $k$ for a given grain species, we adopt the expression given by equation~(11) of \cite{Nozawa2006}. This is the same approach as in \cite{Tielens1994}, except that they use a different formula for the function $\alpha_k$ (their equation~18) that appears in the yield and provides a better agreement with sputtering measurements (for details see \citealt{Nozawa2006, Tielens1994}, and references therein). We neglect dissociative sputtering of very small (carbonaceous) grains by the combination of nuclear interaction, electronic interaction and electronic collisions (\citealt{Micelotta2010b, Bocchio2012}). However, three modifications are made regarding the calculation of the yields: Firstly, the effect of the finiteness of the grains is considered by introducing a factor which multiplies the total yield (\citealt{SerraDiazCano2008}; see Fig.~\ref{fig_finite_sput} and Section~\ref{finitenessyield}). Secondly, the accretion of gas onto the dust is implemented as a negative yield (Section~\ref{sec_gas_acc}). Thirdly, the relative velocities between gas particles and dust grains are calculated taking into account Coulomb forces between charged grains and ionised gas, which have an influence on the impact velocities of the gas particles in the same way as for grain-grain collisions (Section~\ref{sec_col_prob}). The energy of the impinging gas particle in equation~(\ref{energy_gas}) is then replaced by (see Appendix~\ref{sp_ec}) 
\begin{align}
 E = \frac{m_{\text{gas},k}}{2} v^2 +\frac{z\,e\,Q_\text{grain}}{a}. 
\end{align}

Gas particles of species $k$ can cause sputtering of dust grains if their energy is equal to or above the threshold energy 
\begin{align}
E_\text{sp}= \begin{cases}          
\frac{U_0}{4\,\left\langle M_\text{atom}\right\rangle m_{\text{gas},k}}\frac{\left(\left\langle M_\text{atom}\right\rangle + m_{\text{gas},k}\right)^{4}}{\left(\left\langle M_\text{atom}\right\rangle - m_{\text{gas},k}\right)^{2}}\hspace{0.2cm}\text{if}\hspace{0.1cm}\frac{m_{\text{gas},k}}{\left\langle M_\text{atom}\right\rangle}\le0.3,\\
8U_0\left(\frac{m_{\text{gas},k}}{\left\langle M_\text{atom}\right\rangle}\right)^{1/3}\hspace{2.33cm}\text{if}\hspace{0.1cm}\frac{m_{\text{gas},k}}{\left\langle M_\text{atom}\right\rangle} > 0.3
          \end{cases}\label{eq_sp_thr} 
\end{align}
(\citealt{Bohdansky1980, Andersen1981}). 

The adopted sputtering parameters are summarized in Table~\ref{mat_para}. The quantity $U_0$ is the surface binding energy, defined as the minimum energy that is necessary to remove an atom from the top surface layer. $\left\langle Z_\text{atom}\right\rangle$ is the average atomic number of the grain material, and $\left\langle M_\text{atom}\right\rangle$ is the mass of the ejected dust species (average atomic mass). The dimensionless quantity $k_\text{sput}$ enters into one of the terms for the sputtering yield $Y_k(E)$ and has been determined via comparison with laboratory experiments (\citealt{Tielens1994}). Hydrogenation or amorphisation of the sputtered dust grains are neglected and hence the dust is only composed of pure carbon or silicate. Since the initial clump gas in our Cas~A model is composed of pure oxygen, there are only two sputtering gas species ($k\in\mathbb{N}_{\le 2}$), namely oxygen atoms or ions and atoms from the dusty gas. Since the dusty gas is composed of atoms or ions of the same material as the dust, the grains are then sputtered by the same material.

\subsubsection{Size-dependent sputtering}  
\label{finitenessyield}

\begin{figure}
 \centering
    \includegraphics[trim=2.5cm 2.7cm 2.1cm 7.5cm,  clip=true,page=1,width=1.0\linewidth, page=1]{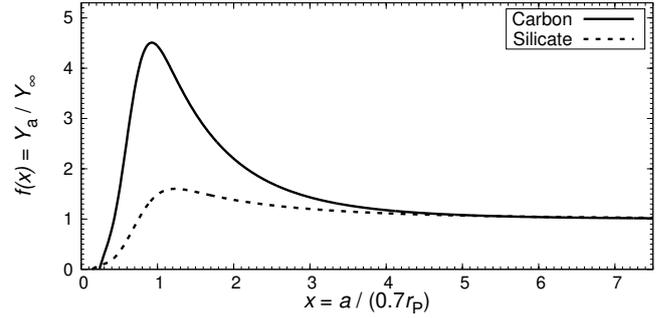}
    \caption{Size-dependence of the sputtering yield as a function of grain radius for carbon and silicate grains. $Y_a$ and $Y_\infty$ are the sputtering yields of a grain of radius $a$ and of a semi-infinite target, respectively, and $r_\text{P}$ is the penetration depth.}  
    \label{fig_finite_sput} 
\end{figure}

The experimental sputtering yields that were used to obtain the analytical function $Y(E) $ have been measured for a semi-infinite target (see e.g.~\citealt{Tielens1994}). However, the finite size of dust grains has a significant effect on the yields. The penetration depth $r_\text{P}$ of energetic gas particles can be comparable to or even larger than the dust grain size, which is especially important for the smallest grains. For grain sizes comparable to the penetration depth $r_\text{P}$, the sputtering yield is increased as the detachment of dust atoms is enhanced by a cascade effect at the grain surface, while for grains much smaller than $r_\text{P}$ the gas particles are mostly transmitted and the sputtering yield approaches 0 (see Fig.~\ref{fig_finite_sput}). On the other hand, for grains much larger than $r_\text{P}$ the finite-size yield approaches that of a semi-infinite target, as the sputtered dust atoms are mainly detached from the region close to the grain surface (\citealt{Jurac1998,SerraDiazCano2008}).

\begin{table}
\centering
\caption{Material parameters $p_i\,(i\in\mathbb{N}_{\le 6}$; \citealt{Bocchio2016}; the data for amorphous carbon and MgSiO${}_3$ are used for carbon and silicate, respectively), the mean excitation energy parameter $E_\text{exc}$  and the slope $\alpha_\text{P}$ (Section~\ref{sec_bethe_bloch}) used for the analytical modelling of the size-dependent sputtering.}
\begin{tabular}{l c c c c c c c c}
\hline\hline\\[-0.25cm]
	&$p_1$ &$p_2$ &$p_3$ &$p_4$ &$p_5$ &$p_6$ &$\!\!E_\text{exc}\,$[eV]$\!\!\!$ &$\alpha_\text{P}$\\[0.05cm]\hline
$\!$carbon$\!$  &$\!4.9\!$ &$\!0.55\!$&$\!0.77\!$&$\!4.7\!$ &$\!3.0\!$ &$\!1.2\phantom{0}\!$ &$\!13.5\!$        	&$\!-4.73\!$	    \\
$\!$silicate$\!$&$\!1.0\!$ &$\!0.5\phantom{0}\!$ &$\!1.0\phantom{0}\!$ &$\!1.8\!$ &$\!2.1\!$ &$\!0.76\!$&$\!13.0\!$          	&$\!-3.34\!$            \\\hline
\end{tabular}
\label{tab_mat_sput}
\end{table}

To take into account the size-dependent sputtering effect, we apply the model of \cite{Bocchio2012,Bocchio2014,Bocchio2016} in which they determine a correction function $f(x)$ between the sputtering yield $Y_\infty$ of a semi-infinite target and the sputtering yield  of a grain of radius $a$, $Y_a=f(x)\,Y_\infty$, with
\begin{align}
 f(x) = 1 + p_1 \exp{\left[-\frac{\left(\ln{\left(x/p_2\right)}\right)^2}{2p_3^2}\right]} -p_4\exp{\left[-\left(p_5 x-p_6\right)^2\right]}.
\end{align}
$x=a/(0.7 r_\text{P})$ is a function of the grain radius $a$ and penetration depth $r_\text{P}$. The factor 0.7 is related to the fact that a projectile has lost most of its energy at $\sim0.7 r_\text{P}$ (\citealt{SerraDiazCano2008}). To avoid negative sputtering yields for small $x$, $f(x)$ is limited to 0 as a lower boundary. The material parameters $p_i,\,i\in\mathbb{N}_{\le 6}$, are given in Table~\ref{tab_mat_sput} and the penetration depth $r_\text{P}$ is discussed in Section~\ref{sec_bethe_bloch}. The differences between $f(x)$ for carbon and silicate dust are shown in Fig.~\ref{fig_finite_sput}. For $x\approx1$ the sputtering yield is increased by a factor of $\sim4.5$ ($\sim1.5$) for carbon (silicate) dust.

\subsubsection{Penetration depth of ions in dust grains}
 \begin{figure}
 \centering
    \includegraphics[trim=2.5cm 2.65cm 2.15cm 7.5cm,  clip=true,page=1,width=1.0\linewidth, page=1]{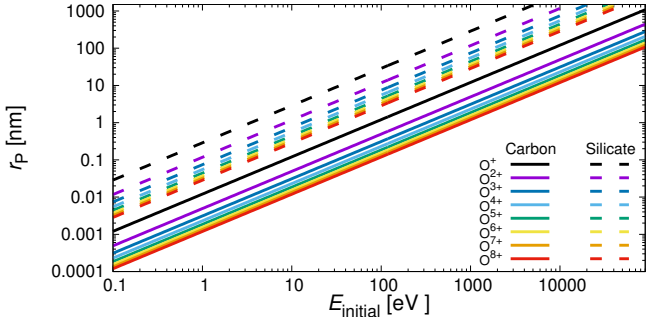}
    \caption{Penetration depth $r_\text{P}$ of oxygen ions in carbon and silicate material  as a function of initial energy $E_\text{initial}$ and the ion charge. The calculations are based on the Bethe-Bloch formula.}  
    \label{fig_penet_depth} 
\end{figure}
\label{sec_bethe_bloch}
For the estimation of the penetration depth $r_\text{p}$ of ions of energy $E_\text{initial}$ into a dust grain\footnote{\cite{Jurac1998} used the code TRIM (TRansport of Ions in Matter) developed by \citealt{Ziegler1985}, and \cite{SerraDiazCano2008} and \cite{Bocchio2014} used the successor SRIM (Stopping and Range of Ions in Matter) to compute the penetration depth $r_\text{p}$.}, we use the Bethe-Bloch formula (\citealt{Bethe1930, Bloch1933}):
\begin{align}
 &\frac{\text{d}E}{\text{d}r} = - C_\text{BB}\,\rho_\text{bulk} \frac{\left\langle Z_\text{atom}\right\rangle }{\left\langle M_\text{atom}\right\rangle } \frac{z^2}{v_\text{ion}^2}\left(\ln{\left[\frac{2m_\text{e}  v_\text{ion}^2c^2}{I\left(c^2-v_\text{ion}^2\right)}\right]} -\frac{v_\text{ion}^2}{c^2}\right),\label{1224}\\
 &\text{with}\nonumber\\
  &C_\text{BB}  = 4\pi \frac{e^4}{(4\pi\epsilon_0)^2m_\text{e}} = \unit[7.34253\times10^{-25}]{J\,m^4\,s^{-2}}\nonumber
 \end{align}
(in SI-units). Here, $\frac{\text{d}E}{\text{d}r}$ is the rate of change of the kinetic energy of the ion per unit length and \mbox{$v_\text{ion}=(2\,E/m_{\text{gas},k})^{0.5}$} is the current velocity of the ion within the grain. \mbox{$I=E_\text{exc} \left\langle Z_\text{atom}\right\rangle$} is the mean excitation energy, where $E_\text{exc}$ is a material constant (Table~\ref{tab_mat_sput}). To better represent the energy loss at low energies, we use the Barkas-equation (\citealt{Barkas1963}) for the effective charge number,
\begin{align}
z_\text{eff}=z\left(1-\exp{\left[-125\frac{v_\text{ion}}{c}z^{-2/3}\right]}\right),\label{1225}
\end{align}
where $c$ is the speed of light, and we replace the charge number $z$ in equation~(\ref{1224}) by $z_\text{eff}$.\\
Using equations~(\ref{1224}) and (\ref{1225}), the penetration length $r_\text{p}$ of an ion penetrating into a solid body is calculated as a function of initial energy $E_\text{initial}$ (in the range $\unit[10^{-1}-10^5]{eV}$) and ion charge number, for oxygen in carbon and silicate dust, respectively (Fig.~\ref{fig_penet_depth}). The penetration depth and initial ion energy follow a relation $\log{(r_\text{P}/\unit[]{nm})} = \log{(E_\text{initial}/\unit[]{eV})} + f(\text{material, }z)$. The function $f(\text{material, }z)=\tilde{c}_1z^{\tilde{c}_2}+\tilde{c}_3$ is then fitted to the data in Fig.~\ref{fig_penet_depth} using a least squares approximation and assuming that the minimum charge number of the ions is $1$ (see Section~\ref{sec_Electrondensity} and Fig.~\ref{Chianti}). We obtain $\tilde{c}_1=2.8$ and $\tilde{c}_2=-0.21$ for both dust materials and a material dependent $\tilde{c}_3$ that is listed, as $\alpha_\text{P}$, in Table~\ref{tab_mat_sput}. The final equation for the penetration depth is then 
\begin{align}                                                                                                                                                                                                                                                                                                                                                                                                                                                                                                                                                                                                                                                                                                                                                                                                                                                                                                                                                                                                                                                       
r_\text{P}=10^{\left(2.8\,z^{-0.21}+\alpha_\text{P}\right)} \frac{E}{\unit[]{eV/nm}}.                                                                                                                                                                                                                                                                                                                                                                                                                                                                                                                                                                                                                                                                                                                                                                                                                                                                                                                                                                                                                                                        \end{align}


\subsubsection{Gas accretion}
\label{sec_gas_acc}
The frequency with which gas particles collide with a dust grain is determined by the skewed Maxwellian distribution given by equation~(\ref{eq_skm}). Gas particles with an energy above the threshold energy $E_\text{sp}$ (equation~\ref{eq_sp_thr}) can cause sputtering. However, those particles with a lower energy are neglected in the studies of \cite{Tielens1994} and \cite{Nozawa2006}. Here, we assume that gas particles can be accreted by the dust grain if their energy is not large enough for sputtering. In this sense gas accretion can be interpreted as negative sputtering that causes negative yields. The grain can even grow if gas accretion dominates over regular sputtering. 

The probability of a gas particle to be accreted is set to 0 for $E=E_\text{sp}$ and $1.0$ for $E=0$. The yield of a gas particle of species $k$ with an energy $E$ below $E_\text{sp}$ is then assumed to linearly decline with decreasing $E$,
\begin{align}
 Y(E) = -(1-E/E_\text{sp})\hspace*{0.2cm}\text{if }\left((E<E_\text{sp})\,\text{and}\,(m_{\text{gas},k}=\left\langle M_\text{atom}\right\rangle\right).
\end{align}
Besides coagulation in a grain-grain collision (Section~\ref{sec_coag}), gas accretion is the second effect included that enables grain growth. Similarly to coagulation, accretion is restricted to the sticking of a gas particle of the same material as the dust grain $(m_{\text{gas},k}=\left\langle M_\text{atom}\right\rangle)$. Therefore, only particles of the dusty gas are accreted. To ensure mass conservation during the accretion process, the number density of the dusty gas particles and of the dust grains are adjusted accordingly. We note that we also tested accretion by the regular gas without significant impact on the dust grain growth.
 
\subsection{Dust motion between spatial cells and grain size bins}
\label{sec_12234}
 \begin{figure*}
\centering
   \fbox{\includegraphics[trim=-1cm 2cm -1cm 2cm, clip=true, width=1.0\linewidth, page=1]{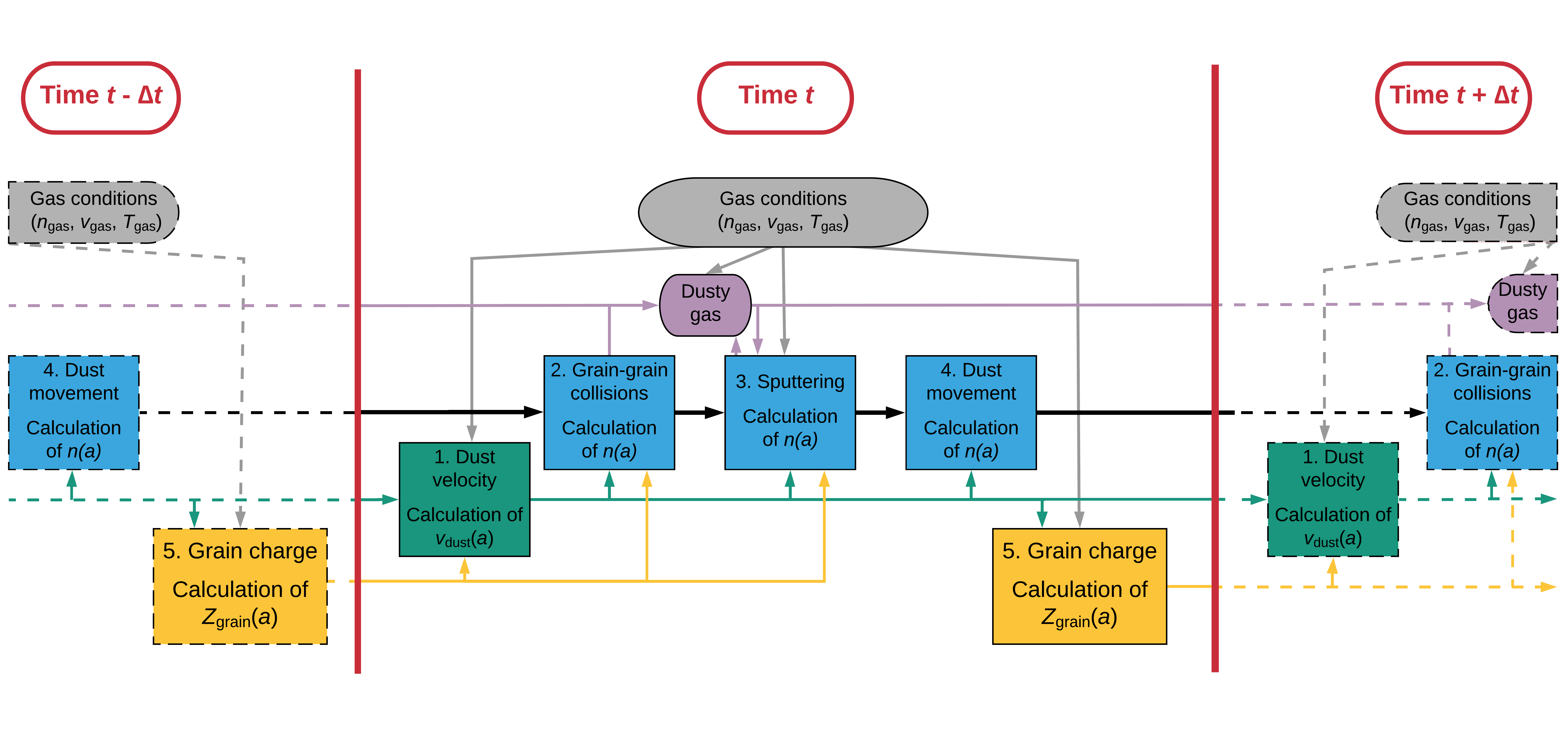}}
   \caption{Flow chart of \textsc{Paperboats} to calculate the number of dust grains for each grain size, dust material, cell and time-step $t_i$.}
\label{flow} 
 \end{figure*}
For the investigation of the processing of dust grains in the SNR it is necessary to understand the temporal evolution of the number density of grains at a certain position in the domain. For this purpose we set $n_{i\Psi}(t)$ as the number density of dust grains with size $a_i$ in cell $\Psi$ at time $t$. Due to advection, dust destruction and dust growth during the time interval $\Delta t$, the $n_{i\Psi}(t)$ particles are transformed to dust across different cells $\Omega \in \mathcal{C}$, where $\mathcal{C}$ is the set of all cells in the domain, and across different dust bin sizes $j\in\mathbb{N}^0_{\le N_\text{grain}}$. As this is the case for all other cells  $\Psi\in \mathcal{C}$ and dust bin sizes $i\in\mathbb{N}^0_{\le N_\text{grain}}$ too, the number density distribution at time $(t+\Delta t)$ is a sum of the processed number densities of all cells and dust bin sizes at time $t$:
\begin{align}
 \forall  j\in\mathbb{N}^0_{\le N_\text{grain}} \forall \Omega \in \mathcal{C}\!:\,n_{j\Omega}(t+\Delta t) = \left(\hspace*{-0.9cm}\sum_{{\Psi\in \mathcal{C}, \above 0pt \hspace*{0.9cm}i\in\mathbb{N}^0_{\le N_\text{grain}}} } \hspace*{-0.9cm}\mathbf{A}_{i\Psi j\Omega}(t)\times n_{i\Psi}(t)\right).\label{eq_theo}
\end{align}

Here, $\mathbf{A}_{i\Psi j\Omega}$ is a $(N_\text{grain}+1)\times N_\mathcal{C}\times(N_\text{grain}+1)\times N_\mathcal{C}$ matrix that fully characterises the change of number density due to the dust-processing, and $N_\mathcal{C}$ is the total number of cells in the domain. Equation~(\ref{eq_theo}) indicates that the dust-processing only changes the number densities at time $t+\Delta t$ and not at time $t$, which is mandatory as otherwise the outcome of the dust-processing would depend on the sequence in which the cells and bins are evaluated. Most of the matrix elements are 0 as the advection will shift the grains during $\Delta t$  only to a restricted number of cells. Because of the discretisation of time $(\Delta t)$ and  space (grid cells), it is necessary to outline in which order the processes described in Sections~\ref{advec}-\ref{sec_sput} are considered (Section~\ref{sec_seq}) and how the dust grains are assigned to the individual grain size bins (Section~\ref{sec_ass_bins}) and spatial cells (Section~\ref{sec_ass_cells}) at time $t + \Delta t$.

\subsubsection{Sequence of processes}
\label{sec_seq}
Fig.~\ref{flow} shows a flow chart for the sequence of processes in \textsc{Paperboats} to calculate the number of dust grains for each grain size $a$, dust material, cell and time-step. The following sequence is conducted for each grain size, each spatial cell and each time-step. 
   \begin{enumerate}
      \item  At the beginning of each time-step, the dust velocities are calculated in each cell based on the present gas density, temperature, and velocity at time $t$, as well as on the dust velocity and grain charge from the previous time-step. The dust grains still remain in their original cells.
      \item Using the dust velocities calculated in (i), the present gas conditions at time $t$ and the grain charges calculated at the previous time-step, the dust grains first undergo grain-grain collisions. The change of their number densities is done for time $t+\Delta t$:   
      The destroyed dust grains are removed from the dust bins and newly produced grains (e.g. by fragmentation or sticking) are assigned to the corresponding size bins. The material from destroyed grains (e.g. by vaporisation) is assigned to the collector bin $i=0$. The newly produced grains instantaneously achieve the velocity (calculated at time $t$) of the size bin they have been assigned to, and the material in the collector bin is assumed to instantaneously achieve the velocity of the regular gas.
     \item In the next step, the sputtering process (including any gas accretion) is evaluated based on the number densities at time $t +\Delta t$. The assignation of sputtered grains to size bins (for $t+\Delta t$) and to their dust velocities (derived at time $t$) is the same as for the grain-grain collisions in (ii). 
    \item After the evaluation of the dust destruction and growth, the dust grains (from the number density at $t+\Delta t$) are shifted to neighbouring cells according to their dust velocities and directions (derived at time $t$). 
    \item Finally, new grain charges are calculated for time $t+\Delta t$ based on the present gas conditions and dust velocities in preparation for the next time-step. 
   \end{enumerate}     
   
\subsubsection{Assigning grains to the grain size bins}
\label{sec_ass_bins}
Grain-grain collisions and sputtering generate dust grains that have to be assigned to the correct grain size bins. However, the newly produced dust grains do not necessarily have exactly the same size as the canonical grain sizes $a_i$ associated with the bins (equation~\ref{eq_177}). In general, the new dust grain with radius $a_\text{new}$ is located between the two dust bin radii $a_i$ and $a_{i+1}$ as $a_i\le a_\text{new}<a_{i+1}, i \in \mathbb{N}_{< N_\text{grain}}$. Assigning it to one of the two dust bin sizes (e.g. the closer one) would result in artificial dust mass destruction or growth. Instead, both bins gain a proportion of the dust mass in grain $a_\text{new}$ taking mass conservation into account.\footnote{In order to maintain mass conservation, the number density of particles is not indicated as an integer but as a float number.} For bin $i$, a proportion 
\begin{align}
 P_\text{pro} = \frac{a_{i+1}^3-a_\text{new}^3}{a_{i+1}^3-a_i^3} \label{eq_388}
\end{align}
of the dust mass and for bin $i+1$ a proportion
\begin{align}
 1- P_\text{pro} = \frac{a_\text{new}^3-a_{i}^3}{a_{i+1}^3-a_i^3} \label{eq_389}
\end{align}
of the dust mass of grain $a_\text{new}$ are considered. For the cases $a_\text{new}<a_1$ (collector bin) and $a_\text{new}>a_{N_\text{grain}}$ one of the constraining grain sizes is missing. Therefore, we adopt here $a_1/\Delta_a$ for the grain radius of bin $i=0$ and  $\Delta_a\,a_{N_\text{grain}}$ for the radius of a imaginary bin for grains with radii larger than the maximum grain radius $a_{\text{max,abs}}$, and calculate the splitting into the bins using equations~(\ref{eq_388}) and (\ref{eq_389}). If $a_\text{new}$ is even smaller (larger) than $a_1/\Delta_a$ ($\Delta_a\,a_{N_\text{grain}}$), all the mass of the grain with $a_\text{new}$ is associated with bin $i=0$ (to the quantity $M_\text{large}$).

\subsubsection{Assigning the grains to the spatial cells}
\label{sec_ass_cells}
After the evaluation of dust destruction and growth the grains are moved across the grid according to their dust velocities. For simplicity, we describe here the 1D-motion of the distribution of dust grains along the x-axis only. With  $\Delta_\text{cell}$ as the cell width, dust grains with size $a_i$ and dust velocity (in the x-direction) $v_{\text{dust},i}$ move in the time interval $\Delta t$  from cell $\Psi$ a number $j_\text{move}=\left(v_{\text{dust},i}\Delta t\right)/\Delta_\text{cell}$ cells in the x-direction. In general, $j_\text{move}$ is not an integer, and in this case $\left(1-\mod{\hspace*{-0.16cm}\left[j_\text{move},1\right]}\right)$ of the dust grains are moved into the cell $\Psi+\left(j_\text{move}-\mod{\hspace*{-0.16cm}\left[j_\text{move},1\right]}\right)$, and $\mod{\hspace*{-0.16cm}\left[j_\text{move},1\right]}$ of the dust grains are moved into the cell $\Psi+(j_\text{move}-\mod{\hspace*{-0.16cm}\left[j_\text{move},1\right)})+1$. 
For 2D and 3D, the dust grains of each size and composition are distributed in up to 4 or 8 cells, respectively. 

After the motion of dust grains of size $i$, each cell can contain grains of the same size that might originate from more than one other cell. In this case, the grain number densities in this cell for size $i$ are just simply summed up while the dust velocities of the grains from the individual cells are averaged according to their frequency in order to assign only one dust velocity for a certain grain size in a certain cell.

\section{Dust advection and destruction in Cas~A}
\label{sec_results}
  In this Section, we investigate the impact of different clump gas densities on the dust advection as well as on the total dust destruction rate in Cas~A. Therefore, we perform 2D hydrodynamical simulations of the cloud-crushing problem, as introduced in Section~\ref{sec_mod_setup}, using the hydrocode \textsc{\mbox{AstroBEAR}}. The number density of the gas in the ambient medium is fixed to $n_\text{am} = \unit[1]{cm^{-3}}$ at the beginning of each simulation while the number density $n_\text{cl}$ of the clump gas is varied to simulate six density contrasts \mbox{$\chi=n_\text{cl}/n_\text{am} \in\left\lbrace 100,200,300,400,600,1000\right\rbrace$.} The simulation time is set in a manner to realise a temporal evolution of three cloud crushing times (eq.~\ref{cloudcrushingtime}) after the first contact of the shock with the clump. Moreover, the size of the domain is chosen to ensure that the dust material (in the form of dust grains or dusty gas) stays in the domain throughout the entire simulation (see Section~\ref{sec_mod_setup} for details). We will focus on the temporal evolution of the gas and dust distribution in Section~\ref{sec_res_adv1} and \ref{sec_res_adv2} and on the dust survival rates in Section~\ref{sec_res_des}.
  
\subsection{Gas advection in Cas~A}
\label{sec_res_adv1}
\begin{figure}
\centering
   \includegraphics[trim=5.6cm 0.1cm 5.5cm 0.1cm, clip=true,height=0.93\textheight, page=2]{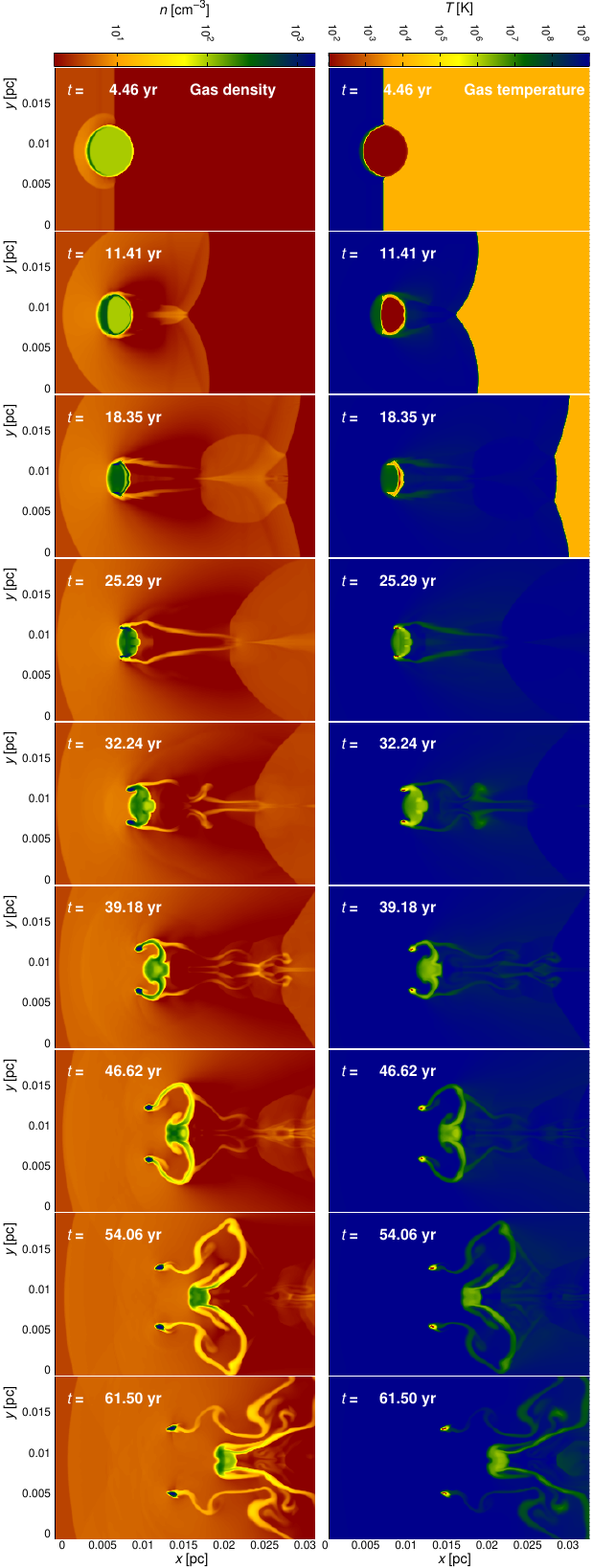}
   \caption{Temporal evolution of the spatial gas density (\textit{left}) and gas temperature (\textit{right}) when the reverse shock impacts the clump. The density contrast is $\chi=100$. The panels show a fixed cutout of the computational domain and the colour scale is fixed for each column \textbf{(Gas advection, $\mathbf{\chi=100}$)}.}
\label{results_adv1} 
 \end{figure}

The temporal evolution of the gas and dust distribution is shown in Figs.~\ref{results_adv1}--\ref{results_cut} for the density contrasts $\chi=100$ and 1000. The simulation time for the density contrast $\chi=1000$ is longer compared to the $\chi=100$ case due to the larger cloud-crushing time.

Figs.~\ref{results_adv1} and \ref{results_adv4} show the spatial distribution of the gas density and temperature for density contrasts $\chi=100$ and 1000, respectively. It can be seen that the clump is impacted by the reverse shock and gets destroyed. During the first cloud-crushing time ($\unit[\sim20]{yr}$ for $\chi=100$ and $\unit[\sim60]{yr}$ for $\chi=1000$), the outer shells of the clump are stripped off and the material is blown away. The density in the inner parts of the clump increases when the shock travels through the clump and compresses it. For  $\chi=100$, the highest densities occurring in the domain are a factor of $~\sim16$ larger than the initial clump densities, while the rise is a factor of $\sim40$ for $\chi=1000$. According to this density enhancement, the shocked clump is compressed to much smaller structures in the case of $\chi=1000$. As the cooling timescale is inversely proportional to the gas density, the gas temperature is much lower in these high-density structures and can reach values down to $\unit[\sim10^2]{K}$, similar to the initial clump gas temperatures. Contrary, the gas temperature in the post-shock ambient medium rises to values of the order of $\unit[\sim10^9]{K}$. For $\chi=100$, the clump starts to disintegrate after the first cloud-crushing time. The low-density components are accelerated and further material is stripped off, while the high-density structures are only slowly accelerated. For $\chi=1000$, most of the material is compressed into a single component which is only slowly accelerated. Gas is stripped off from this highly compressed material and blown away.

In total, the snapshots of the gas advection for $\chi=100$ show that the clump is mostly fragmented and distributed as diffuse material, while  high-density structures occur in the case of $\chi=1000$, which have low gas temperatures and which mostly withstand the disintegration process.

\subsection{Dust advection in Cas~A}
\label{sec_res_adv2}
    \begin{figure*}
  \centering
     \includegraphics[trim=0.8cm 0.0cm 0.6cm 0.0cm, clip=true,height=0.93\textheight,  page=4]{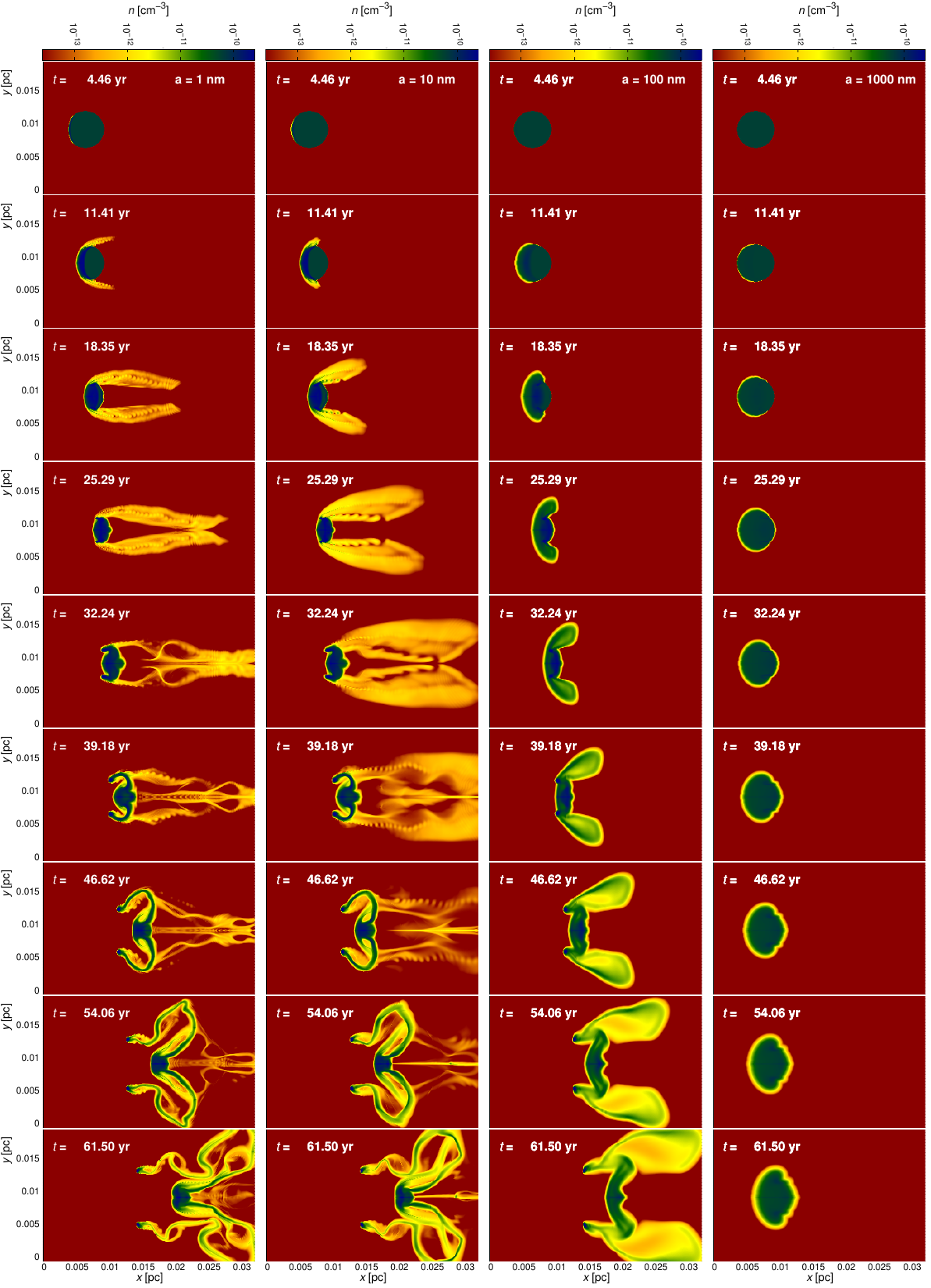}
      \caption{Temporal evolution of the spatial dust density when the reverse shock impacts the clump. The \textit{first}, \textit{second}, \textit{third}, and \textit{fourth} column show the distribution of 1, 10, 100, $\unit[1000]{nm}$ grains, respectively. The density contrast is $\chi=100$. The panels show a cutout of the computational domain and the colour scale is fixed for each column. The dust is only advected, not destroyed \textbf{(Dust advection, $\mathbf{\chi=100}$)}.}    
  \label{results_adv2} 
   \end{figure*}  
   
    \begin{figure*}
  \centering
     \includegraphics[trim=0.8cm 0.0cm 0.6cm 0.0cm, clip=true,height=0.93\textheight,  page=2]{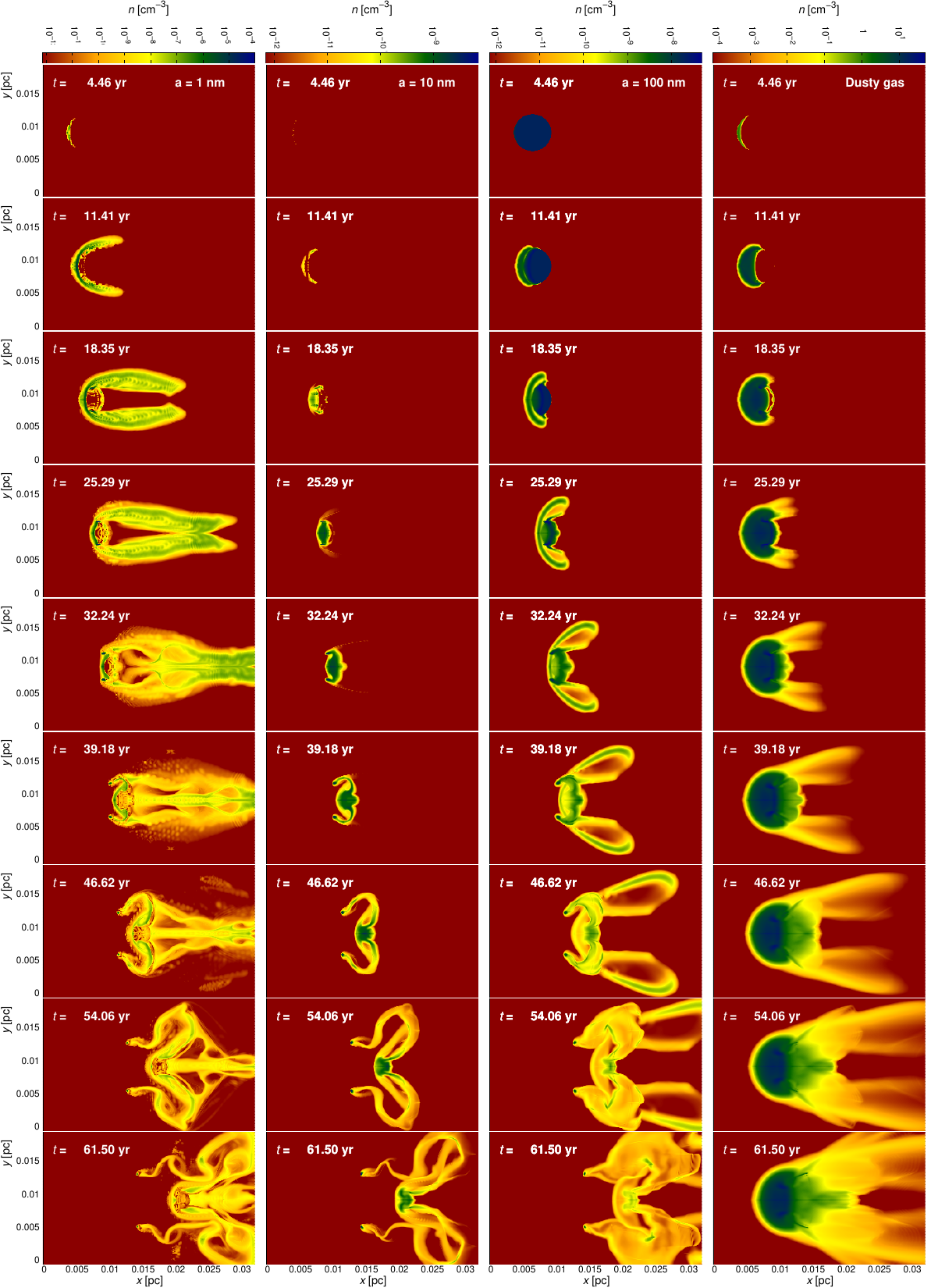}
      \caption{Same as Fig.~\ref{results_adv2}, but with dust destruction \textbf{(Dust advection + destruction, $\mathbf{\chi=100}$)}.}    
  \label{results_adv3} 
   \end{figure*}  
Based on the \textsc{AstroBEAR} hydrodynamical output, we use \textsc{\mbox{Paperboats}} to calculate the evolution of the spatial distribution of the dust density.

We show the results for pure dust advection without dust destruction for density contrasts  $\chi=100$ and 1000 in Figs.~\ref{results_adv2} and \ref{results_adv5}, respectively, to emphasize the different behaviour of carbonaceous grains of different size ($a=1$, 10, 100,  and$\unit[1000]{nm}$). A flat grain size distribution is chosen to compare small and large grains of equal number densities.
One can clearly see that the small grains ($a=1$ and $\unit[10]{nm}$) are quickly accelerated by the shock. While the distribution of dust grains in the inner parts is compressed and forced to higher dust number densities, grains in the outer shells of the clump are swept along with the gas flow and are taken away. In total, the small grains are better coupled to the gas and follow similar density structures and enhancements as outlined in Section~\ref{sec_res_adv1}. Consequently, the dust number density in the shocked clump is strongly increased for $\chi=1000$, while the temperature and velocity of the surrounding gas is low.

This behaviour is different for $a=100$ and $\unit[1000]{nm}$ grains as the grain stopping time roughly increases with dust grain radius. The acceleration is lower and the grains need more time to follow the flow of the shocked gas. As a consequence, they form patterns that significantly differ from the gas density distribution. The $\unit[100]{nm}$ grains for a density contrast $\chi=100$ and the $\unit[1000]{nm}$ grains for a density contrast $\chi=1000$ are widely distributed and smeared out across the domain, while the $\unit[1000]{nm}$ grains are only weakly accelerated for $\chi=100$ and are still located close to the initial position of the clump. In all cases, most of the dust grains are not protected by high-density and thus low temperature gas structures, but are exposed to high-velocity gas streams and high temperatures.

   \begin{figure}
 \centering
    \includegraphics[trim=5.6cm 0.1cm 5.5cm 0.0cm, clip=true,height=0.93\textheight,  page=3]{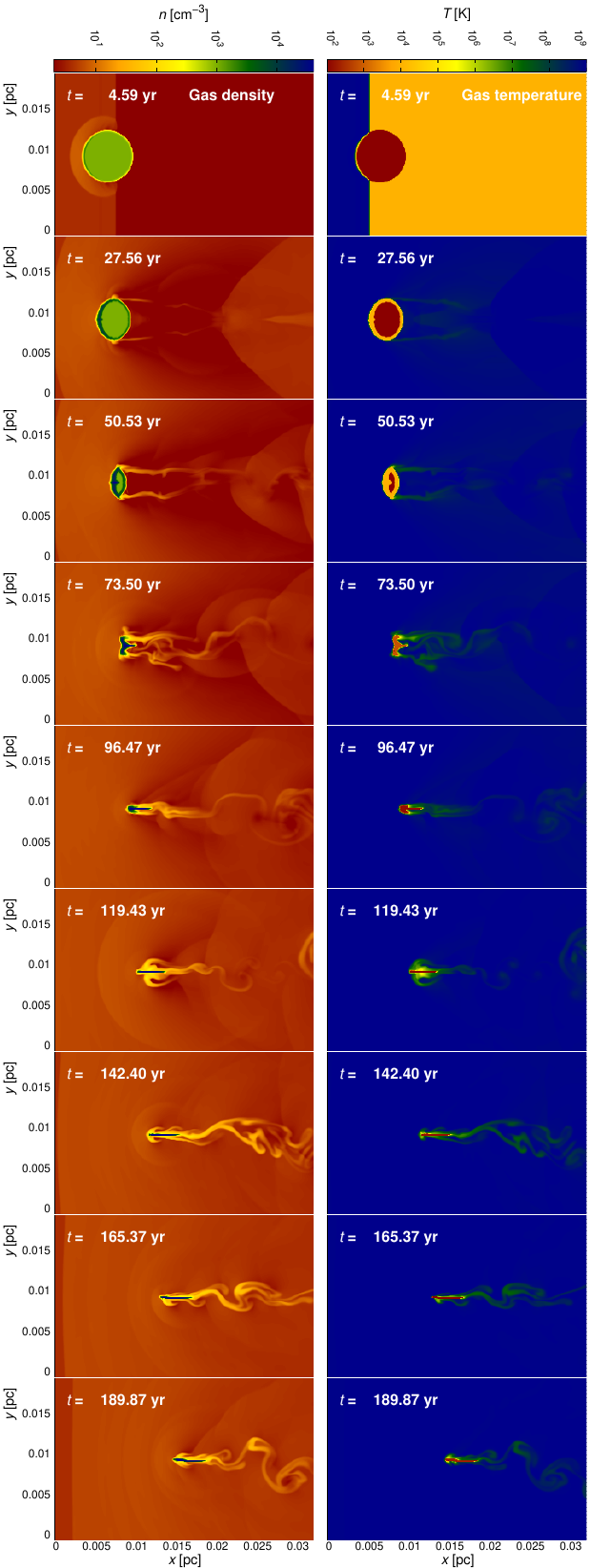}
   \caption{Same as Fig.~\ref{results_adv1}, but for a density contrast $\chi=1000$ \textbf{(Gas advection, $\mathbf{\chi=1000}$)}.}
 \label{results_adv4} 
  \end{figure}
  
Finally, Figs.~\ref{results_adv3} and \ref{results_adv6} show the spatial distribution of the dust density taking into account both dust advection and destruction. The initial grain size distribution is now log-normal, with the maximum of the distribution at $a_\text{peak}=\unit[100]{nm}$ and the distribution width is $\sigma = 0.1$. At the beginning of the simulation, mostly dust grains with size $a_\text{peak}=\unit[100]{nm}$ exist. When the shock impacts the clump, dust grains are destroyed by sputtering and grain-grain collisions, which immediately reduces the number of $\unit[100]{nm}$ grains. Fragmentation produces smaller grains, with more grains with radius $a=\unit[1]{nm}$ than $\unit[10]{nm}$, as the grain size exponent of the fragmentation size distribution is $\gamma_\text{frag}=3.5$ (see Appendix~\ref{sec_frag_app}). These small grains are produced where the shock penetrates into the clump and form a crescent-shaped or circular-shaped pattern around the inner, still unshocked part of the clump. The shock velocity decreases at deeper clump layers and reduces the grain-grain collision rate and thus the production of fragments. As the small dust grains are well coupled to the gas, they follow the gas flow and form  similar patterns as in the case of pure dust advection. However, regions with high number densities of dust grains, which could still be seen at the end of the simulations when the dust was only advected, have vanished once dust destruction is taking into account and the dust number densities of the mass-dominating species (here $\unit[100]{nm}$ grains) are lower. Both effects significantly reduce the total dust mass in the domain. On the other hand, dust grain growth is not efficient and no significant amounts of dust grains of size $a>\unit[100]{nm}$ are built up. The right column in Fig.~\ref{results_adv3} and \ref{results_adv6} also shows the spatial distribution of the dusty gas, which is an indicator for dust destruction. The dusty gas instantaneously follows the gas flow when it is being continuously produced by the ongoing dust destruction processes.

In summary, we can see that there is a strong interplay between the processes of dust advection and dust destruction: dust advection determines the grain velocities as well as the locations of the dust grains and therefore has a strong influence on the dust destruction efficiency. On the other hand, dust destruction rearranges the grain size distribution and triggers, due to the size-dependent collisional and plasma drag, the dust advection.

    \begin{figure*}
 \centering
    \includegraphics[trim=0.8cm 0.0cm 0.6cm 0.0cm, clip=true,height=0.93\textheight,  page=5]{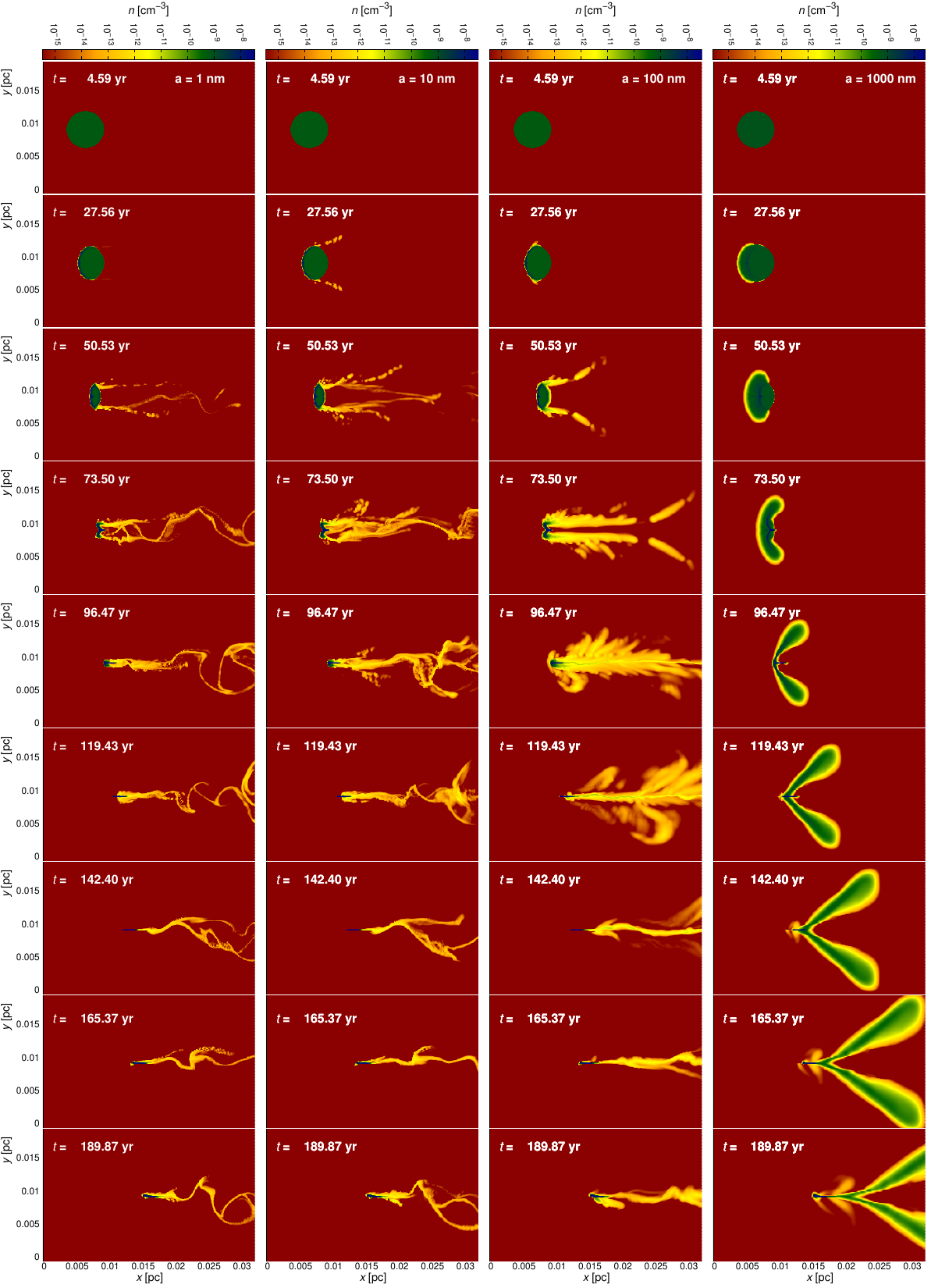}
   \caption{Same as Fig.~\ref{results_adv2}, but for a density contrast $\chi=1000$ \textbf{(Dust advection, $\mathbf{\chi=1000}$)}.}
 \label{results_adv5} 
  \end{figure*}
    \begin{figure*}
 \centering
    \includegraphics[trim=0.8cm 0.0cm 0.6cm 0.0cm, clip=true,height=0.93\textheight,  page=3]{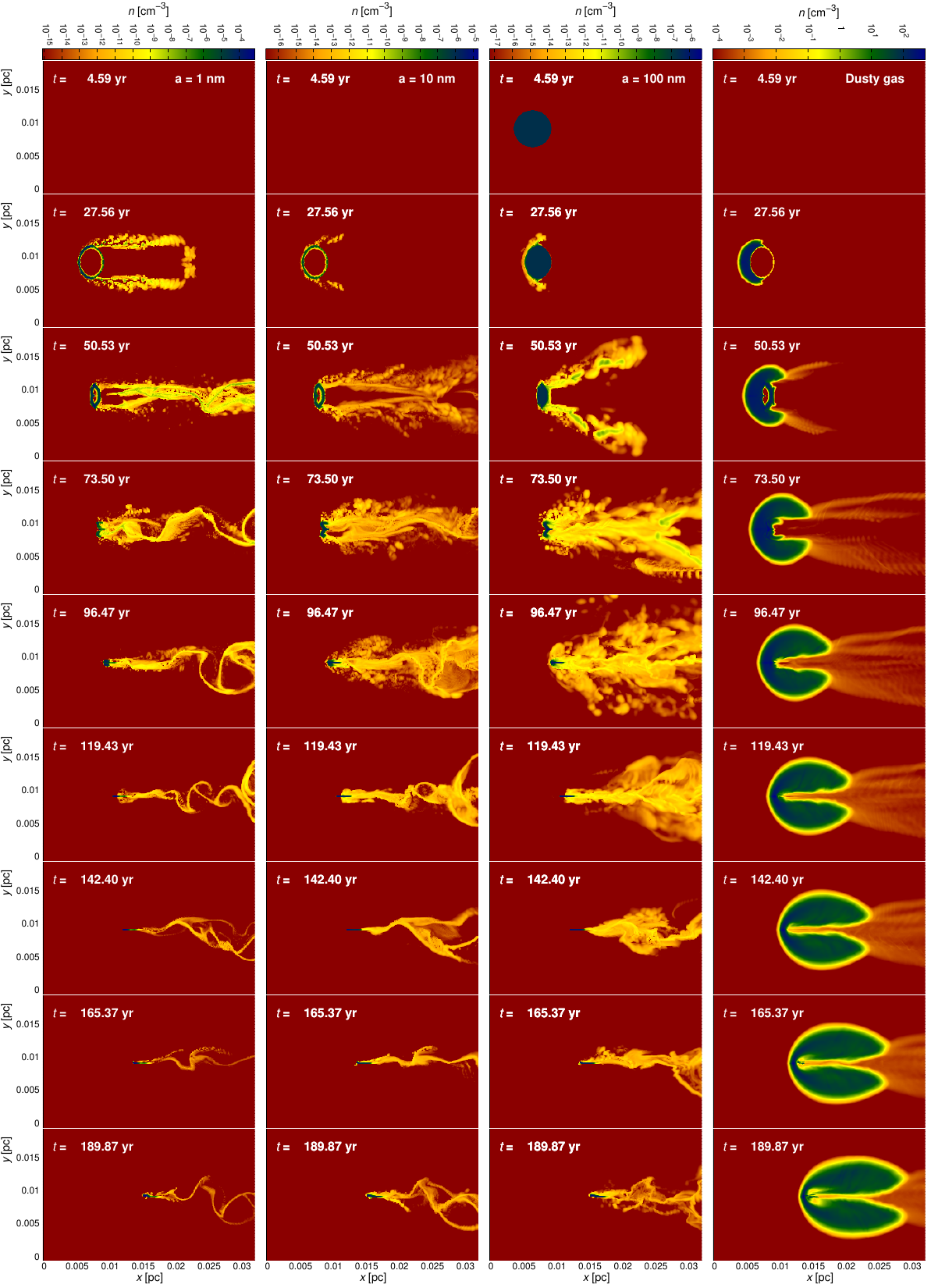}
   \caption{Same as Fig.~\ref{results_adv3}, but for a density contrast $\chi=1000$ \textbf{(Dust advection + destruction, $\mathbf{\chi=1000}$)}.}
 \label{results_adv6} 
  \end{figure*}  
    \begin{figure*}
 \centering
    \includegraphics[trim=1.2cm 7.3cm 1.1cm 0.8cm, clip=true,width=1.0\textwidth,  page=1]{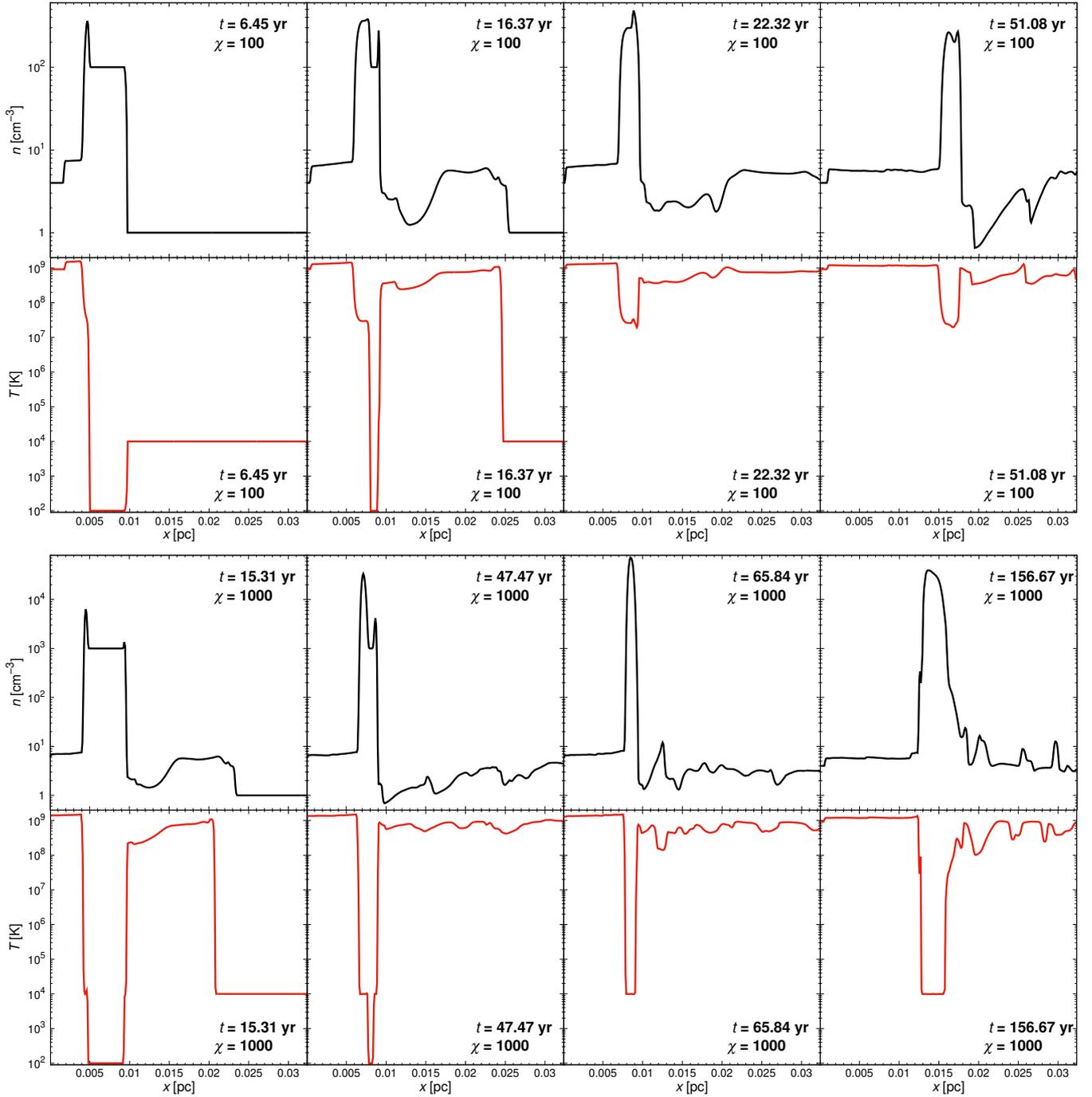}
   \caption{Gas density and gas temperature  profiles for a density contrast $\chi=100$ (\textit{top})  and $\chi=1000$ (\textit{bottom}).  The x-axis is presented in shock direction and through the  midpoint of the original clump. From \textit{left} to \text{right}, the panels correspond to times $0.22\,\tau_\text{cc}, 0.73\,\tau_\text{cc},1.02\,\tau_\text{cc}$, and $2.47\,\tau_\text{cc}$ after the first contact of the shock with the clump, and can be compared to Fig.~5 in \textcolor{blue}{Silvia~et~al.~(2010)}.}
 \label{results_cut} 
  \end{figure*}

\subsection{Dust destruction in Cas~A}
\label{sec_res_des}
 \begin{figure*}
\centering
   \includegraphics[trim=1.4cm 2.5cm 1.2cm 2.8cm, clip=true,height=5.9cm, page=1]{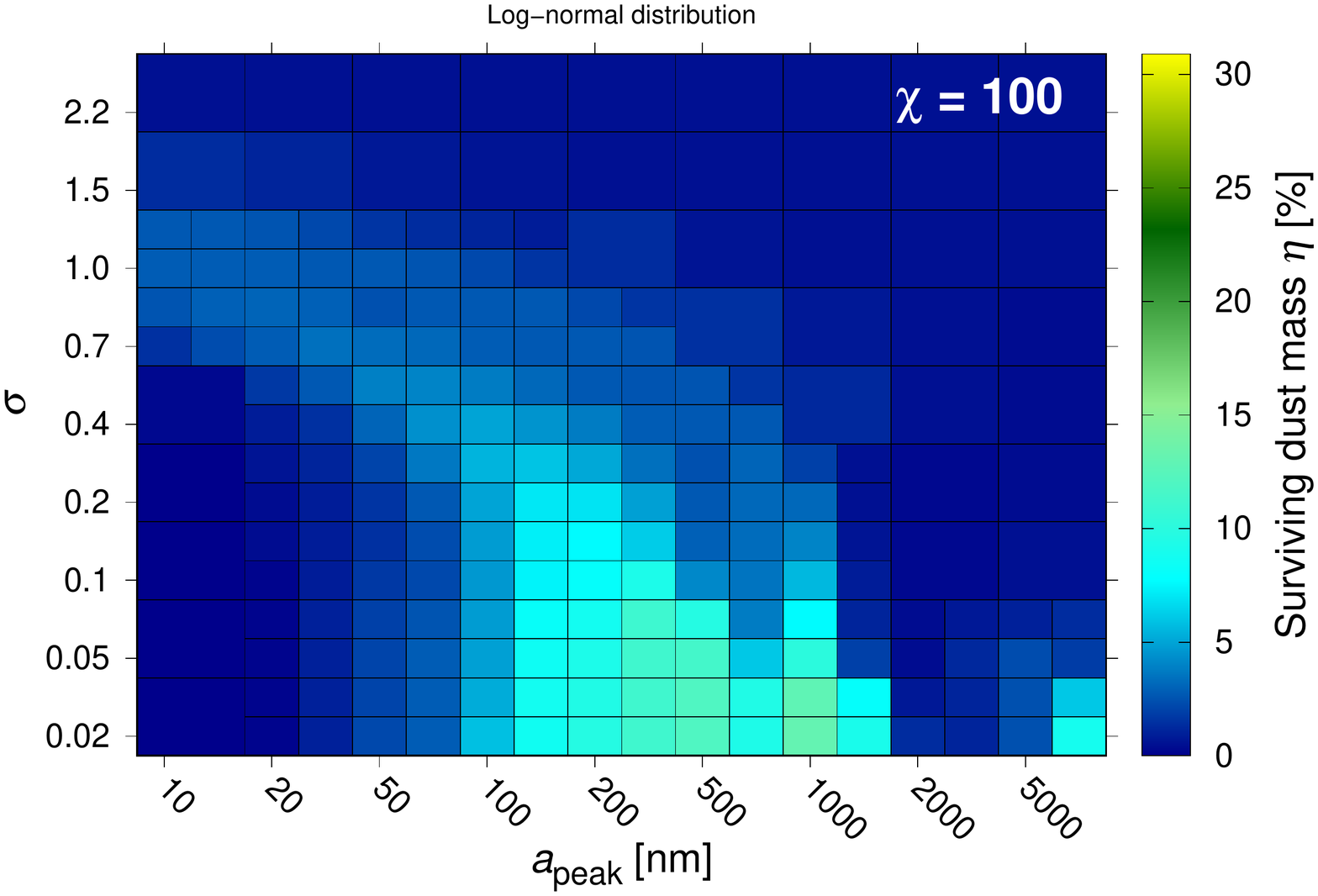}\hspace*{0.1cm} 
   \includegraphics[trim=1.4cm 2.5cm 1.2cm 2.8cm, clip=true,height=5.9cm, page=1]{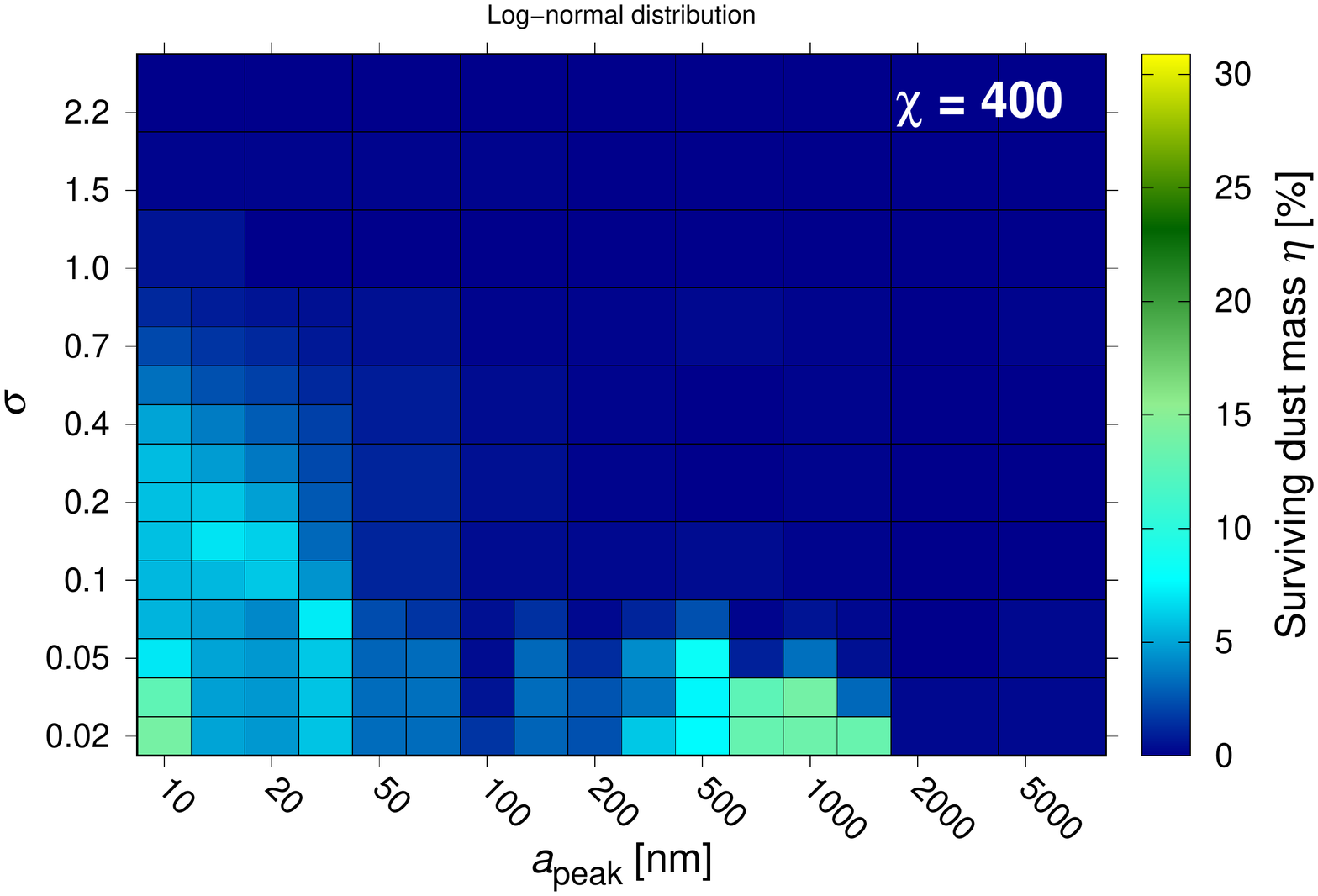}\\
   \includegraphics[trim=1.4cm 2.5cm 1.2cm 2.8cm, clip=true,height=5.9cm, page=1]{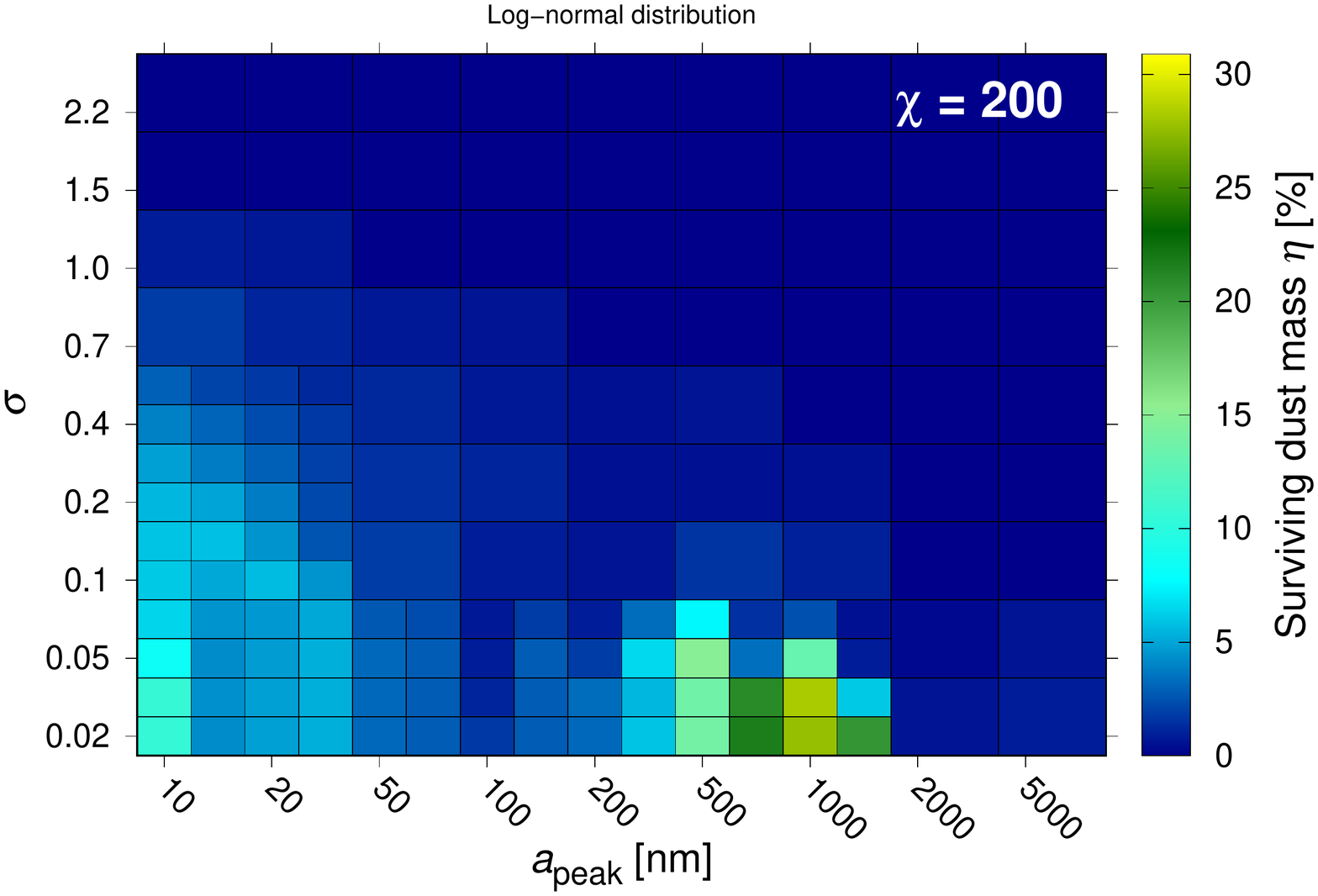}\hspace*{0.1cm} 
   \includegraphics[trim=1.4cm 2.5cm 1.2cm 2.8cm, clip=true,height=5.9cm, page=1]{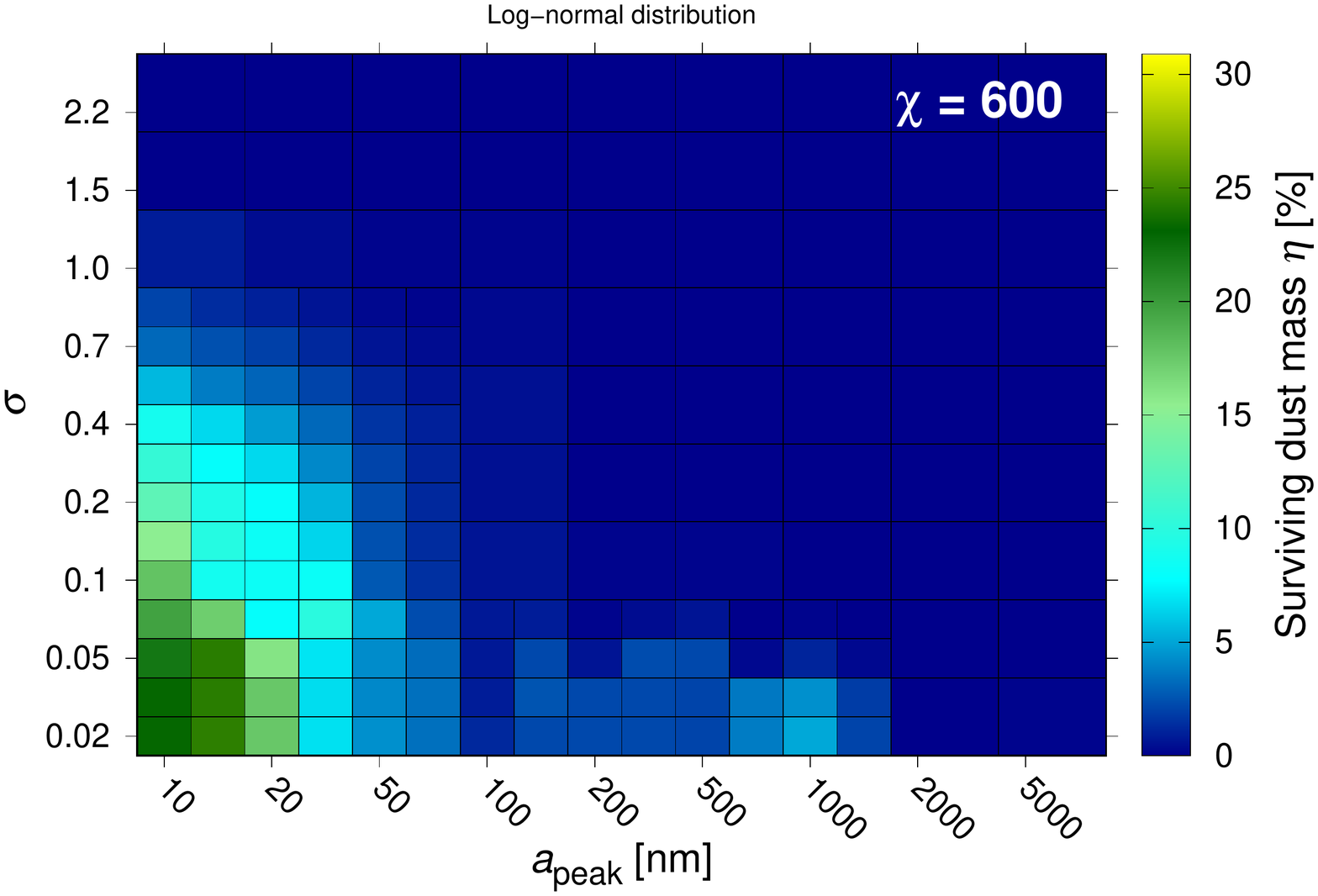} 
   \includegraphics[trim=1.4cm 2.5cm 1.2cm 2.8cm, clip=true,height=5.9cm, page=1]{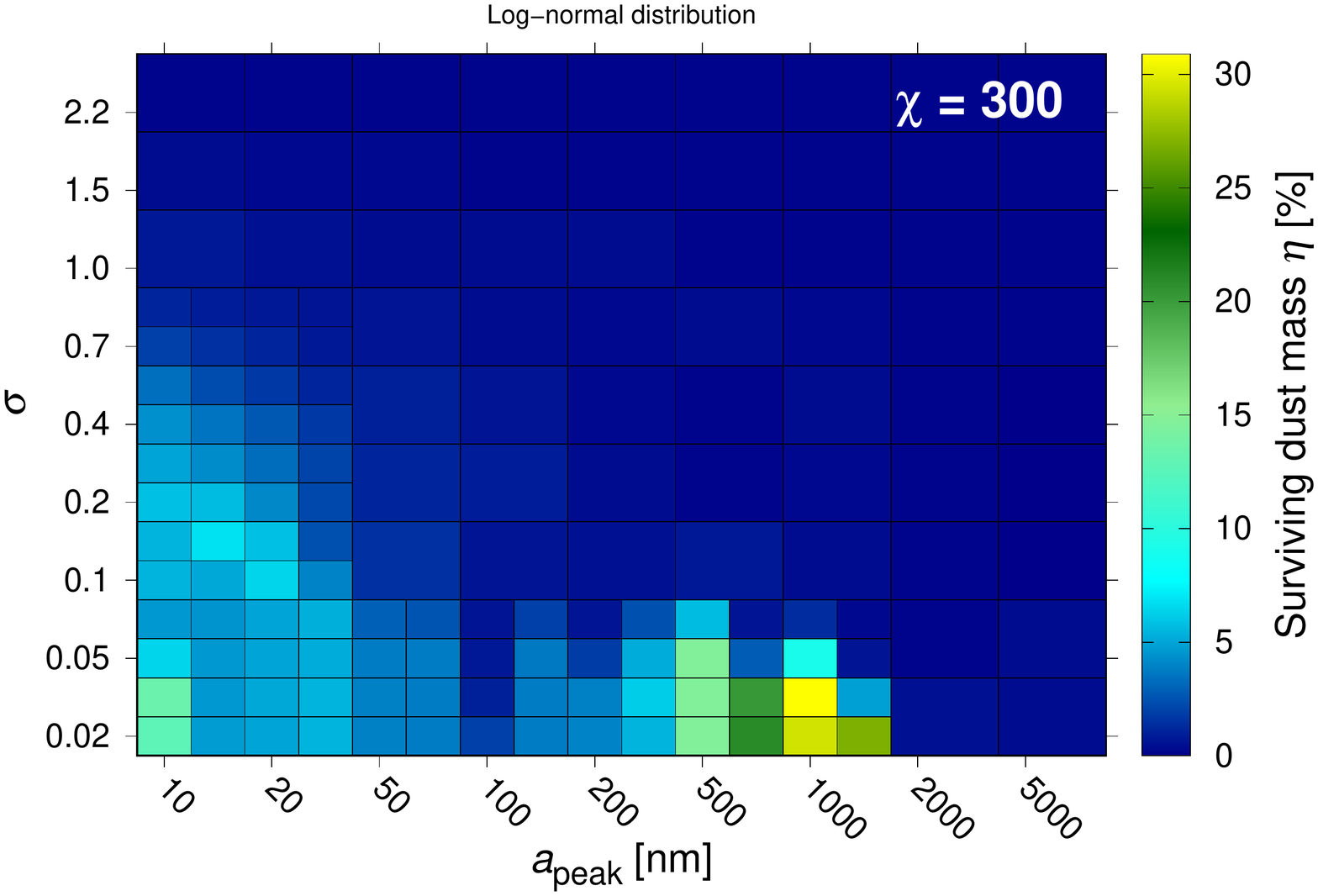}\hspace*{0.1cm} 
   \includegraphics[trim=1.4cm 2.5cm 1.2cm 2.8cm, clip=true,height=5.9cm, page=1]{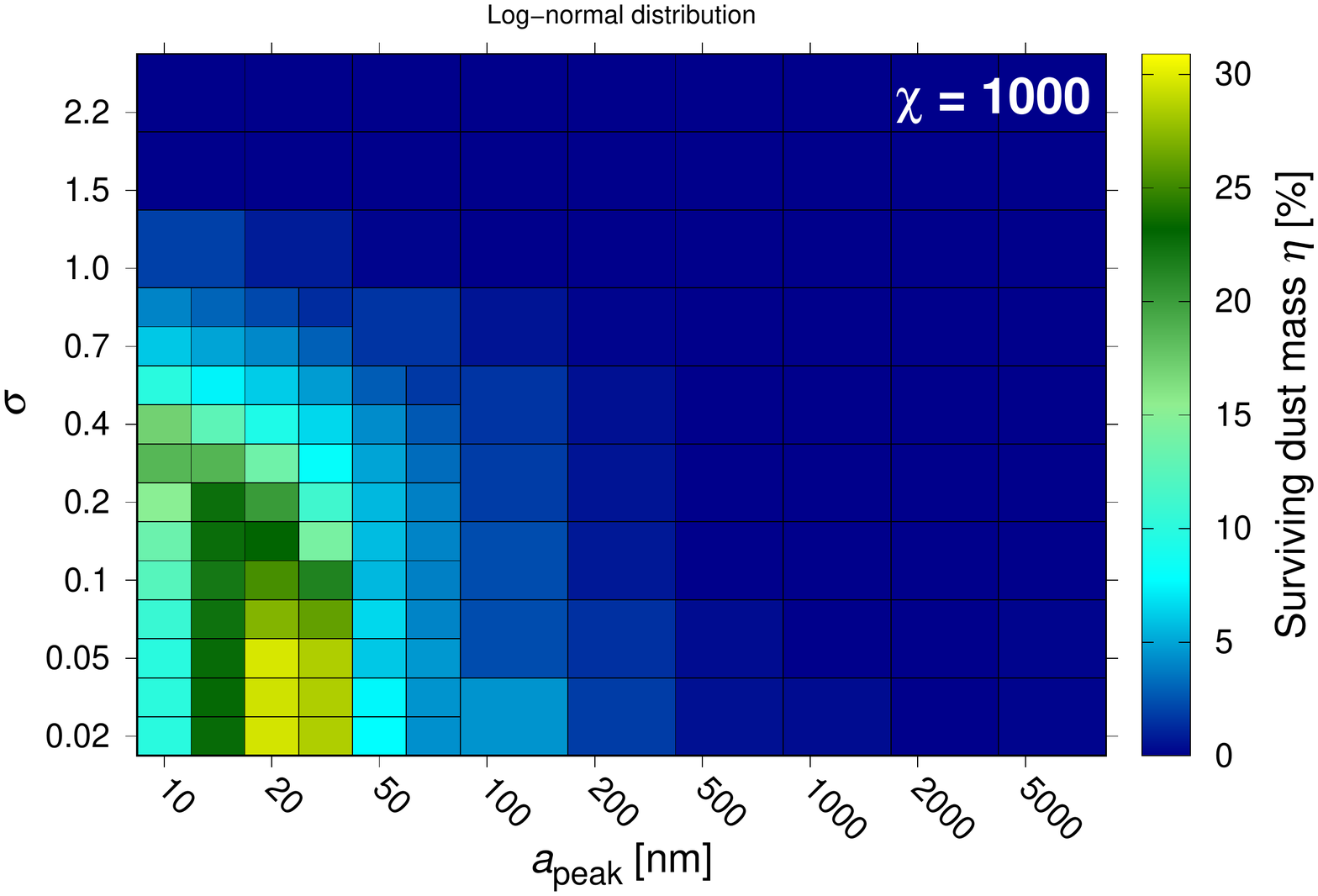}   
   \caption{Carbon dust survival rates as a function of the density contrast $\chi$ between the gas in the clump and the gas in the ambient medium. The parameters $a_\text{peak}$ and $\sigma$ describe the peak as well as the width (increasing from small to large values) of the INITIAL dust grain size distribution (see Fig.~\ref{init_dust_fig}). Each box represents one parameter configuration; in regions of particular interest, the parameter resolution is increased.}
\label{results2} 
 \end{figure*}
 
  \begin{figure}
 \centering
  \includegraphics[trim=1.4cm 2.5cm 1.2cm 2.8cm, clip=true,height=5.9cm, page=1]{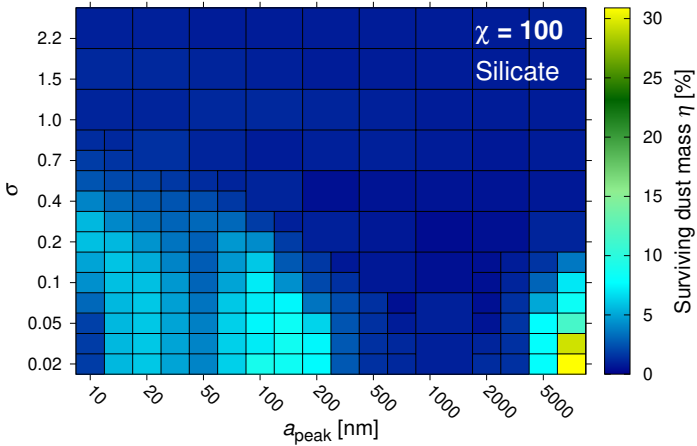}\hspace*{0.1cm} 
   \caption{Same as Fig.~\ref{results2}, only for silicate grains for the case of $\chi=100$.}
\label{results_silicate} 
\end{figure}
 
 \begin{figure*}
 \centering
  \includegraphics[trim=1.4cm 2.5cm 1.2cm 2.8cm, clip=true,height=5.9cm, page=1]{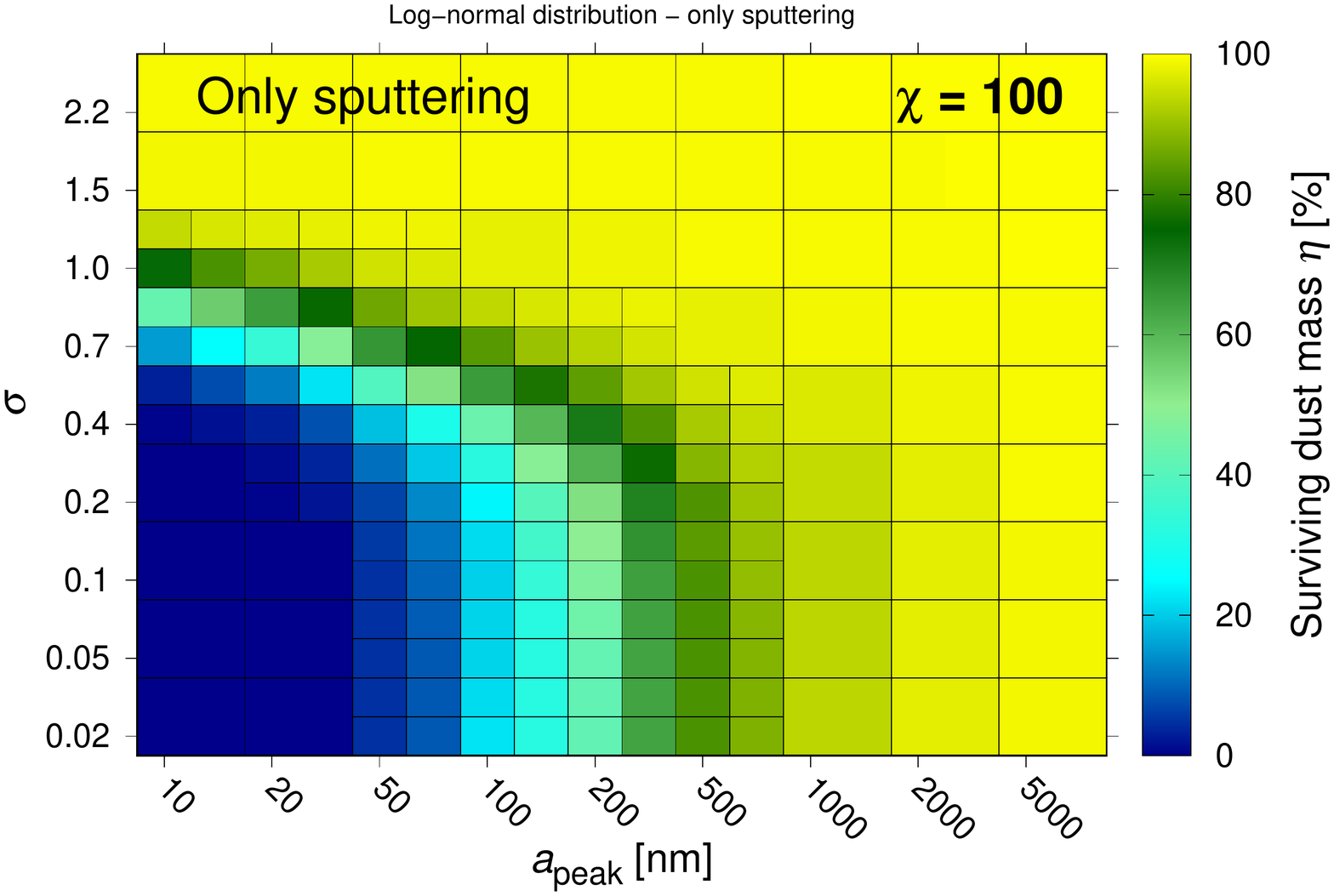}\hspace*{0.1cm} 
  \includegraphics[trim=1.4cm 2.5cm 1.2cm 2.8cm, clip=true,height=5.9cm, page=2]{Pics/Results/Dust_survival_graphite_two_level_cold_sp.pdf}
  \caption{Same as Fig.~\ref{results2}, only for carbon dust destruction by just one of the two processes (\textsl{left}: sputtering, \textsl{right}: grain-grain collisions) for the case of $\chi=100$.}
\label{results_contrib_only} 
\end{figure*}
Based on the hydrodynamical output, we  use \textsc{\mbox{Paperboats}} to determine the dust survival rates as a function of clump densities and initial dust properties. According to studies of dust formation in SN ejecta, the size distribution function of each grain species is predicted to be approximately log-normal with grain sizes in the range of $\sim\unit[1-100]{nm}$ (e.g.~\citealt{Todini2001,Nozawa2003}). In contrast, observations have indicated the presence of grains of sizes around $\unit[1]{\text{\textmu} m}$ in the ejecta of a number of CCSNe (e.g.~\citealt{Gall2014,Owen2015,Wesson2015,Bevan2016,Bevan2017}). We vary the two parameters $a_\text{peak}$ and $\sigma$ for the log-normal initial distribution\footnote{In addition, we calculate in Appendix~\ref{app_powerlaw} the dust survival rate if the initial grain size distribution follows a power-law.} over a range of $\unit[10]{nm}-\unit[7]{\text{\textmu} m}$ and $0.02-2.2$, respectively, and calculate the dust survival rate $\eta$ for the six density contrasts $\chi$ for the case of carbon (Fig.~\ref{results2}) and silicate dust grains (Fig.~\ref{results_silicate}). The survival rate $\eta$ is defined as the ratio of the total mass of all dust grains in all bins (bin 1 to $N_\text{grain}$, plus $M_\text{large}$) at time $t=t_\text{sim}$ to the total dust mass at $t=0$. Material in bin~0 (dusty gas) is denoted as destroyed dust material, while fragments of shattered grains with sizes above $\unit[0.6]{nm}$ are assigned to the surviving dust mass.

The dust destruction is triggered by sputtering, grain-grain collisions and the dust advection, whereby the destruction effects have different impacts for different initial distributions. Furthermore, the influence of sputtering and grain-grain collisions strongly depends on the clump density contrast $\chi$.

For carbon dust and  $\chi=100$, sputtering destroys most of the dust material for initially small dust grains and the dust survival rate is very low for narrow initial distributions with small $a_\text{peak}$ values (see Fig.~\ref{results_contrib_only}, \textit{left}). 
Fig.~\ref{results_contrib_only} (\textit{right}) shows the dust survival rate $\eta$ in the case of grain-grain collisions only (without sputtering, $\chi=100$), reflecting the complexity of the fragmentation and vaporisation processes. However, it can be seen that the dust material is mostly destroyed  in the case of broad initial distributions, while narrow distributions are much less affected by grain-grain collisions, and small dust grains (small $\sigma$ and $a_\text{peak}$) have a larger survival rate than large grains.

The total survival rate of the dust is a function of the interplay of both processes plus the dust advection, if sputtering and grain-grain collisions are combined. In general, narrow initial size distributions (\mbox{$\sigma\lesssim1$}) tend to have higher survival rates than broad initial distributions (\mbox{$\sigma\gtrsim1$}), which is a direct result of the impact of the grain-grain collisions: the broader the distribution, the higher are the relative velocities between small and large grains, which increases the total number of colliding dust grains as well as their collision velocities, both resulting in higher dust destruction rates.

In the following, we will focus on different grain size ranges for narrow initial size distributions, starting with the smallest grains. Grains with radii below $\unit[100]{nm}$ are well coupled to the gas (see Section~\ref{sec_res_adv2}), and thus most of the grains are located in the high-density gas regions. However, even the moderate local gas conditions are sufficient to destroy most of the dust material, as the dust survival rate in the case of pure sputtering indicates (Fig.~\ref{results_contrib_only}, \textit{left}). The presence of grain-grain collisions could further amplify the destruction. Consequently, the dust survival rate $\eta$ of narrow initial distributions with grain radii below $\unit[100]{nm}$ is low. Larger dust grains of a few $\unit[100]{nm}$ are still moderately coupled to the gas but more robust to withstand dust destruction processes. Therefore, initial size distributions with these medium sized dust grains have higher probabilities to survive the passage of the reverse shock. In the case of $\chi=100$, these are $\eta=\unit[13]{\%}$ for $a_\text{peak}=\unit[1000]{nm}$ and $\sigma=0.02$. 
Narrow initial distributions with $a_\text{peak}$ between $\unit[150]{nm}$ and $\unit[1500]{nm}$ show survival rates larger than $\unit[8]{\%}$.

The dust grains get more and more decoupled from the gas flow with increasing grain size, which is accompanied by an exposure to higher gas temperatures, larger gas velocities, and a larger dust velocity spreading. While grains of a few micrometers radius have a significant survival rate if either sputtering or grain-grain collisions are considered (Fig.~\ref{results_contrib_only}), the combined destruction effects erode and process these grains to smaller particles, which again are then easily destroyed. In total, the dust survival rate $\eta$ drops at grain sizes of a few micrometers.

Finally, the largest considered grains ($\unit[7]{\text{\textmu} m}$) again show an increase in the survival rate. As the total gas mass, and thus the total dust mass, is constant at the beginning of each simulation, an increase of the grain size results also in a decrease of the product of the number densities and cross sections of the grains. Therefore, the collision probability (equation~\ref{lab_P2}) becomes lower and the largest considered grains can survive. For \mbox{$a_\text{peak}=\unit[7]{\text{\textmu} m}$} and $\sigma=0.02$, $ \eta=\unit[9]{\%}$ of the initial carbon dust mass survives. This is consistent with the increased collisional timescale for large grains as outlined in Section~\ref{coltime}.

\begin{figure}
 \centering
 \includegraphics[trim=2.6cm 2.1cm 0.9cm 2.1cm, clip=true,page=1,width=\linewidth]{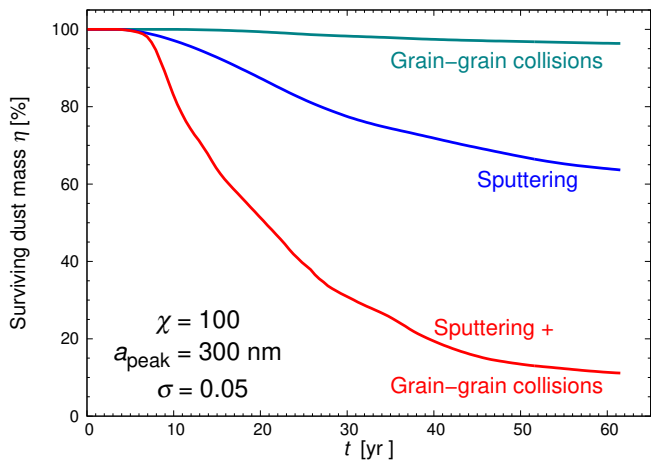} 
 \caption{Comparison between carbon dust destruction by sputtering (blue), grain-grain collisions (green) and by their combined effects (red) as a function of time for an example simulation ($\chi=100, a_\text{peak}=\unit[300]{nm}, \sigma=0.05$). Since sputtering and grain-grain collisions are synergistic, the total destruction rate is higher than that of the single processes.}
\label{results_synergy}
\end{figure}

We wish to highlight that grain-grain collisions and sputtering are synergistic processes. When a grain-grain collision results in fragmentation, and the smallest fragments are larger than  $\unit[0.6]{nm}$, no dust is destroyed in the sense of our definition of the dust survival rate  $\eta$. However, the fragments can then be eroded by sputtering which is more efficient than sputtering of the original, larger grains. Therefore, grain-grain collisions take over the preliminary work in dust-processing, with or without vaporising dust material, and sputtering can then erode the resulting fragments. Consequently, the total dust destruction rate by sputtering and grain-grain collisions can be significantly higher than their individual contributions acting alone (Fig.~\ref{results_synergy}). \cite{Slavin2015} also outlined the importance of grain-grain collisions for altering the grain size distribution and for the sputtering of the resulting fragments for the case of SN shocks impacting the ISM.
 
 \begin{figure*}
 \centering
  \includegraphics[trim=2.3cm 2.1cm 2.0cm 2.2cm, clip=true,page=1,width=0.49\textwidth]{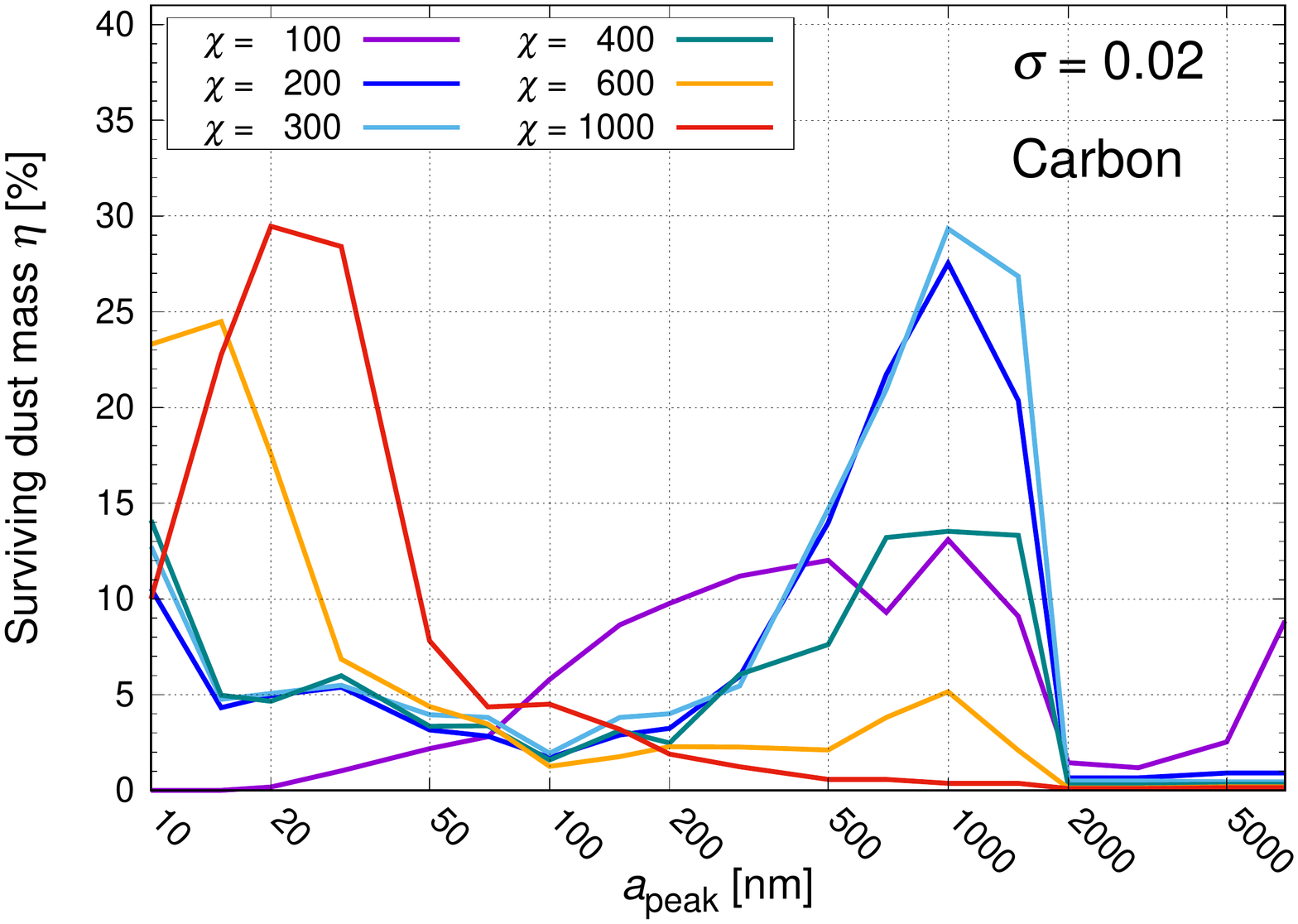} 
  \includegraphics[trim=2.3cm 2.1cm 2.0cm 2.2cm, clip=true,page=2,width=0.49\textwidth]{Pics/Results/Final_dust_changing_chi.pdf}   
  \caption{\textit{Left}: Cuts of Fig.~\ref{results2} for a fixed width $\sigma=0.02$ of the initial distribution of carbon grains. The plot shows that the dust survival rate  $\eta$ is enlarged if the initial size distribution is composed only of small dust grains ($\unit[\sim10-50]{nm}$), in the case of density contrasts $\chi\gtrsim600$, or of medium-sized grains ($\unit[\sim500-1500]{nm}$) for $\chi<600$. \textit{Right}: Same as \textit{left}, only for silicate dust.}
\label{results_changing_chi}
\end{figure*}

The survival fractions $\eta$ change for other density contrasts $\chi$ as sputtering, grain-grain collisions and the dust advection are affected. The shock velocity in the clump scales as $\chi^{-1/2}$ and thus decreases with increasing $\chi$ while the cooling timescale is inversely proportional to the gas density. Both mitigate kinematic and thermal sputtering for the case of $\chi=200$ and larger density contrasts.  Small dust grains follow the gas flow and are then better protected in the denser clumps and less exposed to the hot post-shock gas. As a result, small grains can more easily survive and the dust survival rate increases for narrow initial distributions with small $a_\text{peak}$ values. Simultaneously, the enhanced gas density in the clump is equivalent to an enhanced number density of dust grains, which increases the collision probability and reduces the chances of survival for the medium sized dust grains.

Fig.~\ref{results_changing_chi} (\textit{left}) shows cuts of Fig.~\ref{results2} for a fixed distribution width ($\sigma=0.02$) for carbon dust. It can clearly be seen that two grain size ranges exists for which the dust survival rate is up to $\unit[30]{\%}$. For low and medium density contrasts ($\chi=100-400$), a large proportion of the dust material can survive if the initial dust grain radii peak around $\unit[\sim500-1500]{nm}$, whereas, high density contrasts ($\chi>400$) enable small dust grains with sizes around $\unit[\sim10-50]{nm}$ to survive the passage of the reverse shock in the ejecta clump. We want to highlight that the former values match very well the grain sizes derived from observations \mbox{($\sim\unit[1]{\text{\textmu} m}$; e.g.~\citealt{Wesson2015})} and the sizes predicted by dust formation studies ($\sim\unit[1-100]{nm}$; e.g. \citealt{Nozawa2003}), respectively.

The dust survival rate  $\eta$ of silicate grains is shown in Fig.~\ref{results_silicate} for the density contrast $\chi=100$. We find that silicate grains with initial radii around $\unit[100]{nm}$ have a survival rate of up to $\unit[9]{\%}$ and a lower rate at smaller grain sizes. However, the highest survival rates  $\eta$ exist for narrow distributions with grain sizes of a few micrometers (up to $\unit[31]{\%}$) where the collision probabilities are reduced due to the reduced number densities at these large grain sizes. Similarly to carbon dust, this effect vanishes for larger $\chi$ (Fig.~\ref{results_changing_chi}, \textit{right}). It can be further seen, that also for silicate dust two grain size ranges exist for which the survival rate is increased. For low density contrasts ($\chi=100-200$), dust can survive if the initial dust grain radii peak around $\unit[\sim100]{nm}$, though this survival peak is with  $\eta=\unit[9]{\%}$ not as significant as for carbon dust. 
In addition, medium and high density contrasts ($\chi\gtrsim200$) enable small dust grains with sizes around $\unit[\sim10-30]{nm}$ to survive the passage of the reverse shock with a survival fraction of up to $\unit[40]{\%}$.

Two dust growth processes have been considered in our study: gas accretion as ``negative'' sputtering and the sticking of dust grains in low-velocity collisions. Both effects are found to be minimal which is a consequence of the high velocities in our simulations. As a result, the contribution to the total dust budget of $M_\text{large}$, the dust mass of all grains in the domain with radii larger than $a_\text{max, abs}$, is negligible.
 
\begin{figure}
 \centering
  \includegraphics[trim=2.4cm 2.1cm 2.1cm 2.2cm, clip=true,page=3,width=\linewidth]{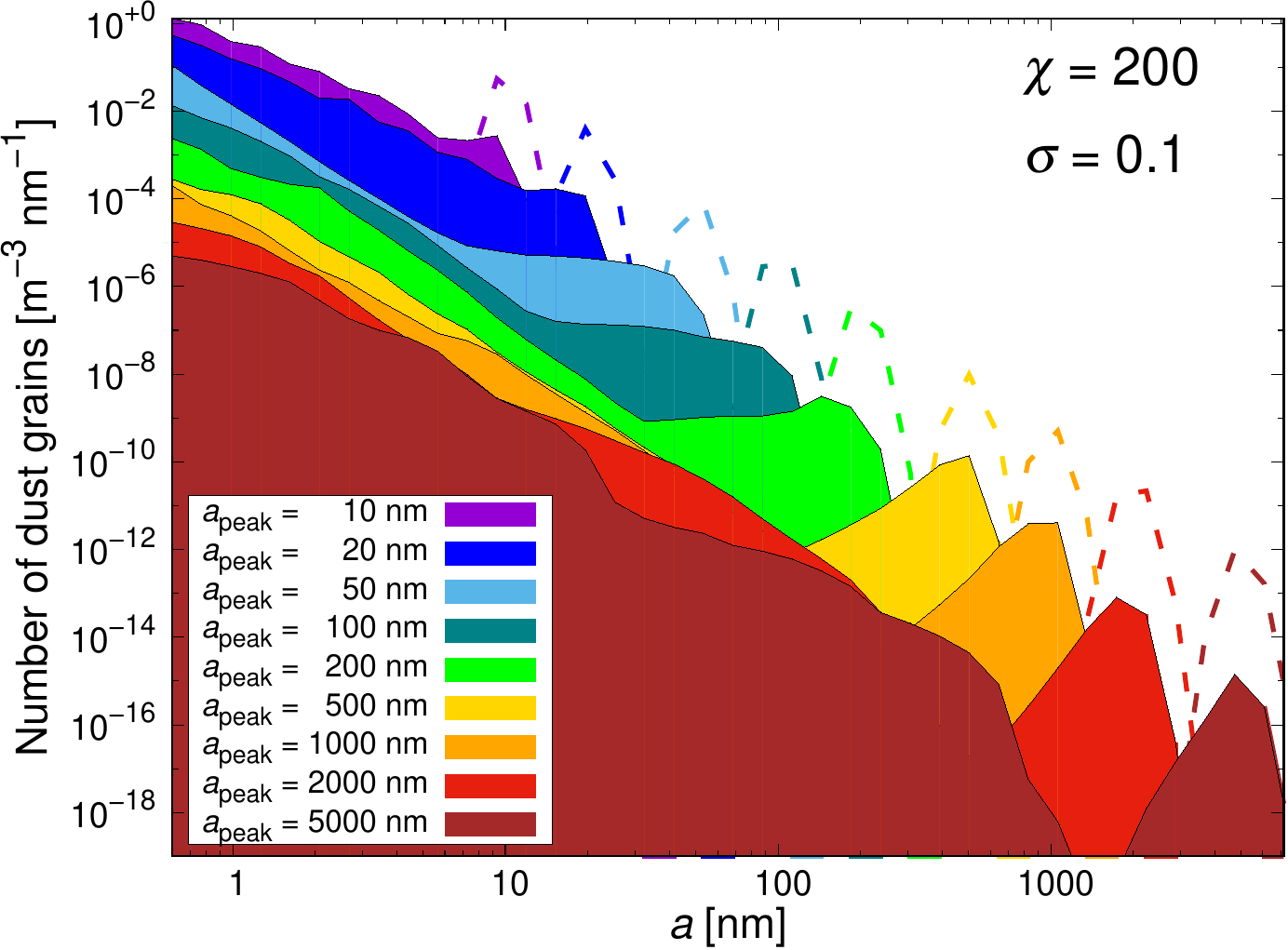} 
  \caption{Final grain size distributions of carbon dust after processing by sputtering and grain-grain collisions (coloured areas). The number of particles per unit volume and grain size is shown as a function of grain size $a$ for different $a_\text{peak}$ of the initial distribution. The initial distributions are shown as dashed lines, and the coloured areas of the final distribution cover each other. Two components can be differentiated for each final distribution: a power-law distribution of small grains, and the remnant of the initial distribution at larger grain sizes.}
\label{results_contrib_onlyfinal_distrib}
\end{figure} 
 
We have discussed so far only the initial dust properties. However, when the reverse shock has passed the clump and processed the dust, the remaining dust mass has been rearranged into a new grain size distribution (Fig.~\ref{results_contrib_onlyfinal_distrib}). This new, final distribution is essentially composed of two components: The first is the remnant of the initial distribution, though reduced in grain number density due to sputtering and collisions. The distribution of this component is smeared out to lower  grain sizes, as the sputtering has reduced the dust grain sizes. The second component is a power-law distribution of smaller dust grains that reflect the fragments of shattering collisions, and is defined by equation~(\ref{fragdistrib}). When the initial size distribution is narrow, the final size distribution shows a gap between the two components, which is a consequence of the fact that the largest fragments are significantly smaller than the original grains (except for partial destruction or cratering). However, as the first component is smeared out by sputtering to smaller dust grain sizes and since sputtering is more efficient for smaller grains, the gap is most pronounced for narrow initial distributions with large dust grains.

The final grain size distributions indicate that the grain sizes of the initial distribution are still present after the dust-processing, but reduced in number. As a consequence, if micrometer sized carbon grains are able to form in the SN ejecta, some of them will survive the passage of the reverse shock while it is harder to explain the presence of micrometer sized silicate grains.


\section{Comparison to previous studies}
\label{sec_compar}
A number of previous studies have investigated dust destruction rates caused by the passage of an SNR reverse shock. Their formalisms, approach and models are quite different, and some of them examined the temporal evolution of the clump-free remnants up to $\sim\unit[10^4-10^5]{yr}$. However, we attempt to verify our results by comparing with appropriately chosen cases. In most works, only sputtering without grain-grain collisions has been considered. Since many investigated the impact of pure sputtering on carbon dust (Fig.~\ref{results_contrib_only}), we will mainly compare to this case. We start with several works on SNRs in general before focussing on Cas~A.

\cite{Bianchi2007} investigated the dust destruction rate for a uniform, clump-free density distribution inside the ejecta of a SN with kinetic energy $\unit[1.2\times 10^{51}]{erg}$. They re-evaluated the initial grain size distribution from the study of \cite{Todini2001} for a progenitor with mass $\unit[12-40]{M_\odot}$, resulting in an initial log-normal size distribution for carbon grains which peaked at $a_\text{peak}\sim\unit[10]{nm}$, with a width of $\sigma\sim0.3$. Using the semi-analytical model of \cite{Truelove1999} to describe the dynamics of the reverse shock, they computed the dust destruction from both thermal and kinematic sputtering, but neglected gas drag and grain charge.  In their model the velocity of the reverse shock, and thus the dust survival rate, is a function of the density of the surrounding ISM ($\rho_\text{ISM}=\unit[10^{-25},10^{-24},10^{-23}]{g \, cm^{-3}}$) and of the reverse shock velocity. They found a survival rate $\eta=\unit[2-20]{\%}$ of the total ejecta dust mass, depending on the ISM density, however, they did not distinguish between the rates for the contributing dust compositions, carbon or silicate. Compared to that, we find for an initial log-normal carbon distribution with a peak-size of $a_\text{peak}=\unit[10]{nm}$ and a width of $\sigma=0.3$, plus a density contrast $\chi=100$, a survival rate below $\unit[1]{\%}$ (Fig.~\ref{results_contrib_only}). However, the comparison suffers due to the adoption of a uniform, clump-free medium in the study of \cite{Bianchi2007}. 

\cite{Nozawa2007} evaluated the time evolution of the clump-free gas density and gas temperature of spherically symmetric shocks, adopting 1D hydrodynamic models of the ejecta of Population III SNe from \cite{Umeda2002}. For each grain size they calculated the dust motion due to the gas drag to evaluate the velocity relative to the gas. The initial carbon grain size distribution was log-normal and was adopted from \cite{Nozawa2003}, with $a_\text{peak}\sim\unit[10]{nm}$ and $\sigma\sim0.8$ in the case of a progenitor mass of $\unit[20]{M_\odot}$. Thermal and  kinematic sputtering were considered as dust destruction processes. Depending on the density of the surrounding ISM ($n_\text{ISM}=\unit[0.1-10]{cm^{-3}}$), they found a survival rate of between 10 and $\unit[85]{\%}$ for the carbon dust component, and $\unit[1-61]{\%}$ for the total dust mass. In particular, grains with initial sizes below $\unit[50]{nm}$ were completely destroyed by sputtering in the post-shock gas. Considering our sputtering only case (Fig.~\ref{results_contrib_only}), with $a_\text{peak}=\unit[10]{nm}$ and $\sigma=0.85$, we find in our simulations a carbon dust survival rate of $\eta=\unit[43]{\%}$. Our results depend strongly on the width of the initial distribution. Slightly broader distributions ($\sigma=1.0$) result in a survival rate of $\eta=\unit[74]{\%}$, while only $\unit[15]{\%}$ of the initial dust material survives for $\sigma=0.7$. Taking into account the clump-free model in \cite{Nozawa2007} and the uncertainty in the reverse shock velocity due to the variation of the ISM density, our values broadly match the dust survival rates of \cite{Nozawa2007}. 

\cite{Nath2008} investigated the dust destruction rate for a 1D, clump-free, analytical evolution model for a SN with explosion energy $\unit[1\times10^{51}]{erg}$. Considering only thermal sputtering they found survival rates of $\eta=\unit[80-99]{\%}$ for carbon and silicate dust. These relatively high survival rates are a consequence of the disregard of further sputtering of grains in the hot plasma between the forward and reverse shock, the neglect of dust motions and kinematic sputtering, the assumption of a gas with solar abundances, and the use of a maximum grain size of $\unit[300]{nm}$ for a power-law distribution for which most of the mass is in the form of large dust grains (which are more robust against sputtering). 

\cite{Silvia2010} conducted 3D hydrodynamical simulations of the cloud-crushing scenario, for different shock velocities and density contrasts between the clump gas and the gas in the inter-clump medium, and evaluated the corresponding dust destruction in a post-processing routine. The dust was directly coupled to the gas (no drag), which is why only thermal sputtering and not kinematic sputtering or grain-grain collisions were considered. Similarly to the study of \cite{Nozawa2007}, the initial grain size distribution was adopted from \cite{Nozawa2003} for a progenitor with mass $\unit[20]{M_\odot}$ ($a_\text{peak}\sim \unit[10]{nm}, \sigma\sim0.8$). For $\chi=100$ and $v_\text{sh}=\unit[1000]{km/s}$ ($\unit[3000]{km/s}$), they found $\eta=\unit[96]{\%}$ ($\unit[95]{\%}$) for the survival rate of carbon dust. Considering our sputtering only case (thermal and kinematic; Fig.~\ref{results_contrib_only}), with $a_\text{peak}=\unit[10]{nm}$ and $\sigma=0.85$, we get from our simulations a carbon survival rate of $\eta=\unit[43]{\%}$. Slightly narrower or broader distributions result in survival rates of $\eta=\unit[15-74]{\%}$. Taking into account the different dust processes (e.g. no kinematic sputtering, size-dependent dust drag or grain charging) and a much lower gas molecular weight ($\sim$~solar abundance), we discern no clear disagreements between our results and those of \cite{Silvia2010}, although the results for the models that are best suited for comparison diverge.  \cite{Silvia2012} considered higher metallicity cases but found no significant deviations if the shock velocity was $\unit[\le3000]{km/s}$.

While the above studies treated the ejecta of general SNe, the following focussed on Cas~A in detail:

\cite{Nozawa2010} applied the method of \cite{Nozawa2003} to  model dust formation in the ejecta of a Type IIb SN. Compared to a Type~II-P SN, the gas densities are lower which causes the formation of smaller dust grains with average radii below $\unit[10]{nm}$ for carbon and silicates. Following the method of \cite{Nozawa2007}, dust destruction rates were determined on the basis of 1D~hydrodynamic models simulating a clump-free ejecta. The reverse shock velocity is again a function of the density of the circumstellar medium. For all scenarios, the dust was completely destroyed ($\eta<\unit[0.1]{\%}$) as the relatively small grains are easily sputtered in the clump-free ejecta.

\cite{Biscaro2016} modelled kinematic sputtering within over-dense clumps in the ejecta of a \mbox{Type II-n} SN in its remnant phase. Further, they considered thermal sputtering in the inter-clump medium after destruction of the clump by the reverse shock. The initial grain size distribution was adopted from \cite{Biscaro2014} where the carbon grains peak at $a_\text{peak}\sim\unit[0.9]{nm}$ ($\sigma\sim0.25$) and the silicate grains at  $a_\text{peak}\sim\unit[2]{nm}$ ($\sigma\sim0.25$). These small grains are very easily sputtered. The silicate dust was completely destroyed while a population of very small carbon grains $(\unit[0.4-0.8]{nm}$) could survive the  passage of the reverse shock. Considering further dust materials such as alumina (Al${}_2$O$_3$) and silicon carbide (SiC), a total fraction of $\eta\sim\unit[6-11]{\%}$ of the dust mass could survive. Their high destruction rates for silicate and carbon dust are matched by our results when we assume initial dust grain sizes below $\unit[10]{nm}$. \cite{Biscaro2016} also investigated dust destruction in Type II-P SNe as well as for a model for SN1987A (\citealt{Sarangi2015}), both with higher gas densities and thus larger initial grain sizes. Coupled with high over-densities ($\chi\ge1000$) in the ejecta, they found significantly higher dust survival rates of $\eta\sim\unit[14-45]{\%}$ and $\sim\unit[42-98]{\%}$ of the total initial dust mass, respectively.

\cite{Micelotta2016} generalized the analytical model of \cite{Truelove1999} for the dynamics of Cas~A. For an explosion energy of $\unit[2\times10^{51}]{erg}$  and an ejecta mass of $\unit[2]{M_\odot}$  they reproduced the dynamics and the  evolution of the density and temperature within the ejecta of Cas~A. Over-dense clumps ($\chi=100$) were added to the ejecta and it was assumed that they do not affect the dynamics of the reverse shock in the inter-clump medium. Silicates and amorphous carbon were adopted as dust components, initially following a MRN power-law distribution with $a_\text{min}=\unit[5]{nm}$ and $a_\text{max}=\unit[250]{nm}$. Kinematic sputtering by pure oxygen gas then eroded the dust grains within the clump while thermal sputtering was limited to the phase when the dust grains are ejected into the hot post-shock gas of the ambient medium. Neglecting grain-grain collisions, they found a survival rate of $\eta\approx\unit[13-17]{\%}$ for carbon dust and $\unit[10-13]{\%}$ for the silicate component. A comparison to our study is difficult as their initial size-distribution followed a MRN distribution. We investigate a power-law distribution in Appendix~\ref{app_powerlaw}  and find for the MRN grain sizes a carbon survival rate below $\unit[1]{\%}$, however, grain-grain collisions are considered which contribute significantly to the destruction of the larger grains. Considering sputtering only, we find for the log-normal distribution a carbon dust survival rate of $\unit[63]{\%}$ for grains of initial peak radius $a_\text{peak}=\unit[250]{nm}$, while the survival rate is substantially lower for smaller grain radii (e.g. $\unit[5]{\%}$ for $\unit[50]{nm}$ grains). Therefore, our simulations constrain the carbon survival rate to $\unit[0-63]{\%}$ if the conditions given in the study of  \cite{Micelotta2016} are taken into account, and we discern no disagreement between their results and ours.
 
Finally, the study of \cite{Bocchio2016} is to our knowledge the only previous ejecta dust study that simultaneously considered sputtering and grain-grain collisions. They extended the semi-analytical model of \cite{Bianchi2007} by including the full dynamics of dust grains within the ejecta of Cas~A.  Vaporisation and fragmentation processes were implemented following the treatment of \cite{Jones1994, Jones1996}. Their applied dust-formation model (\citealt{Marassi2015}) resulted in initial grain sizes that were significantly larger compared to previous studies, with log-normal distributions for carbon ($a_\text{peak}\unit[\sim 120]{nm}, \sigma\sim0.3$) and silicate dust ($a_\text{peak}\unit[\sim 50]{nm}, \sigma\sim0.35$). Since their models were clump-free, the dust grain number densities were low and grain-grain collisions were rare events and contributed little to the dust destruction. However, they found a survival rate of $\eta\approx\unit[1]{\%}$ of the total dust mass, whereby mainly carbon dust had a survival rate of $\unit[9]{\%}$ while the silicate dust components were completely destroyed. Compared to our clumped study we would predict a carbon survival rate of  $\eta=\unit[40]{\%}$ if only sputtering is considered, but $\unit[6]{\%}$ if sputtering and grain-grain collisions are considered together. The differences can be explained by the presence of clumps: In the case of sputtering only, the grains are sheltered in the clumps from the high gas velocities caused by the shock and from the high gas temperatures in the inter-clump medium, reducing the kinematic and thermal sputtering rates and thus increasing the survival rates of the dust material. On the other hand, if grain-grain collisions are taken into account the over-dense clumps increase the grain number density and thus the collision probability, which decreases the dust survival rate.

In summary, a direct comparison of our results with previous studies is complicated by the different approaches that have been used (numerical/hydrodynamical or (semi-)analytical), the diverse morphologies of the ejecta that have been considered (clumpy or smooth, evolutionary or static) and the various dust physics implemented (from gas drag to grain-grain collisions). A significant difference is present in studies with clumps: firstly, clumpy ejecta generally tend to form larger dust grains compared to grains in smooth ejecta, as a consequence of their higher gas densities. Considering only sputtering, such grains are harder to destroy. Secondly, the dust is protected in clumps from the high gas velocities arising from the reverse shock passage, mitigating the efficiency of kinematic sputtering. Thirdly, the grains are not exposed to the harsh conditions in the post-shock ambient gas, reducing the efficiency of thermal sputtering. Therefore, the presence of clumps significantly reduces dust destruction rates if only sputtering processes are considered. On the other hand, we have seen that grain-grain collisions can destroy significant amounts of dust for the case of large grains in high-density clumps.

The present study is the first work to consider grain-grain collisions in clumpy ejecta, and also the first to treat gas and plasma drag, kinematic sputtering and further dust processing such as gas accretion or grain-grain sticking as part of a hydrodynamical simulation of the reverse shock in a SNR.

\section{Conclusions}
\label{sec_conc}
We have investigated the effects of a range of clump densities on dust survival rates during passage through the reverse shock in Cas~A. For this purpose, we have developed the dust post-processing code \textsc{Paperboats} to calculate dust advection and dust destruction in the ejecta of the SNR based on the output of the hydrodynamical simulations using \textsc{AstroBEAR}. We summarise here the code description as well as the results of our dust destruction simulations for Cas~A.

\subsection{Paperboats}
\textsc{\mbox{Paperboats}} is a post-processing code to calculate the dust destruction and dust growth in a streaming gas. The dust is accelerated by the moving gas via collisional and plasma drag. The calculation of the grain charge is performed for a moving grain in an ionised gas with respect to impinging electrons and ions, secondary electron emission,
transmission/tunnelling effects, and field emission. 

Dust destruction and dust growth processes occur in the form of sputtering and grain-grain collisions. For the sputtering, thermal and kinematic sputtering are considered, as well as a size-dependent factor to correct the sputtering yields of semi-infinite targets. The penetration depths of ions into the dust material are calculated using the Bethe-Bloch formalism. Gas accretion onto the dust grains is realised in the form of ``negative'' sputtering and Coulomb interaction between charged grains and the ionised gas are considered. 

Along with thermal and kinematic sputtering, grain-grain collisions are also considered as a major component of the dust-processing. Collisions occur due to the relative velocities between grains of different sizes, caused by the size-dependent gas and plasma drag.  The collisions are calculated assuming a homogeneous distribution of dust grains with an isotropic velocity field for each grain size, dust material, cell, and time point, respectively. Depending on the collision energy, the dust grains can vaporise, shatter into smaller fragments, bounce, or stick together.  Coulomb interactions between charged grains as well as between the ionised gas and charged grains have an effect on the collision or impact velocity as well as on the collision cross sections, and are taken into account.

Using the described formalisms, we are able to track with time the spatial distribution of the dust grain density, for each dust grain size and dust material. This allows us to follow the evolution of the grain size distribution and particularly the total dust mass, as well as to investigate the enrichment of metals in the gas due to the destruction of dust grains. In general, the dust survival rate in various CCSNe remnants can potentially be determined by adjusting the shock velocity, the gas and dust properties, the gas-to-dust mass ratio and the clump size, as well as the gas density and temperature in the clump and the ambient medium.

\subsection{Results for dust destruction in Cas~A}
In order to examine the dust survival rate in Cas~A, we have simulated the impact of the reverse shock on an oxygen-rich, cooled clump of gas and dust embedded in a low-density ambient medium of gas. We find that dust survival rates strongly depend on the grain sizes and the widths of the initial grain size distributions, as well as on the gas density contrast between the clump and the ambient (inter-clump) medium. Density contrasts between 100 and 1000 have been investigated. 

Low and medium gas density contrasts ($\chi<600$) tend to preserve carbon dust material if the initial grain sizes are around $\sim \unit[0.5-1.5]{\text{\textmu}m}$, while large density contrasts \mbox{($\chi\gtrsim600$)} enable distributions with initial grain sizes around \mbox{$\sim \unit[10-50]{nm}$} to survive. We find the highest dust survival rates (up to $\eta=\unit[30]{\%}$) for narrow initial size distributions with grain radii around $\unit[20]{nm}$ radius (density contrast $\chi= 1000$) or $\unit[1]{\text{\textmu}m}$ \mbox{($\chi= 300$).} 

Silicate grains with initial radii around $\unit[100]{nm}$ show survival rates of up to $\eta=\unit[9]{\%}$ for low gas density contrasts \mbox{($\chi\lesssim200$)}. Medium and high density contrasts \mbox{($\chi\gtrsim200$)} enable silicate distributions with initial radii around \mbox{$\unit[\sim10-30]{nm}$} to survive the reverse shock with a surviving fraction of up to $\unit[40]{\%}$.

For both silicate and carbon grains, an enhanced survival rate exists for low density contrasts ($\chi\sim100$) and initial grain sizes of a few micrometre. The enhancement can be explained by the low number density of these grain sizes and in this environment, which mitigates the importance of grain-grain collisions, as well as by the negligible impact of sputtering for large grains.

We find that grain-grain collisions are crucial for dust destruction by the reverse shock and have to be taken into account. Moreover, sputtering and grain-grain collisions are synergistic. The surviving dust material is rearranged into a new size distribution that can be approximated by two components: a power-law size distribution of small grains and a log-normal distribution of grains with the same size range as the initial distribution. The rate of dust growth by gas accretion or grain sticking is very low which is a consequence of the high velocities occurring in our simulations. 

Dust formation theories favour the formation of dust grains of radii $\unit[\sim1-100]{nm}$ in the ejecta of SNe (e.g. \citealt{Nozawa2003}). When the density contrast between clump and ambient medium is of the order of \mbox{$\sim600-1000$,} carbon grains of that size show a relatively high survival rate and should be able to contribute to the dust budget of the ISM. For silicate grains, even lower gas density contrasts \mbox{($\chi\gtrsim200$)} enable the survival of a significant fraction of the dust mass. However, several observational studies have indicated the presence of dust grains in the micrometre size range. Our study indicates that if such large grains of carbonaceous material are able to form in the SN~ejecta, some of them are able to survive in clumps with density contrasts of \mbox{$\sim100-400$}. In contrast, silicate material having initial distributions with grain sizes around $\unit[1]{\text{\textmu} m}$ is completely destroyed.  
 
\subsection{Outlook}
We are able to follow the temporal and spatial density distribution of dust grains of different sizes during the destruction of an over-dense clump by a reverse shock. We will focus on this evolution in a future work and investigate the potential impact of 3D simulations instead of the current  2D simulations. Moreover, the presence of magnetic fields has been proven in SNRs, which will affect the dust trajectories of charged grains. In particular, the gyro-motions of charged grains due to betatron acceleration will change the frequency of collisions. We intend to make \textsc{\mbox{Paperboats}} available in the public domain after corresponding development. In a future study we will implement the dust-processing directly into the magneto-hydrodynamical code \textsc{AstroBEAR} in order to increase the accuracy of our modelling, in particular to increase the spatial and temporal resolution in order to investigate small-scale and feedback effects.


\section*{Acknowledgements}
We would like to thank Chris L. Fryer, Chris M. Mauney and Chris M. Malone from the Center for Theoretical Astrophysics at the Los Alamos National Laboratory for fruitful discussions. FK, FS, MJB, AB and FP were supported by European Research Council Grant SNDUST ERC-2015-AdG-694520. FP acknowledges funding by the STFC. This work was in part supported by the US~Department of Energy through the Los Alamos National Laboratory. Los Alamos National Laboratory is operated by Triad National Security, LLC, for the National
Nuclear Security Adminstration (Contract No.~89233218CNA000001).\\
Simulations were performed using the UCL HPC RC cluster {\textsc{GRACE}} and the data intensive {\textsc{Peta4-Skylake}} service at Cambridge. {\textsc{Peta4-Skylake}} usage is supported through DiRAC project ACSP190 (SNDUST) using the Cambridge Service for Data Driven Discovery (CSD3), part of which is operated by the University of Cambridge Research Computing on behalf of the STFC DiRAC HPC Facility (\href{www.dirac.ac.uk}{www.dirac.ac.uk}). The DiRAC component of CSD3 was funded by BEIS capital funding via STFC capital grants ST/P002307/1 and ST/R002452/1 and STFC operations grant ST/R00689X/1. DiRAC is part of the National \mbox{e-Infrastructure.} We thank Clare Jenner (UCL), Lydia Heck (Durham University), UCL RC support, and Cambridge HPCS support for their assistance.

  \appendix
  \setcounter{secnumdepth}{+2}
  \setcounter{section}{0} 
  \renewcommand\thesection{\Alph{section}}

   
\section{Grain potential in an ionised gas} 
\label{app_grain_pot}
Here, we summarise the fitting function for the grain potential $\Phi_\text{total}$, for full details we refer to \cite{Fry2018}.

The total grain potential $\Phi_\text{total}$ is the sum of six potentials $\Phi_\text{x}$ which correspond to different charging regimes  and which are weighted by six scaling functions $w_i, i\in\mathbb{N}_{\le6}$,
\begin{align}
 \Phi_\text{total} &= \left[\Phi_\text{imp} \left(1-w_2\right) + \Phi_\text{sta} + \Phi_\text{se1} w_2 + \Phi_\text{se2}w_2w_3 \right] \times\nonumber\\ 
 & w_1w_6\left(1-w_4\right) + \Phi_\text{tra}w_4w_5 + \Phi_\text{the} w_2(1-w_4)(1-w_6),\\
 &\text{where}\nonumber\\
 \Phi_\text{imp} & = - 0.084 + 1.112\times10^{-3}\, v_7^2 + \left(\frac{T_\text{rel}}{T_5}\right)^{0.75},\\
 \Phi_\text{sta} & = 1-\sqrt{\frac{m_\text{ion}}{m_\text{e}}}\exp{\left[\Phi_\text{sta}\right]},\\
 \Phi_\text{se1} & = 1.74\,\left(1-\exp{\left[-0.1037 v_7^2\right]}\right)+1.005,\\
 \Phi_\text{se2} & = \max{\left(0,-0.2267v_7^2+1.43\right)},\\ 
 \Phi_\text{tra} & =0.1953\,T_5^{\,-0.162} ,\\
 \Phi_\text{the} & = 0.1862 \ln \left[T_5\right]-1.756,\\
 &\text{and}\nonumber\\
 w_1 &=  \left(1.0+\left(T_\text{imp}/T_5\right)^{36.99}\right)^{-1} \Theta\!\left(T_\text{tra}-T_\text{imp}\right),\\
 w_2 &= \left(1.0+ \left(T_\text{se1}/T_5\right)^{38.48}\right)^{-1},\\
 w_3 &= \left(1.0+ \left(T_\text{se2}/T_5\right)^{(1.563\,+\,0.3545\,\ln{\left[v_7\right]})}\right)^{-1},\\
 w_4 &= \exp{\left[-\left(\frac{T_\text{cri}}{T_5}\right)^4\right]},\\
w_5 &= \exp{\left[-\left(\frac{a}{10\,\lambda_\text{esc}}\right)^4\right]},\\
 w_6 &=  \exp{\left[-\left(\frac{T_5}{T_\text{the}}\right)^4\right]},\\
&\text{with}\nonumber\\
 v_7 &=       v_\text{rel}/\left(\unit[10^7]{cm/s}\right),\\
 T_5 &=       T_\text{gas}/\left(\unit[10^5]{K}\right),\\ 
 T_\text{rel} &=0.2506\,v_7^2,\\
 T_\text{imp} &=0.3433\,v_7^2,\\
 T_\text{tra} &=10.57\left(1.0-\exp{\left[-\left(\frac{\lambda_\text{esc}}{a}\right)^{0.75}\right]}\right)^{-1},\\  
 T_\text{se1} &=3.404\,v_7^2,\\ 
 T_\text{se2} &=34.82\,v_7^{1.223},\\ 
 T_\text{cri} &=\max{\left(T_\text{tra}, T_\text{imp}\right)},\\  
 T_\text{the} &=\max{\left(703.8,9.964\,v_7^2\right)},  
 \end{align}
 and $m_\text{ion}, m_\text{e}$ and $\Theta$ are the ion and electron mass and the Heaviside step function, respectively. Following \cite{Fry2018}, the escape length for electrons is\break $\lambda_\text{esc}=\,R_\text{e}(E_\text{max})/\left(R_\text{m}(\mathcal{L}_\text{e})\right)^{\mathcal{L}_\text{e}}$, with  $R_\text{e}(E_\text{max}) = \tilde{R}\left(\frac{E_\text{max}}{\unit[1]{keV}}\right)^\beta$ as the stopping range of electrons at energy $E_\text{max}=\unit[0.4]{keV}$. For the dust material\footnote{We selected Fe$_2$O$_3$ as a substitute for silicate from a list comprising Fe, FeO, Fe$_2$O$_3$ and Fe$_3$O$_4$, for which the parameters were derived using the CASINO software (\citealt{Drouin2007}). However, the differences between these four materials for $\mathcal{L}_\text{e}, R_\text{m}(\mathcal{L}_\text{e}), \tilde{R}$ and $\beta$  are small.}
 Fe$_2$O$_3$, the escape length parameters are $\mathcal{L}_\text{e}=1.5935$ and $R_\text{m}(\mathcal{L}_\text{e}) =1.1611$, and the stopping range parameters of electrons are $\tilde{R}=\unit[7.1477]{nm}$ and $\beta = \mathcal{L}_\text{e}=1.5935$, which gives the escape length for electrons as $\lambda_\text{esc}=\unit[1.308]{nm}$.
 
 To account for field emission, $\Phi_\text{total}$ for a grain of radius $a$ is limited by (\citealt{McKee1987})
 \begin{align}
 &\Phi_\text{min} \le \Phi_\text{total} \le  \Phi_\text{max},\\[0.1cm]
 &\phantom{phim}\text{where}\nonumber\\
 &\Phi_\text{min} = -\frac{116}{T_5}\,\left(\frac{a}{\unit[1]{\text{\textmu} m}}\right),\\
 &\Phi_\text{max} = \frac{3480}{T_5}\,\left(\frac{a}{\unit[1]{\text{\textmu} m}}\right). 
 \end{align}  

 
\section{The impact of grain charge on grain-grain collisions} 
\label{gg_ec}
Grain charging causes a repulsion or attraction of the grains during a grain-grain collision. This additional force has an impact on the collision velocity and cross section and the actual values are calculated here.


\subsection{Collision velocity}
At large distances ($r=\infty$), the Coulomb force between two grains $i$ and $j$ with charges $Q_i$ and $Q_j$, respectively, can be ignored. The velocity difference between $i$ and $j$ is \mbox{$|\mathbf{v}_{\text{dust},i} - \mathbf{v}_{\text{dust},j}|$} and the energy of the reduced mass is just the kinetic energy 
\begin{align}
 E_{\text{kin},\infty}=\frac{1}{2}\frac{m_im_j}{m_i+m_j}|\mathbf{v}_{\text{dust},i} - \mathbf{v}_{\text{dust},j}|^2.\label{0901a}
\end{align}
When the two dust grains collide, their separation is $r=a_i+a_j$ and the potential energy is (in cgs-units)\footnote{In SI-units, equation~(\ref{0901}) would transform to
 $E_{\text{pot,col}}=\frac{1}{4\pi\epsilon_0}\frac{Q_i\,Q_j}{a_i+a_j}$.}:
\begin{align}
E_{\text{pot,col}}=\frac{Q_i\,Q_j}{a_i+a_j}.\label{0901}
\end{align}
We assume here, that the grain charges are located in the centres of the dust grains. The kinetic energy at the moment of the collision is
\begin{align}
E_{\text{kin,col}}&=\frac{1}{2}\frac{m_im_j}{m_i+m_j}v_\text{col}^2.\label{0901b} 
\end{align}
As a consequence of energy conservation, and after dividing by $ E_{\text{kin},\infty}$, it follows that:
\begin{align}
 1= E_{\text{pot,col}}/ E_{\text{kin},\infty} +  \left(v_\text{col}/|\mathbf{v}_{\text{dust},i} - \mathbf{v}_{\text{dust},j}|\right)^2.
\end{align}
Introducing 
\begin{align}           
\alpha_q=E_{\text{pot,col}}/ E_{\text{kin},\infty} = \frac{2\,Q_i\,Q_j\,(m_i+m_j)}{(a_i+a_j)m_im_j|\mathbf{v}_{\text{dust},i} - \mathbf{v}_{\text{dust},j}|^2},
 \end{align} we get:
\begin{align}
v_\text{col} = (1-\alpha_q)^{0.5}|\mathbf{v}_{\text{dust},i} - \mathbf{v}_{\text{dust},j}|.\label{vcol}
\end{align}
The dust grains collide if $\alpha_q<1$, otherwise the charge repulsion is so large that a collision is prevented.


\subsection{Collision cross section}
 \begin{figure}
 \centering
 \fbox{\includegraphics[trim=1.5cm 2.1cm 0cm 9.9cm, clip=true,page=1,width=1.0\linewidth]{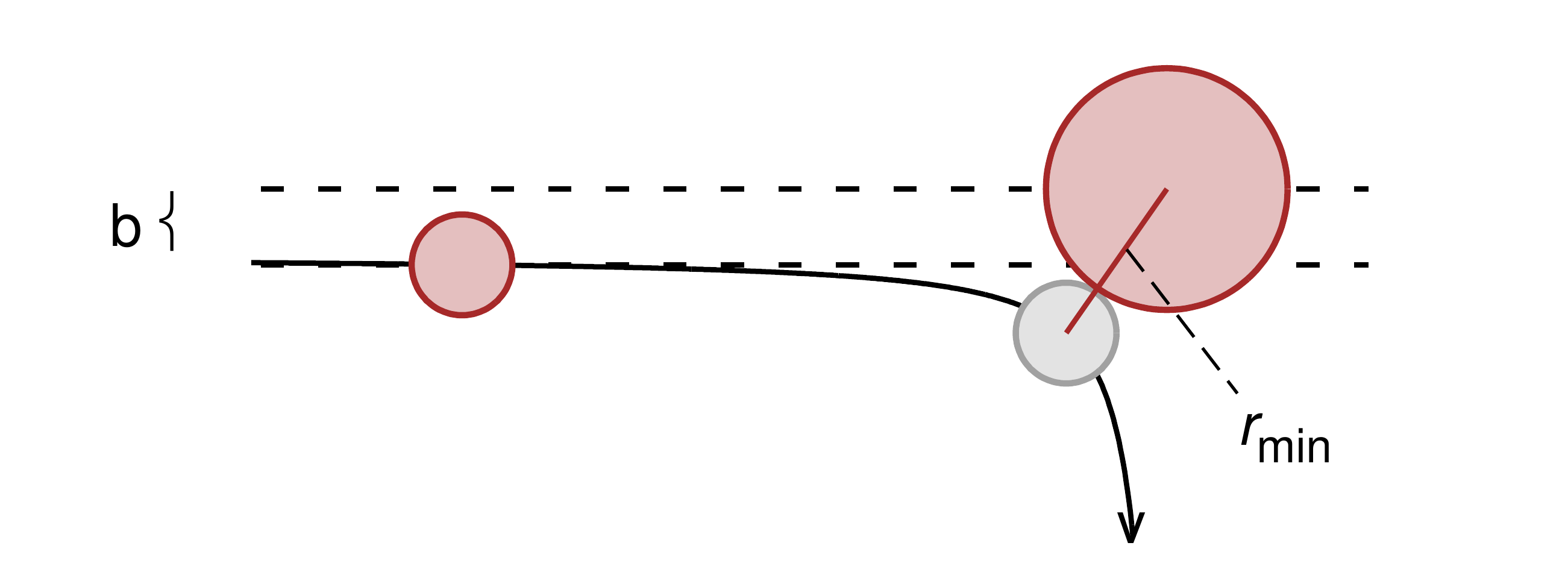}}
\caption{Depiction of the collision of charged grains and in particular the two scattering parameter $b$ and $r_\text{min}$. Here, the grains are both positively or negatively charged, which causes a repulsion, and $\sigma_\text{col}=\pi b^2<\pi r^2_\text{min}$.}
\label{fig_Rutherford} 
\end{figure}
The collision of charged grains $i$ and $j$ is related to Rutherford's scattering experiment in which the particles are elastically scattered by the Coulomb interaction (\citealt{Rutherford1911}). For the scenario in which the two charged grains just touch each other, one can derive\footnote{see e.g. \href{https://physics.stackexchange.com/questions/208304/minimum-hyperbolic-distance-for-rutherford-scattering}{https://physics.stackexchange.com/questions/208304/ minimum-hyperbolic-distance-for-rutherford-scattering}.}
\begin{align}
 \frac{1}{2}\frac{m_i m_j}{m_i+m_j}v^2_\text{min}= E_\text{tot} \frac{b^2}{r^2_\text{min}}.\label{aint}
\end{align}
Here, $v_\text{min}$ is the velocity at the minimum distance \mbox{$r_\text{min}=a_i+a_j$,} which is just the distance at the collision, and thus $v_\text{min}=v_\text{col}$. $E_\text{tot}$ is the total energy, which is equal to the kinetic energy at large distances (equation~\ref{0901a}) and $b$ is the distance between a grain path for a non-central collision from a grain path for a central collision (Fig.~\ref{fig_Rutherford}). Furthermore, $b$ defines the cross section of the collision, if the grain charges are considered: $\sigma_\text{col}=\pi b^2$. It follows from equation~(\ref{aint}) that:
\begin{align}
& \frac{1}{2}\frac{m_i m_j}{m_i+m_j}v^2_\text{col}= \frac{1}{2}\frac{m_i m_j}{m_i+m_j}|\mathbf{v}_{\text{dust},i} - \mathbf{v}_{\text{dust},j}|^2 \frac{b^2}{\left(a_i+a_j\right)^2},\\
&\text{and hence}\nonumber\\
& \left(\frac{v_\text{col}}{|\mathbf{v}_{\text{dust},i} - \mathbf{v}_{\text{dust},j}|}\right)^2 = \frac{b^2}{\left(a_i+a_j\right)^2}.\label{aint2}\\
& \text{Combining equations~(\ref{vcol}) and~(\ref{aint2}), we get}\nonumber\\
&1-\alpha_q = \frac{\sigma_\text{col}}{\pi\left(a_i+a_j \right)^2}, \, \text{and finally}\\
&\sigma_\text{col} = \left(1-\alpha_q\right)\pi\left(a_i+a_j \right)^2.
\end{align}


\section{Fragmentation theory} 
\label{sec_frag_app}

We follow the fragmentation description of \cite{Tielens1994}, \cite{Jones1996} and \cite{Hirashita2009}. As for vaporisation, the collisional outcome is evaluated for grain $i$ only; fragmentation of grain $j$ will be considered if $i$ and $j$ are exchanged. Two cases can be distinguished then:

\subsubsection*{I. Target $a_i$, projectile $a_j$}
At first, the case $a_i\ge a_j$ is considered. The mass of dust grain $i$ that is shocked to the critical pressure for fragmentation by a collision with $j$ is given by
\begin{align}
M_\text{shocked} = \frac{m_j}{2\sigma_\text{l}^{8/9} \sigma_{r}^{1/9}} \frac{1+2\mathcal{R}}{(1+\mathcal{R})^{9/16}} \left(\frac{\mathcal{M}_\text{r}}{\mathcal{M}_\text{l}}\right)^{16/9}.\label{fragmshock}
\end{align}
Here, 
\begin{align}
 &\mathcal{R} = \sqrt\frac{s_i \rho_{\text{bulk,}i}}{s_j \rho_{\text{bulk,}j}}
 \end{align}
 is a quantity determined by the ratio of the dimensionless material constants $s_i$ and $s_j$ that give the relation between the shock velocity and the velocity of the shocked material in grain $i$ and $j$, respectively. As collisions of different dust types are allowed, the bulk densities $\rho_{\text{bulk},i}$ and $\rho_{\text{bulk},j}$ are distinguished. In addition, the terms $\mathcal{M}_\text{r}$ and $\mathcal{M}_\text{l}$\footnote{The indices `l' and `r' in $M_\text{r},M_\text{l}, \sigma_\text{r},\sigma_\text{l},P_\text{l}$ and $\Phi_\text{l}$ correspond to the notation in \cite{Jones1996} and \cite{Hirashita2009}.} are defined as 
 \begin{align}
 &\mathcal{M}_\text{r}= \frac{v_\text{col}}{c_{0,i}} 
 \end{align}
 is the Mach number of the collision velocity $v_\text{col}$ corresponding to the speed of sound $c_{0,i}$ in the material, while
 \begin{align}
 &\mathcal{M}_\text{l}=\frac{2\Phi_\text{l}}{1+(1+4s_i \Phi_\text{l})^{1/2}}
  \end{align}
 is the Mach number corresponding to the critical pressure $P_{\text{l},i}$ of the material with the dimensionless quantity
 \begin{align}
 &\Phi_\text{l} = \frac{P_{\text{l},i}}{\rho_{\text{bulk,}i} c_{0,i}^2}.
 \end{align}
  $\sigma_\text{l}$ and $\sigma_\text{r}$ are Mach number related quantities,\\   
 \begin{align} 
 &\sigma_\text{l} = \frac{0.3 (s_i +1/\mathcal{M}_\text{l} - 0.11)^{1.3}}{s_i+1/\mathcal{M}_\text{l} - 1},\\
 &\sigma_\text{r} = \frac{0.3 (s_i +(1+\mathcal{R})/\mathcal{M}_\text{r} - 0.11)^{1.3}}{s_i+(1+\mathcal{R})/\mathcal{M}_\text{r} - 1}.\label{fragsigmar}
 \end{align}
 The parameters $s_i, c_{0,i}$ and $P_{\text{l},i}$ are listed in Table~\ref{mat_para} for carbon and silicate materials.
 
It is assumed that if more than half of the target mass is shocked the entire target is shattered (total fragmentation), and otherwise only a fraction of the shocked material mass is shattered (partial fragmentation, including cratering). The finally ejected and fragmented mass from grain $i$ is then
 \begin{align}
  m_\text{frag}= \begin{cases}
           0.4 M_\text{shocked}\hspace{0.3cm}\text{if}\hspace{0.1cm}M_\text{shocked} \le 0.5\,m_i,\\
           m_i\hspace{1.63cm}\text{if}\hspace{0.1cm}M_\text{shocked} > 0.5\,m_i.
          \end{cases}
 \end{align}
For the sake of simplicity, the fragmented mass is assumed to follow a size distribution of compact, spherical grains,
 \begin{align}
  n_\text{frag}\,\text{d}a= C_\text{frag}a^{-\gamma_\text{frag}}\,\text{d}a.\label{fragdistrib}
 \end{align}
The grain size exponent $\gamma_\text{frag}$ commonly takes a value between $2$ and $4$ (e.g.\,\citealt{Dohnanyi1969,Jones1996}) and we set $\gamma_\text{frag}=3.5$. The normalization factor $C_\text{frag}$ is determined by the fragmented mass,
\begin{align}
C_\text{frag}=\frac{m_\text{frag}}{\int_{a_\text{min,frag}}^{a_\text{max,frag}} 4/3\pi\rho_{\text{bulk},i} a^{3-\gamma_\text{frag}}\,\text{d}a}, \label{fragcfrag}                                                                                                                                                                                                                                   
\end{align}
where $a_\text{min,frag}$ and $a_\text{max,frag}$ denote the minimum and maximum radius of the size distribution of the fragments, given by
\begin{align}
 a_\text{max,frag} = 
 \begin{cases}
 0.168363\,\left(\frac{m_\text{frag}}{\rho_{\text{bulk},i}}\right)^{1/3}\hspace{0.3cm}\text{if}\hspace{0.1cm}M_\text{shocked} \le 0.5\,m_i,\\
0.22\,a_i \frac{c_{0,i}}{v_\text{col}} \left(\frac{m_i}{m_j}\right)^{9/16}\left(\frac{1+\mathcal{R}}{(1+2\,\mathcal{R})^{9/16} }  \right)\sigma_\text{l}^{1/2} \sigma_\text{r}^{1/16} \mathcal{M}_\text{l}\\\hspace*{3.26cm}\text{if}\hspace{0.1cm}M_\text{shocked} > 0.5\,m_i,
\end{cases}\label{maxfrag}
\end{align}
and
\begin{align}
a_\text{min,frag} = 
\begin{cases}
\left(\frac{P_{\text{l},i}}{P_{\text{v},i}}\right)^{1.47}a_\text{max,frag}\hspace{0.3cm}\text{if}\hspace{0.1cm}M_\text{shocked} \le 0.5\,m_i,\\ 0.03\,a_\text{max,frag}\hspace*{1.1cm}\text{if}\hspace{0.1cm}M_\text{shocked} > 0.5\,m_i.
 \end{cases}\label{minfrag}
\end{align}
Here, $P_{\text{v},i}$ is the critical pressure for vaporisation (Table~\ref{mat_para}). In addition to equations~(\ref{maxfrag}) and (\ref{minfrag}), a grain remnant is left if the fragmentation is partial ($M_\text{shocked} \le 0.5\,m_i$). The remnant grain is assumed to be spherical with radius  
\begin{align}
a_\text{rem}=\left(a_i^3-\frac{0.3}{\pi}\frac{M_\text{shocked}}{\rho_{\text{bulk},i}}\right)^{1/3}                                                                                                                                                                                                                                                                                                                                                                                                                                                                                                                                                                                                                \end{align}
 with respect to mass conservation.

If the fragmentation condition is fulfilled (equation~\ref{fragthre}), $n_\text{col}$ dust grains are removed from bin $i$.  Using the fragmentation size distribution from equation~(\ref{fragdistrib}) within  the boundaries $a_\text{min,frag}$ and $a_\text{max,frag}$ as well as considering the dust grain remnant $a_\text{rem}$ (if applicable), the fragments of the $n_\text{col}$ dust grains are placed in the corresponding bins with mass conservation taken into account (see Section~\ref{sec_ass_bins} for the description of assigning grains to a bin). 

If $a_\text{min,frag}<a_\text{min,abs}$, the fragmentation distribution is calculated as in equations~(\ref{fragdistrib}) and~(\ref{fragcfrag}) but only bins $i\ge1$ are filled up with the corresponding number of dust grains. The missing mass is assumed to be destroyed and transformed into the collector bin $0$ (dusty gas). If even $a_\text{max,frag}<a_\text{min,abs}$ (or $a_\text{rem}<a_\text{min,abs}$, if applicable), the whole fragmented mass is destroyed and removed from bin $i$, and the corresponding number of atoms/averaged atoms are placed in the collector bin.

\subsubsection*{II. Projectile $a_i$, target $a_j$}
 The second case is $a_i< a_j$ for which the whole projectile is assumed to fragment, $m_\text{frag}=m_i$. The fragments follow the same grain size distribution as in equations~(\ref{fragdistrib}) and (\ref{fragcfrag}) with minimum and maximum radius according to equations~(\ref{maxfrag}) and (\ref{minfrag}), respectively, based on the collision quantities of equations~(\ref{fragmshock})$-$(\ref{fragsigmar}). Note that  contrary to \cite{Hirashita2009} $i$ and $j$ do not have to be exchanged in these equations to consider this case.

\section{The impact of grain and gas charge on the sputtering process} 
\label{sp_ec}
For the collision of a gas particle of species $k$ with a dust grain, the energy equations are similar to equations~(\ref{0901}) \& (\ref{0901b}) for grain-grain collisions. However, one of the  dust grains is replaced here by a gas particle. The distance of the colliders at the moment of collision is the dust grain radius $a$, and the reduced mass in equations~(\ref{0901a}) \& (\ref{0901b}) is approximated by $m_{\text{gas},k}$ because $m_\text{grain}~\gg~m_{\text{gas},k}$. With $Q_\text{grain}$ as the grain charge, $z$ as the average charge number of the gas particles, and $v_\text{rel}$ and $v$ as the relative velocity between gas particle and dust grain at large distances and at the moment of collision, respectively, the energy equation (in SI-units) is given by
\begin{align}
 E = \frac{m_{\text{gas},k}}{2} v_\text{rel}^2 =\frac{m_{\text{gas},k}}{2} v^2 +\frac{z\,e\,Q_\text{grain}}{(4\pi\epsilon_0)\,a}. 
\end{align}
For cgs-units, the right hand term has to be multiplied by $(4\pi\epsilon_0)$.


\section{Dust survival for an initial power-law distribution} 
\label{app_powerlaw}
\begin{figure}
 \centering
  \includegraphics[trim=4.9cm 2.5cm 2.8cm 2.9cm, clip=true,height=5.9cm, page=1]{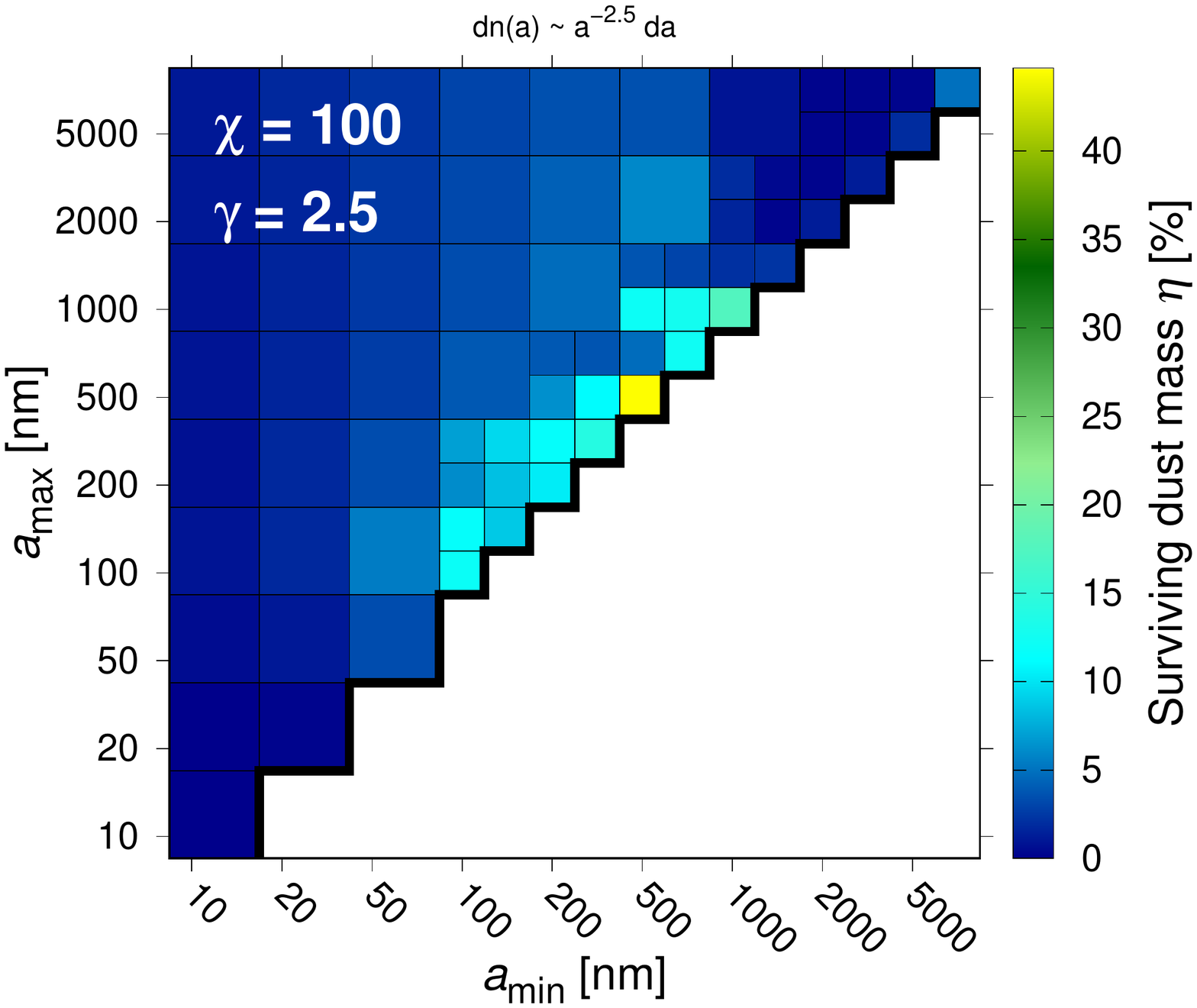}\\
   \includegraphics[trim=4.9cm 2.5cm 2.8cm 2.9cm, clip=true,height=5.9cm, page=2]{Pics/Results/Dust_survival_graphite_cold_2,5_3,5.pdf} 
   \caption{Same as Fig.~\ref{results2}, only for power-law grain size distributions with minimum and maximum grain size $a_\text{min}$ and $a_\text{max}$, respectively, and grain size exponent $\gamma=2.5$ (\textsl{top}) and $3.5$ (\textsl{bottom}).}
\label{results_powerlaw} 
\end{figure}

In Section~\ref{sec_results} we investigate the dust survival rates for log-normal initial grain size distributions. Although log-normal-distributions are favoured by dust formation theories (\citealt{Todini2001,Nozawa2003}), we consider here a power-law distribution as defined in equation~(\ref{120519}). Power-law distributions have been previously used in the studies of \cite{Nath2008} and \cite{Micelotta2016}. We vary the minimum and maximum grain size, $a_\text{min}$ and $a_\text{max}$, over a range of $\unit[10]{nm}-\unit[7]{\text{\textmu} m}$, respectively, with $a_\text{min}\le a_\text{max}$, and calculate the carbon dust survival rate $\eta$ for the grain size exponent $\gamma=2.5$ and $3.5$ (Fig.~\ref{results_powerlaw}). The gas density contrast between clump and ambient medium is set to $\chi=100$.

The highest dust survival rates occur again for narrow distributions ($a_\text{min}\sim a_\text{max}$) which are located close to the diagonal in Fig.~\ref{results_powerlaw}, where the diagonal ($a_\text{min}= a_\text{max}$) represents single grain size distributions. Most of the dust material is destroyed by sputtering if the grain sizes of the initial distribution are small ($a<\unit[100]{nm}$). Initial distributions with grain sizes larger than $\unit[\sim100]{nm}$ and in particular broad distributions are subject to grain-grain collisions. Consequently, initial size distributions with medium sized dust grains ($\unit[100-700]{nm}$) have the highest probability to survive the passage of the reverse shock. The largest survival fraction exists for the single grain size distribution with $a=\unit[500]{nm}$ ($\eta=\unit[44]{\%}$). In addition, single grain size distributions with $a>\unit[1]{\text{\textmu} m}$ show an increased dust survival rate due to the low number densities of these grains.

Compared to $\gamma=3.5$, grain size distributions with \mbox{$\gamma=2.5$} have slightly smaller chances to survive, as grain-grain collisions are more frequent due to the greater number of large grains for the flatter distribution, however, the differences in the survival rates are small. In summary, the survival rates of the power-law distributions peak at similar dust grain sizes as the rates for log-normal distributions ($\unit[150-1500]{nm}$; Fig.~\ref{results2}) and show the same trends.
 
  \bibliographystyle{mnras}
{\footnotesize
  \bibliography{Literature}
}
\label{lastpage}

\bsp	

\end{document}